\pdfoutput=1
\documentclass[mnsc,nonblindrev]{informs3_hide} 

\OneAndAHalfSpacedXI 

\usepackage{natbib}
 \bibpunct[, ]{(}{)}{,}{a}{}{,}%
 \def\bibfont{\small}%
\usepackage{color}
\usepackage{algorithm}
\usepackage{algorithmic}
\usepackage{setspace}
\usepackage{bm}
\usepackage{tabu}
\usepackage{tikz}
\newcommand{\Cross}{\mathbin{\tikz [x=1.4ex,y=1.4ex,line width=.2ex] \draw (0,0) -- (1,1) (0,1) -- (1,0);}}%

\usepackage{array}
\usepackage{multirow}
\usepackage{comment}
\usepackage[toc,page]{appendix}
\usepackage{appendix}
\usepackage{setspace}
\usepackage{hyperref}

\TheoremsNumberedThrough     
\ECRepeatTheorems

\EquationsNumberedThrough    

\newcommand{\mathbbm}[1]{\text{\usefont{U}{bbm}{m}{n}#1}}
\newcommand{\bI}{\mathbbm{1}}

\newcommand{\ALG}{\mathsf{ALG}}
\newcommand{\Dis}{\mathsf{Dis}}

\newcommand{\cS}{\mathcal{S}}

\newcommand{\sC}{\mathsf{C}}

\newcommand{\cP}{\mathcal{P}}
\newcommand{\bX}{\mathbb{X}}
\newcommand{\Var}{\mathrm{Var}}

\newcommand{\bE}{\mathrm{E}}

\begin{document}

\RUNAUTHOR{Zhao and Zhou}
\RUNTITLE{Pigeonhole Design}

\TITLE{Pigeonhole Design: Balancing Sequential Experiments from an Online Matching Perspective}

\ABSTRACT{
Practitioners and academics have long appreciated the benefits of covariate balancing when they conduct randomized experiments. For web-facing firms running online A/B tests, however, it still remains challenging in balancing covariate information when experimental subjects arrive sequentially. In this paper, we study an online experimental design problem, which we refer to as the ``Online Blocking Problem.'' In this problem, experimental subjects with heterogeneous covariate information arrive sequentially and must be immediately assigned into either the control or the treated group. The objective is to minimize the total discrepancy, which is defined as the minimum weight perfect matching between the two groups. To solve this problem, we propose a randomized design of experiment, which we refer to as the ``Pigeonhole Design.'' The pigeonhole design first partitions the covariate space into smaller spaces, which we refer to as pigeonholes, and then, when the experimental subjects arrive at each pigeonhole, balances the number of control and treated subjects for each pigeonhole. We analyze the theoretical performance of the pigeonhole design and show its effectiveness by comparing against two well-known benchmark designs: the match-pair design and the completely randomized design. We identify scenarios when the pigeonhole design demonstrates more benefits over the benchmark design. To conclude, we conduct extensive simulations using Yahoo! data to show a $10.2\%$ reduction in variance if we use the pigeonhole design to estimate the average treatment effect.
}


\ARTICLEAUTHORS{
\AUTHOR{Jinglong Zhao}\AFF{Boston University, Questrom School of Business, Boston, MA, 02215,
\EMAIL{jinglong@bu.edu}}
\AUTHOR{Zijie Zhou}\AFF{Massachusetts Institute of Technology, Operations Research Center, Cambridge, MA, 02139, \EMAIL{zhou98@mit.edu}}
}

\maketitle

\section{Introduction}

The design of online controlled experiments, or ``A/B tests,'' has long provided tremendous benefits to firms, especially in the technology sector \citep{bakshy2014designing, johari2021always, johari2022experimental, kohavi2007practical, kohavi2009online, kohavi2020trustworthy, li2021interference, lewis2015unfavorable, wager2021experimenting}.
In an A/B test, the experimenter compares the standard offering of some policy ``A'' to a new version of the policy ``B,'' by splitting subjects into the ``control'' and the ``treated'' groups.
By comparing the outcomes from these two groups, the experimenter discovers how much the new version is better or worse than the standard version, which is referred to as the ``treatment effect.''
Such a simple approach has been widely adopted by a variety of online web-facing companies, including ``search engines (e.g., Bing, Google, Yandex), online retailers (e.g., Amazon, eBay, Etsy), media service providers (e.g., Netflix), social networking services (e.g., Facebook, LinkedIn, Twitter), and travel services (e.g., Airbnb, Booking.com, Lyft, Uber)'' \citep{gupta2019top}.
A/B tests have brought tremendous value to the firms throughout their product development processes \citep{koning2019experimentation}.

To obtain trustworthy results, experimenters in an A/B test must properly handle heterogeneity.
Heterogeneity arises when the same treatment triggers different levels of effects across subjects with different covariates.
For example, a newer color of an online display advertisement may have a different effect on click through rates for junior people compared to senior people.
One method to address heterogeneity is through ``blocking,'' which is also referred to as ``stratification'' \citep{athey2017econometrics, chase1968efficiency, cochran1957experimental, cox1958planning, fisher1936design, imbens2015causal, imbens2011experimental, matts1988properties}.
Typically, in a block experiment, all subjects are partitioned into several different blocks, such that the covariates of different subjects are similar within each block.
Then, after all the blocks are fixed, the experimenter randomly assigns half of the subjects from each block to the control group, and the other half to the treated group.
This simple approach has been considered to be ``the gold standard in handling heterogeneity in randomized control trials'' \citep{rubin2008comment}, and has been extensively studied from an optimal experimental design perspective \citep{bai2022optimality, greevy2004optimal, harshaw2019balancing, higgins2016improving, imai2009essential, lu2011optimal, rosenbaum1989optimal}.

For online web-facing firms, however, such a simple block experiment is not always feasible.
In an online field experiment, the covariates of the subjects are not known in advance.
Instead, the covariates are sequentially revealed as subjects arrive in an online fashion.
Upon a subject's arrival, the experimenter must immediately decide if the subject is assigned to the control or the treated group based on this subject's covariate information, without knowing the covariate information of future arriving subjects.
It is the uncertainty of future subjects' covariate information that hinders the usage of the well-studied designs of block experiments.

In this paper, we study an online experimental design problem, which we refer to as the ``Online Blocking Problem.''
To solve this problem, we propose a randomized experimental design approach, which we refer to as the ``Pigeonhole Design.''
We analyze the theoretical performance of the pigeonhole design against two well-known benchmark designs of experiments and show the effectiveness of the pigeonhole design.

\subsection{The Online Blocking Problem}
In the online blocking problem, the experimenter is given in advance a fixed budget of experimental subjects.
During the experiment, a fixed number of subjects with unknown covariate information arrive one by one.
Upon the arrival of each subject, the experimenter must immediately assign each subject into either the control or the treated group.
The objective is to minimize the total discrepancy between the covariates of these two groups, which is defined as the size of a minimum weight perfect matching \citep[Chapter 7.8]{bertsimas1997introduction} between these two groups.
A smaller discrepancy leads to a more accurate estimation, e.g., it leads to a smaller ex-post bias in estimating the average treatment effect \citep{bai2022optimality, cytrynbaum2021designing}.

To the best of our knowledge, the online blocking problem is related to, but distinct from the following problems in the literature.
\begin{enumerate}
\item \textit{Online experimental design under $D_A$-optimality}.
This line of literature follows the work of \citet{atkinson1982optimum, atkinson1999optimum}.
Recently, \citet{bhat2020near} revisits this problem by articulating the tension between online algorithms and experimental design, which has greatly motivated our work.
In this problem, there are several prognostic factors that are assumed to impact the potential outcomes through a linear model, including the treatment status.
We refer to such prognostic factors as covariate information, and treatment status is one of them.
Subjects with unknown covariate information arrive one by one, and the covariate information is revealed upon the arrival of one subject.
The experimenter must immediately assign each subject into the control or treated group, which determines the value of a single binary indicator among the covariates.
The objective is the $D_A$-optimality, which refers to minimizing the variance of the Ordinary Least Squares (OLS) estimator on some certain covariates.

Our problem is closely related to the online $D_A$-optimal experimental design problem as well, with the following three distinctions.
First, the online $D_A$-optimal experimental design problem allows the number of subjects in the control and treated groups to be different whereas our online blocking problem requires the numbers to be the same.
Second, the objective functions are different.
The online $D_A$-optimal experimental design problem uses an OLS estimator to estimate the effects of treatment by postulating a linear dependence of the potential outcomes on the covariates. The objective is to minimize the variance of the OLS estimator on some certain covariates, \emph{e.g.}, on the treatment indicator.
In contrast, our objective is the size of the minimum weight perfect matching.
Third, the models of the arrival sequence are different.
The model in \citet{atkinson1982optimum} assumes that the covariate information is part of the decision making.
The model in \citet{bhat2020near} assumes a stochastic arrival sequence drawn from an elliptical distribution with finite second order moment.
An implication of choosing this model is that the optimal design will be deterministic, i.e., there is no randomization in the optimal design.
In contrast, we model the arrival sequence to be adversarial, and our design of experiment is a randomized design.
Our model has the flexibility to extend to the stochastic arrival setting, which potentially improves our results.
\item \textit{Online vector balancing}.
This line of literature follows the work of \citet{spencer1977balancing}.
Recent works have studied a variant of this problem under the stochastic arrival model \citep{alweiss2021discrepancy, bansal2020online, bansal2021online}.
In an online vector balancing problem, subjects with unknown vector information arrive one by one.
Upon arrival, the vector information is revealed, and the decision maker must immediately assign a positive or negative sign $\pm$.
The objective is to minimize the $L_\infty$ norm of the total signed prefix-sum.

\begin{figure}[!tb]
\center
\includegraphics[width=0.45\textwidth]{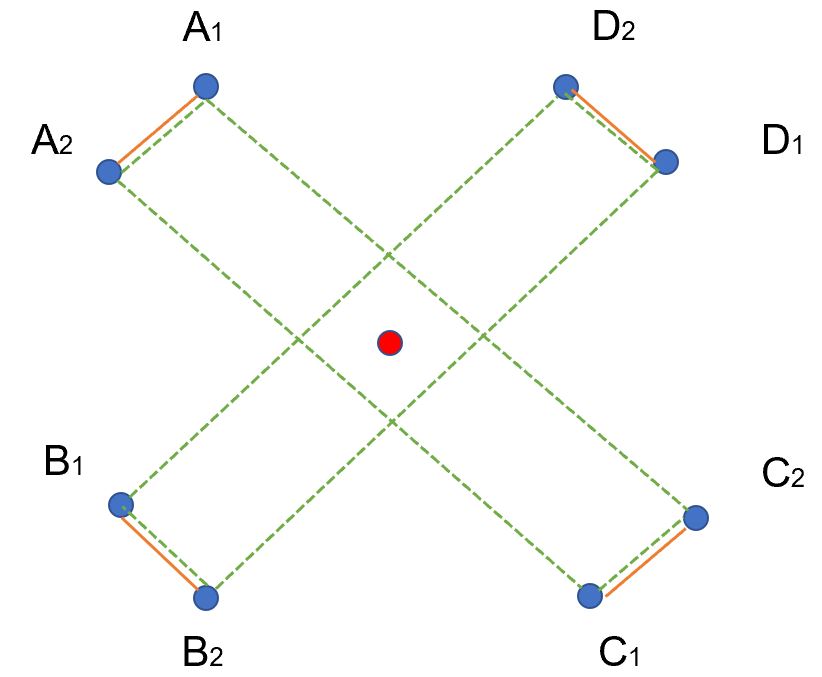}
\caption{An illustrator of the difference between the online vector balancing objective and the online blocking objective.}
\label{fig:DifferentObjectives}
\end{figure}

Our problem is related to the online vector balancing problem, but with the following major distinction in the objective functions.
The online vector balancing problem minimizes the total signed prefix-sum, whereas our online blocking problem minimizes the size of the minimum weight perfect matching.
See Figure~\ref{fig:DifferentObjectives} for an illustration.
As shown in Figure~\ref{fig:DifferentObjectives}, there are $8$ subjects ($A_1$, $A_2$, $B_1$, $B_2$, $C_1$, $C_2$, $D_1$, $D_2$) whose covariates are in a two-dimensional space.
Under our online blocking objective, one optimal design could be matching $(A_1, A_2)$ into a pair, ..., $(D_1, D_2)$ into a pair.
Thus, the optimal design selects exactly one from each pair into the control group, and exactly the other from each pair into the treated group.
On the other hand, under the online vector balancing objective, one optimal design (not necessarily unique) could be assigning negative signs to $A_1, A_2, C_1, C_2$, i.e., into the control group, and positive signs to $B_1, B_2, D_1, D_2$, i.e., into the treated group.
Such a design could possibly lead to imbalance between the two groups, as the geometric center may not fully capture the locations of all subjects.
\item \textit{Online bipartite matching}.
This line of literature follows the works of \citet{karp1990optimal, mehta2007adwords}, and has been summarized in the textbook of \citet{mehta2013online}.
In an online bipartite matching problem, one side of the bipartite graph, usually referred to as the resources, is known in advance and endowed with fixed capacities.
The other side of the bipartite graph, usually referred to as the subjects, arrives in a sequential fashion.
Upon arrival, one subject must be immediately matched to one of the adjacent resources, and earns an immediate reward that is equal to the weight of the created edge.
The objective is to maximize the total sum of rewards accumulated over the horizon.
While the online bipartite matching problem typically has a maximization objective, some works have discussed finding the minimum weight matching \citep{kalyanasundaram1993online}.
Recent works have extensively studied a number of other variants, \emph{e.g.}, under the stochastic arrival model \citep{alaei2012online, devanur2013randomized, feldman2009online, jaillet2014online}, under the random order model \citep{goel2008online}, and allowing for delays \citep{ashlagi2019edge}.

Our problem is intrinsically different from the online bipartite matching problem in the following two aspects.
First, in an online bipartite matching problem, one side of the bipartite graph (the resource side) is fixed in advance \citep{kanoria2021dynamic, chen2023feature}.
But in the online blocking problem, both sides are the experimental subjects and arrive one by one.
Instead of being a two-sided matching, the online blocking problem is an one-sided matching in nature.
The closest related works in one-sided matching include kidney exchange \citep{ashlagi2012new, roth2004kidney} and game matchmaking \citep{gan2023essays}, but the objective functions are much different.
Second, the rewards in an online bipartite matching problem are generated immediately when subjects are matched to resources.
The core trade-off lies in balancing immediate rewards and potential future rewards.
An assignment executed in a previous period only makes an impact on the future rewards through the remaining resources.
But in the online blocking problem, while we make an assignment decision immediately upon the arrival of a subject, the discrepancy is calculated at the end of the horizon.
As a result, the impact of an assignment executed in an early period will not be fully understood until the end of the horizon.
\end{enumerate}



\subsection{The Pigeonhole Design}
To solve the online blocking problem, we propose a pigeonhole design that works in the following.
First, the pigeonhole design partitions the covariate space into smaller spaces, which we refer to as pigeonholes.
When each subject arrives, we observe its covariates and route it to its respective pigeonhole.
Next, we assign each subject to the control or treated group based on how many subjects are there already in the same pigeonhole.
When there is an equal number of control and treated subjects, we randomly assign the arriving subject into either the control or the treated group, with half probability each;
when there is more subjects in one group, we assign the arriving subject to the opposite group.
By doing so, we make the number of treated and control subjects as balanced as possible, and sequentially match the arrived subjects in pairs.

To the best of our knowledge, the pigeonhole design is similar but distinct to a few experimental design methods in the literature.

\begin{enumerate}
\item \textit{Matched-pair design}. 
The idea of the matched-pair design, or more generally, blocking, has been well studied from the beginning of the experimental design literature \citep{fisher1936design}.
The matched-pair design is a limiting case in which each block only has two experimental subjects \citep{athey2017econometrics, chase1968efficiency, imbens2015causal}.
In a matched-pair design, there is an even number of subjects, and all the subjects are grouped into size-two pairs according to some criterion.
From each pair of the two subjects, one is randomly assigned control and the other is assigned treatment.
Other than the criterion considered in this work of finding the minimum weight perfect matching, there are many other criteria in deciding how to pair the subjects, including minimizing the maximum within-pair distance \citep{higgins2016improving}, minimizing the (generalized) Mahalanobis distance \citep{diamond2013genetic}, and minimizing the risk function \citep{bai2022optimality}.
Since the matched-pair design requires information about all the subjects, this design has not been well adapted to the online setting when subjects arrive one by one.

In the clinical trial literature, practitioners and trialists have taken the above block design idea and applied it to the sequential setting.
In the survey by \citet{kernan1999stratified}, the authors described a two-stage procedure that first groups all the subjects into blocks, where each block usually comprises four or six subjects.
Then, subjects are randomly assigned control or treated using a randomization list, which is created before the trial is begun.
Conceptually, the pigeonhole design is similar to this experimental design idea: Each pigeonhole plays a role of blocking on the covariates.
Yet the pigeonhole design does not limit each pigeonhole to have a size of four or six.
More importantly, this paper discusses how to design the number of pigeonholes and the sizes of each pigeonhole, whereas the previous works have not discussed how to choose the blocks.
\item \textit{Completely randomized design}.
The idea of completely randomized design has also been well studied from the beginning of the experimental design literature \citep{fisher1936design}.
It is one of the most common designs of experiments where randomization serves as the basis of validity.
In a completely randomized design, a fixed number of experimental subjects (usually half of total population) are randomly assigned to the control group while the remaining subjects to the treated group.
Since randomization can be determined before running the experiment, this design can be easily implemented in the online setting \citep{efron1971forcing}. 
But a completely randomized design does not take into account the covariate information.
\item \textit{Biased coin design and its covariate-adaptive versions}. This line of literature is pioneered by the work of \citet{efron1971forcing}, and has been subsequently generalized to covariate-adaptive versions as in \citet{atkinson1982optimum, atkinson1999optimum, pocock1975sequential}.
In a biased coin design, each time the next subject arrives, the experimenter flips a biased coin to determine whether the next subject receives control (if tails) or treated (if heads).
The head-up probability of the coin can depend on the number of control and treated subjects so far.
One special case of the biased coin design is the completely randomized design, where the biased coin flipping always tends to balance the number of control and treatment assignments.

In the generalization to covariate-adaptive versions, there are a fixed number of prognostic covariates and the head-up probability depends on a generic function of the covariate imbalance.
This is a general framework, as the generic function could almost capture any covariate-adaptive design of experiment.
Yet when it comes to the executable experimental design, as commented in \citet{bhat2020near}, the biased coin designs and their covariate-adaptive versions ``can be regarded as myopic policies ... that only consider the immediate impact of an allocation decision but not the impact on future decisions.''
Specifically for the covariate-adaptive versions, as commented in \citet{rosenberger2008handling} and \citet{rosenberger2015randomization}, ``very little is known about their theoretical properties.''
\item \textit{Matching-on-the-fly}.
Recently, \citet{kapelner2014matching} proposes a creative idea of maintaining a reservoir to match subjects.
Whenever a subject arrives, if there is no good match with the reservoired subjects, measured in the Mahalanobis distance, then flip a fair coin to determine the assignment of this arriving subject.
If there is a good match with one of the reservoired subjects, then both of them are matched into a pair and then removed from the reservoir.
Such a simple idea has been shown to be useful in practice, and it also provides a framework to study delayed assignment problems.
The difference between \citet{kapelner2014matching} and this paper is that, \citet{kapelner2014matching} focuses on the large sample statistical property of their matching-on-the-fly method, and we formulate it as a discrepancy minimization problem.
\end{enumerate}

\subsection{Performance Guarantees} 

\begin{table}[tb]
\centering
\TABLE{Summary of results measured in discrepancies.
\label{tbl:Summary}}
{\small
\tabulinesep=1mm
\begin{tabu}{|l|cl|cl|cl|}
\hline
Dimension of Covariates            & \multicolumn{2}{c|}{$p = 0, q \geq 1$}                                                     & \multicolumn{2}{c|}{$p = 1, q \geq 0$}                                                                              & \multicolumn{2}{c|}{$p \geq 2, q \geq 0$}                                                           \\ \hline
Matched-Pair Design                & $\Theta(1)$                       & \textbf{Theorem~\ref{thm:MatchedPair:p=0}}          & $\Theta(1)$                         & \textbf{Theorem~\ref{thm:MatchedPair:p=1}}           & $\Theta\big(T^{\frac{p-1}{p}}\big)$    & \textbf{Theorem~\ref{thm:MatchedPair:p>=2}}             \\ \hline
Completely Randomized Design       & $\Theta\big(T^{\frac{1}{2}}\big)$ & \textbf{Theorem~\ref{thm:CompletelyRandomized:p=0}} & $\Theta\big(T^{\frac{1}{2}}\big)$   & \textbf{Theorem~\ref{thm:CompletelyRandomized:p=1}}  & $\Omega\big(T^{\frac{p-1}{p}}\big)$    & \textbf{Theorem~\ref{thm:CompletelyRandomized:p>=2}}    \\ \hline
Pigeonhole Design                  & $\Theta(1)$                       & \textbf{Theorem~\ref{thm:Pigeonhole:p=0}}           & $\Theta\big(T^{\frac{1}{4}}\big)$   & \textbf{Theorem~\ref{thm:Pigeonhole:p=1}}            & $\Theta\big(T^{\frac{p-1}{p}}\big)$    & \textbf{Theorem~\ref{thm:Pigeonhole:p>=2}}              \\ \hline
\end{tabu}
}
{In this table, $T$ stands for the total number of experimental subjects; $p$ and $q$ stand for the dimensions of continuous and discrete covariates, respectively.
All Landau notations here ignore logarithmic factors. 
We consider the case when the number of covariates $p+q$ is much smaller than the number of experimental subjects $T$. 
In other words, we only consider the low-dimensional online blocking problem.}
\end{table}

\begin{table}[tb]
\centering
\TABLE{Summary of results measured in discrepancies, when the arrival sequence is highly clustered.
\label{tbl:SummaryCluster}}
{\small
\tabulinesep=1mm
\begin{tabu}{|l|cl|cl|}
\hline
            & \multicolumn{2}{c|}{$\gamma$ known}                                                              & \multicolumn{2}{c|}{$\gamma$ unknown}                                                                                    \\ \hline
Matched-Pair Design                & $\Theta\big(T^{\max\{1-\frac{1}{p}-\gamma, 0\}}\big)$           & \textbf{Theorem~\ref{thm:Pigeonhole:clustered}} & $\Theta\big(T^{\max\{1-\frac{1}{p}-\gamma, 0\}}\big)$           & \textbf{Theorem~\ref{thm:Pigeonhole:clustered}}    \\ \hline
Completely Randomized Design       & $\Omega\big(T^{\max\{1-\frac{1}{p}-\gamma, \frac{1}{2}\}}\big)$ & \textbf{Theorem~\ref{thm:Pigeonhole:clustered}} & $\Omega\big(T^{\max\{1-\frac{1}{p}-\gamma, \frac{1}{2}\}}\big)$ & \textbf{Theorem~\ref{thm:Pigeonhole:clustered}}    \\ \hline
Pigeonhole Design                  & $O\big(T^{\frac{p-1}{p}(1-\gamma)}\big)$                        & \textbf{Theorem~\ref{thm:Pigeonhole:clustered}} & $O\big(T^{\frac{p-1}{p} (1-\underline{\gamma})}\big)$   & \textbf{Theorem~\ref{thm:Pigeonhole:clustered}}    \\ \hline
\end{tabu}
}
{In this table, $\gamma \geq \underline{\gamma} > 0$ is a constant parameter such that $T^{-\gamma}$ stands for the diameter of the clusters. 
All Landau notations here ignore logarithmic factors. 
We distinguish two cases when $\gamma$ is known and when $\gamma$ is unknown. 
In both cases, the pigeonhole design outperforms the completely randomized design when the arrival sequence is highly clustered, i.e., $\underline{\gamma} > \frac{1}{2}$.}
\end{table}

In this paper, we analyze the performance of the matched-pair design, the completely randomized design, and the pigeonhole design whose discrepancies are summarized in Table~\ref{tbl:Summary}.
First, the matched-pair design serves as a benchmark that any online design of experiments cannot achieve.
This is because the matched-pair design uses all the covariate information which would not be available to the online designs, as it is sequentially revealed over time.
Second, the completely randomized design serves as a naive benchmark by not using any covariate information at all.
As a result, the expected discrepancy for a completely randomize design is large.
Third, the pigeonhole design as we propose in this paper has a performance in between the matched-pair design and the completely randomized design.

Table~\ref{tbl:Summary} also summarizes the discrepancies of the three designs under three cases: 
(1) all the covariates are discrete, i.e., $p = 0, q \geq 1$;
(2) there is one continuous covariate and possibly many discrete covariates, i.e., $p=1, q \geq 0$;
(3) there are more than one continuous covariates and possibly many discrete covariates, i.e., $p \geq 2, q \geq 0$.

When $p = 0, q \geq 1$, the matched-pair design has a discrepancy on the order of $\Theta(1)$; the completely randomized design has an expected discrepancy on the order of $\Theta(T^{\frac{1}{2}})$; the pigeonhole design has an expected discrepancy on the order of $\Theta(1)$.
This case is more common than it seems to be, as many online platforms keep their user demographics using categorical, or even binary, records.

When $p=1, q \geq 0$, the matched-pair design has a discrepancy on the order of $\Theta(1)$; the completely randomized design has an expected discrepancy on the order of $\Theta(T^{\frac{1}{2}})$; the pigeonhole design has an expected discrepancy on the order of $\Theta\big(T^{\frac{1}{4}}\big)$ (ignoring logarithmic factors).
The first and the second cases are where the pigeonhole design demonstrates the most benefits.

When $p \geq 2, q \geq 0$, even the matched-pair design has a non-negligible discrepancy on the order of $\Theta\big(T^\frac{p-1}{p}\big)$.
The completely randomized design has an expected discrepancy at least $\Omega\big(T^\frac{p-1}{p}\big)$, as it cannot be better than the matched-pair design.
But the upper bound on the discrepancy is still an open question in the optimal transport literature \citep{fournier2015rate}.
The pigeonhole design has an expected discrepancy on the order of $\Theta\big(T^\frac{p-1}{p}\big)$ (ignoring logarithmic factors).
The performance of the pigeonhole design closely matches that of the matched-pair design.

We further show that, in scenarios where the arrival sequence is highly clustered (in contrast to being weakly clustered, see Figure~\ref{fig:Clustered}), the pigeonhole design brings substantial benefits over the completely randomized design (See Table~\ref{tbl:SummaryCluster}). 
Suppose the diameter of each cluster does not exceed $T^{-\gamma}$, where $\gamma \geq \underline{\gamma} > 0$. 
We say that the arrival sequence is highly clustered when $\underline{\gamma} > \frac{1}{2}$.
If $\gamma$ is known, the pigeonhole design has an expected discrepancy on the order of $O\big(T^{\frac{p-1}{p}(1-\gamma)}\big)$. 
If $\gamma$ is unknown, the pigeonhole design has an expected discrepancy on the order of $O\big(T^{\frac{p-1}{p} (1-\underline{\gamma})}\big)$. 
Meanwhile, in both scenarios, the completely randomized design has an expected discrepancy of $\Omega\big(T^{\max\{1-\frac{1}{p}-\gamma, \frac{1}{2}\}}\big)$.
These findings suggest that, when the arrival sequence is highly clustered, the pigeonhole design outperforms the completely randomized design.

\subsubsection*{Roadmap.}
The paper is structured as follows. 
In Section~\ref{sec:Definitions} we introduce the online blocking problem.
In Sections~\ref{sec:MatchedPair} --~\ref{sec:PigeonholeDesign} we introduce the matched-pair design, the completely randomized design, and the pigeonhole design, and show their performances in the single continuous dimension case.
In Section~\ref{sec:MultiDimension} we show the performances of three designs in the general cases, and demonstrate the benefits of pigeonhole design over the completely randomized design.
In Section~\ref{sec:Simulations:ATE} we use Yahoo! data to show a reduction in variance if we use the pigeonhole design to estimate the average treatment effect.
In Section~\ref{sec:Conclusion} we conclude with three practical suggestions, and point out two limitations of this work.

\section{The Online Blocking Problem}
\label{sec:Definitions}

Consider the following experimental design problem for an online platform.
Let there be a discrete, finite time horizon of $T$ periods, where $T$ is assumed to be an even number.
In practice, $T$ is typically known and given in advance, which reflects the size of the experimental budget.
For example, a typical experimental budget ranges between $100K \sim 1M$, and this number is typically determined in advance by the product managers who are testing their new products.

At any time $t \in [T] := \{1,2,...,T\}$, one experimental subject arrives at the platform.
Each subject is associated with some covariates $\bm{x}_t$ from a $(p+q)$-dimensional space $\cS$, where $p$ stands for the number of continuous covariates and $q$ stands for the number of discrete covariates.
In this paper, we normalize the covariate space to be $\cS = [0,1]^{(p+q)}$ by re-scaling the support of each dimension to $[0,1]$, as such a re-scaling does not change the dependence on $T$.
Each dimension reflects an important covariate that the experimenter needs to consider.
For example, at Yahoo! front page \citep{lewis2010s}, the covariates could be a combination of several user demographics and several key performance indices such as ``browsing type.''

Upon the arrival of each subject $t$, the experimenter observes the covariate information of the subject, and then must immediately and irrevocably assign an intervention $W_t \in \{0,1\}$ to this subject without knowing the covariates of the future arriving subjects.
For example, at Yahoo! front page \citep{lewis2010s}, the assignment of control or treatment must be determined within $10$ milliseconds when one subject arrives.
Following convention, we say that subject $t$ is assigned to the control group if $W_t=0$, and treated group if $W_t=1$.
We use $W_t$ to stand for a random assignment, and $w_t$ for a realization.

In this context, a design of randomized experiment refers to a sequential decision rule $\bm{\mu} = (\mu_1, \mu_2, ..., \mu_T)$, such that
$\mu_1(x_1): \cS \to [0,1]$ for $t=1$, and $\mu_t(x_1,x_2,...,x_t,w_1,w_2,...,w_{t-1}): \cS^t \times \{0,1\}^{t-1} \to [0,1]$ for $t \geq 2$.
This decision rule takes as input the covariate information of all subjects that arrived up to the current period, as well as the treatment assignments to all subjects that arrived up to the previous period.
It then outputs a treatment probability, i.e., $\Pr(W_t = 1) = \mu_t(\cdot)$, and $\Pr(W_t = 0) = 1 - \mu_t(\cdot)$.

The design of randomized experiment must respect the following constraint: 
Out of a total of $T$ experimental subjects, there must be a half of them that are assigned to the control group, and the other half assigned to the treated group.
This half-half assignment 
has been widely assumed in the experimental design literature \citep{bai2022optimality, basse2019minimax, bojinov2020design, candogan2021near, greevy2004optimal, harshaw2019balancing, li1983minimaxity, lu2011optimal, rosenbaum1989optimal, wu1981robustness, xiong2019optimal}.
As \citet{pocock1975sequential} also commented, ``it is desirable to have the treated groups\footnote{In the original paper they refer to``control" as one version of treatment. So ``two treated groups" refers to the control and treated groups in our paper.} of equal sizes.''
For modern online platforms, even though each new intervention is tested on only a tiny portion of total population, the experiment is still conducted in a half-half manner.
For example, \citet{huang2023estimating} documented a search bar experiment on WeChat, where the new intervention is tested on less than $0.01\%$ of their total active users.
The way they conduct the experiment is by first designating $1.3M$ users into a holdout experiment, and then randomly selecting a half of them to receive the new intervention and the other half to receive the status-quo.
We would like to comment that such a constraint only serves as the foundation for comparisons between different experimental designs.
In practice, the pigeonhole design still remains valid if the above constraint is violated; we will discuss more details in Section~\ref{sec:PigeonholeDesign}.

To formally incorporate such a constraint, we denote $A = \{t \vert W_t = 0\}$ as the set of subjects in the control group, and $B = \{t \vert W_t = 1\}$ as the set of subjects in the treated group.
The two groups form a partition of all the subjects, $A,B \subseteq [T], A \cup B = [T], A \cap B = \emptyset$, and each group must have one half of the total subjects, $\left|A\right| = \left|B\right| = T/2$.
Denote $\mathcal{M}$ to be the set of all such possible partitions.

The quality of any design of experiment is measured by discrepancy, which we define as follows.
For any pair of subjects $t \in A$ and $t' \in B$, the discrepancy between $t$ and $t'$ is their $L_2$ distance, i.e.,
\begin{align*}
d_{t, t'} = \left\| \bm{x}_{t} - \bm{x}_{t'} \right\|_2.
\end{align*}
For two fixed groups of subjects $A$ and $B$, the discrepancy between $A$ and $B$ is the minimum weight perfect matching \citep[Chapter 7.8]{bertsimas1997introduction} between groups $A$ and $B$, i.e.,
\begin{align}
d_{A, B} = \min_{\bm{y}} & \ \sum_{t \in A} \sum_{t' \in B} d_{t,t'} y_{t,t'} && \label{eqn:Matching} \\
s.t.  \ & \ \sum_{t' \in B} y_{t,t'} = 1, && \forall \ t \in A, \nonumber \\
& \ \sum_{t \in A} y_{t,t'} = 1, && \forall \ t' \in B, \nonumber \\
& \ y_{t,t'} \in \{0,1\}, && \forall \ t \in A, \ t' \in B, \nonumber
\end{align}
where $y_{t,t'}$ is a binary variable that takes value $1$ if subjects $t$ and $t'$ are matched.
For any design of experiment, the design itself induces a distribution of the two groups of subjects $(A,B) \in \mathcal{M}$.
We define the expected discrepancy to be the expectation of $d_{A,B}$, where randomness comes from the distribution induced by the randomized design of experiment.

The minimum weight perfect matching has been extensively studied in both the operations research literature \citep{ahuja1988network, bertsimas1997introduction, schrijver2003combinatorial} and the statistics literature 
\citep{abadie2012martingale, dehejia2002propensity, greevy2004optimal, lu2011optimal, rosenbaum1989optimal}.
It is a simple and meaningful notion of discrepancy.
If an experimenter is interested in estimating the average treatment effect between the two groups of subjects, recent works of \citet[Assumption~2.3]{bai2021inference} and \citet[equation~(2.13)]{cytrynbaum2021designing} argued that a smaller discrepancy as measured by the minimum weight perfect matching leads to a smaller ex-post bias of the estimator.
Using the Landau notation, if a design of experiment leads to a discrepancy on the order of $O(T^r)$, the associated difference-in-means estimator suffers from an ex-post bias on the order of $O(T^{r/2})$.

The primary focus of this paper is to find designs of randomized experiments that minimize the expected discrepancy under the worst-case arrival sequence of length $T$.
We model the covariates of each subject to be adversarially chosen from the covariate space $\cS$, irrespective to the treatment assignments of the previous subjects.
Such a model is often referred to as a non-adaptive or oblivious adversarial arrival model, and is widely adopted in the online algorithms literature \citep{buchbinder2009design, mehta2013online}.

The subsequent sections will proceed as follows: 
We introduce two benchmark experimental designs, namely the matched-pair design and the completely randomized design, and analyze their performances in a simple, single (continuous) dimensional case.
Following this, we introduce the pigeonhole design and analyze its performance in the same single (continuous) dimensional case. 
We then analyze the performances of these designs in the more general cases.

\section{The Matched-Pair Design}
\label{sec:MatchedPair}

If there was a clairvoyant oracle that could reveal the covariates of all subjects before they arrive, the experimenter can leverage such information to design experiments.
Experimental designs under such clairvoyant information are typically in the form of matched-pair designs \citep{bai2022optimality, greevy2004optimal, lu2011optimal, rosenbaum1989optimal}.
In this section, we describe one version of the well-known matched-pair design.

\begin{definition}[Matched-Pair Design]
\label{defn:MatchedPairDesign}
In a matched-pair design, the experimenter endowed with clairvoyant information first solves the following problem,
\begin{align}
(A^*, B^*) = \argmin_{(A, B) \in \mathcal{M}} d_{A,B}, \label{eqn:SmallestDiscrepancy}
\end{align}
and finds the solution $\bm{y}^*$ to the minimum weight perfect matching between $A^*$ and $B^*$ as in \eqref{eqn:Matching}.
Second, recall that $y^*_{t,t'}$ is a binary variable that takes value $1$ if subjects $t$ and $t'$ are matched.
For each pair of subjects $t, t' \in [T]$ such that $y^*_{t,t'} = 1$, randomly assign control to one subject and treatment to the other, with probability $1/2$ each.
\end{definition}

\begin{figure}[!tb]
\centering
\includegraphics[width=0.5\textwidth]{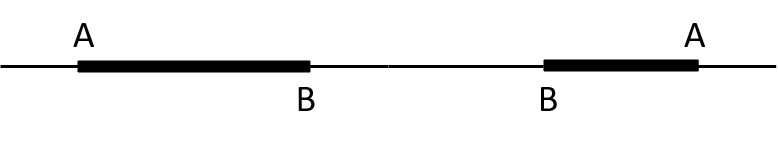}
\caption{An illustrator of the matched-pairs in a single continuous dimension. The $A$ stands for subjects in the control group; whereas $B$ stands for subjects in the treated group.}
\label{fig:MatchedPairDesign}
\end{figure}

The matched-pair design, despite its infeasibility for the online blocking problem due to its usage of clairvoyant information, has the smallest discrepancy compared to any other design of experiment.
To illustrate the small discrepancy that it incurs, we focus on one special case when there is $p=1$ continuous covariate and $q=0$ discrete covariate.
In this special case, the matched-pair design easily solves \eqref{eqn:SmallestDiscrepancy} and proceeds in the following way: first, rearrange all subjects from the smallest to the largest in the covariate space and group them in size-two pairs; second, for the two subjects in each pair, randomly assign control to one subject and treatment to the other, with probability $1/2$ each.
See Figure~\ref{fig:MatchedPairDesign} for an illustration.
This design has a discrepancy of no more than $1$, which we state in Lemma~\ref{lem:MatchedPair:p=1}.

\begin{lemma}
\label{lem:MatchedPair:p=1}
When $p=1, q=0$, the matched-pair design has a discrepancy less than or equal to $1$.
\end{lemma}

The proof of Lemma~\ref{lem:MatchedPair:p=1} is essentially following the rearrangement argument that we made above. 
The complete proof is deferred to Appendix~\ref{sec:Proof:MatchedPair:p=1}.



\section{The Completely Randomized Design}
\label{sec:CompeltelyRandomzied}

In Section~\ref{sec:MatchedPair} we have seen that the matched-pair design achieves the smallest discrepancy, by knowing the covariates of all the subjects in advance.
In the online blocking problem, however, this is not feasible.
Nonetheless, there is another well-known design of experiments, the completely randomized design, that is feasible for the online setting.

\begin{definition}[Completely Randomized Design]
\label{defn:CompeltelyRandomizedDesign}
Out of a total of $T$ subjects, a completely randomized design randomly selects $T/2$ subjects as the control group $A$, and the other $T/2$ subjects as the treated group $B$.
Upon the arrival of a subject $t$, assign control if $t \in A$, and treatment if $t \in B$, regardless of subject $t$'s covariate information.
\end{definition}

Since a completely randomized design does not use any covariate information in assigning control and treatment to each subject, it is a feasible design for the online blocking problem.
As expected, it has a larger discrepancy.
We illustrate using the following example.

\begin{example}
\label{exa:p=1}
Let there be $4$ subjects, each endowed with a single-dimensional covariate $x_1 = 0.1, x_2 = 0.7, x_3 = 0.4, x_4 = 0.9$.

Under a matched-pair design, we rearrange the four subjects from the smallest to the largest in the covariate space, $x_1 \leq x_3 \leq x_2 \leq x_4$.
Then we match subjects $\{1,3\}$ in one pair and $\{2,4\}$ in one pair.
Within each pair, we randomly assign one to the control group and the other to the treated group.
The above matched-pair design has a discrepancy of $0.5$.

Under a completely randomized design: with probability $1/6$, $A = \{1,2\}, B = \{3,4\}, d_{A,B} = 0.5$;
with probability $1/6$, $A = \{1,3\}, B = \{2,4\}, d_{A,B} = 1.1$; 
with probability $1/6$, $A = \{1,4\}, B = \{2,3\}, d_{A,B} = 0.5$; 
with probability $1/6$, $A = \{2,3\}, B = \{1,4\}, d_{A,B} = 0.5$; 
with probability $1/6$, $A = \{2,4\}, B = \{1,3\}, d_{A,B} = 1.1$; 
with probability $1/6$, $A = \{3,4\}, B = \{1,2\}, d_{A,B} = 0.5$.
In expectation, the discrepancy of this completely randomized design is equal to $0.7$, which is larger than the discrepancy of a matched-pair design.
\end{example}

Next, we formally show in Lemma~\ref{lem:CompletelyRandomized:p=1} that the expected discrepancy of a completely randomized design is indeed larger than that of a matched-pair design when $p=1, q=0$.

\begin{lemma}
\label{lem:CompletelyRandomized:p=1}
When $p=1, q=0$, the completely randomized design has an expected discrepancy on the order of $\Theta(T^{\frac{1}{2}})$.
\end{lemma}

We explain the main idea of the proof of Lemma~\ref{lem:CompletelyRandomized:p=1} here in an unrigorous way, and defer the detailed proof to Appendix~\ref{sec:Proof:CompeltelyRandomzied:p=1}.

\proof{Sketch Proof of Lemma~\ref{lem:CompletelyRandomized:p=1}.}
We first show the $\Omega(T^{\frac{1}{2}})$ part by constructing an instance $\bm{x}^*$ of the arrival sequence. 
Let the covariates of the first $\frac{T}{2}$ subjects be equal to $0$, i.e., $x^*_t = 0, \forall \ 1 \leq t \leq \frac{T}{2}$;
let the covariates of the last $\frac{T}{2}$ subjects be equal to $1$, i.e., $x^*_t = 1, \forall \ \frac{T}{2}+1 \leq t \leq T$.
In Lemma~\ref{lem:lowerfluid}, we show that on this instance $\bm{x}^*$, the completely randomized design has an expected discrepancy that is approximately $\frac{1}{\sqrt{\pi}}\sqrt{T}$.

Next, we turn to the $O(T^{\frac{1}{2}})$ part.
We introduce Lemma~\ref{lem:permutation} to show that the completely randomized design in the $p=1$ case is invariant to permutations of the arrival sequence.
For any two sequences $\bm{x}', \bm{x}''$, if there exists a one-to-one correspondence $\sigma: [T] \to [T]$, such that $\forall \ t \in [T]$, $x'_t = x''_{\sigma(t)}$, then the expected discrepancy of the completely randomized design will be the same for these two arrival sequences $\bm{x}'$ and $\bm{x}''$.
Intuitively, this is because the completely randomized design, when determining the assignment of each subject, completely randomizes the indices of the subjects without using any covariate information of the other subjects.
With this permutation-invariant observation, and given that we focus on the $p=1$ case, we are able to focus only on the arrival sequences that are ordered from the smallest to the largest.

Finally, to conclude the $O(T^{\frac{1}{2}})$ part, we introduce Lemma~\ref{lem:upperfluid} to show that, among all possible arriving sequences, the sequence $\bm{x}^*$, with the first half subjects equal to zero and the second half subjects equal to one, is the one that has the largest expected discrepancy.
This shows that the expected discrepancy of the completely randomized design is on the order of $O(T^{\frac{1}{2}})$.
\Halmos
\endproof

We emphasize here that Lemma~\ref{lem:CompletelyRandomized:p=1} suggests an expected discrepancy exactly on the order of $\Theta(T ^{\frac{1}{2}})$.
This is both an upper bound and a lower bound.
By comparing the matched-pair design to the completely randomized design, we make the following two observations.
First, the matched-pair design, though infeasible to the online blocking problem, fully leverages the covariate information of all the subjects.
As a result, the discrepancy is very small.
Second, the completely randomized design, feasible to the online blocking problem, does not use any covariate information at all.
These two observations motivate the following question: is there a design that adaptively uses the covariate information to achieve a performance in between?
We will provide an affirmative answer in Section~\ref{sec:PigeonholeDesign}.

\section{The Pigeonhole Design}
\label{sec:PigeonholeDesign}

In this section we introduce the Pigeonhole Design, a new experimental design method, that adaptively assigns control and treatment assignments to the online subjects.
We will first introduce what a pigeonhole design is and how to design the pigeonholes.
Then we will provide intuitions regarding why and how it improves the completely randomized design.
We conclude this section by analyzing the discrepancy of the pigeonhole design.

\subsubsection*{Definition of the pigeonhole design.} 
We start with intuitions of the pigeonhole design.
The pigeonhole design partitions the covariate space.
We refer to the smaller, partitioned spaces as pigeonholes.
In each pigeonhole, we wish to make the number of control and treated subjects as balanced as possible, by sequentially matching the subjects that arrive at the same pigeonhole in pairs.
Due to the unknown nature of the arrival sequence, it is impossible to guarantee that the number of control and treated subjects are exactly equal in each pigeonhole.
So at the moment when either the control or the treated group reaches $T/2$ in size, it is possible that some of the pigeonholes contain an odd number of subjects, and there is one subject that cannot be matched from within the same pigeonhole.
We will then have to match these extra subjects across different pigeonholes.
This would require some extra balancing efforts in the last few periods, which we refer to as the balancing periods.

Now we are ready to formally describe the pigeonhole design.
Let the covariate space be $\cS$.
Define a $K$-partition of $\cS$ to be $\mathcal{P}(\cS) = \{\cS_1, \cS_2, ..., \cS_K\}$, such that $\cup_{k=1}^K \cS_k = \cS$ and that $\cS_k \cap \cS_{k'} = \emptyset, \forall k, k' \in [K]$.
These sets $\{\cS_k\}_{k \in [K]}$ are referred to as the pigeonholes.
We only consider time-invariant pigeonholes as they are much easier to implement for practitioners.
Since a pigeonhole design adaptively assigns subjects into control and treated groups, we use the following notations, $n^0_{k,t}$ and $n^1_{k,t}$, to denote the number of control and treated subjects in the $k$-th pigeonhole after the arrival of subject $t$.
Denote $n^0_{k,0} = n^1_{k,0} = 0, \forall k \in [K]$, to reflect that no subjects have arrived yet at the beginning of the entire horizon.
For any partition $\mathcal{P}(\cS)$, define the pigeonhole design as follows.

\begin{definition}[Pigeonhole Design]
\label{defn:PigeonholeDesign}
A pigeonhole design takes as input a partition of the covariate space $\mathcal{P}(\cS)$.
\begin{enumerate}
\item Upon the arrival of subject $t$, if either the control or the treated group reaches $T/2$ in size, assign subject $t$ to the opposite group.
\item If no group reaches $T/2$ in size, find the associated pigeonhole $k$ such that $x_t \in \cS_k$.
If there are fewer control subjects in this pigeonhole than treated subjects, i.e., $n^0_{k,t-1} < n^1_{k,t-1}$, assign control to subject $t$;
if there are fewer treated subjects than control subjects, i.e., $n^0_{k,t-1} > n^1_{k,t-1}$, assign treatment to subject $t$;
if there are equal numbers of control and treated subjects, i.e., $n^0_{k,t-1} = n^1_{k,t-1}$, assign control or treatment with probability $1/2$ each.
\end{enumerate}
\end{definition}

It is worth noting that Step~1 in the above definition ensures that a pigeonhole design satisfies the constraint that a half of the subjects are assigned to the control group, and the other half are assigned to the treated group.
Even if we do not have Step~1, the pigeonhole design is still a valid design of experiment, with the number of control and treated subjects possibly being unequal.
In this paper, we choose to maintain this constraint for the purpose of making comparisons with the two benchmark designs discussed, especially when analyzing their performances.

\subsubsection*{Analysis of the pigeonhole design.}
Below we give Example~\ref{exa:p=1:continued}, which illustrates an execution of the pigeonhole design when there is one single continuous dimension.
We show that a pigeonhole design has a smaller expected discrepancy compared to the completely randomized design.

\begin{example}[Example~\ref{exa:p=1} Continued]
\label{exa:p=1:continued}
Let there be $4$ subjects, $x_1 = 0.1, x_2 = 0.7, x_3 = 0.4, x_4 = 0.9$.
In a pigeonhole design, suppose we choose the pigeonholes to be $\cS_1 = [0, 0.5), \cS_2 = [0.5, 1]$.

This pigeonhole design works as follows.
When $x_1 = 0.1 \in \cS_1$ arrives, there is no subject in $\cS_1$. So we randomly assign control or treatment. Suppose we assign treatment, i.e., $w_1 = 1$.
When $x_2 = 0.7 \in \cS_2$ arrives, there is no subject in $\cS_2$. So we randomly assign control or treatment. Suppose we assign control, i.e., $w_2 = 0$.
When $x_3 = 0.4$ arrives, there is one treated subject and zero control subject in $\cS_1$. Since the number of control subjects is fewer, we deterministically assign control, i.e., $w_3 = 0$.
The last period belongs to the balancing period, because at this moment there is $1$ treated subject and $2$ control subjects.
Since there are more subjects in the control group than in the treated group, we deterministically assign the last subject to the treated group, i.e., $w_4 = 1$.

Under this trajectory of randomness, the discrepancy of this pigeonhole design is $0.5$.
\end{example}

The performance of a pigeonhole design critically depends on how the pigeonholes are devised.
We make the following observations.
First, since the arrival sequence is adversarially chosen, the pigeonholes should be uniformly split.
Otherwise, the design will suffer from large discrepancy when the arrival sequence puts more subjects to the largest pigeonhole.
Second, the number of pigeonholes should be neither too small nor too big. 
If there is only one single pigeonhole that is the entire covariate space, i.e., $K=1$ and $\cS_1 = \cS$, then this is the \textit{biased coin design} with parameter equal to $1$ as proposed in \citet{efron1971forcing}.
However, 
when there is only one pigeonhole, the pigeonhole design does not consider covariate information at all, so it will lead to a large discrepancy.
On the other hand, when there are too many pigeonholes, the design suffers from large discrepancy when each pigeonhole has an odd number of subjects.
So each pigeonhole will have one subject unmatched.
These unmatched subjects will have to be matched across different pigeonholes, thus causing a large discrepancy.

Now we illustrate how to choose the correct number of pigeonholes.
We consider a very simple design that achieves $O(T^{\frac{1}{2}})$ expected discrepancy in the $p=1, q=0$ special case.
In this design, we equally divide the covariate space $[0,1]$ into $T^{\frac{1}{2}}$ pigeonholes with the length of each pigeonhole being $T^{-\frac{1}{2}}$.
Consider the first $T-T^{\frac{1}{2}}$ arriving subjects.
For every two subjects that arrive at the same pigeonhole, we match them in pairs.
For each pair of subjects matched in each pigeonhole, the discrepancy is no more than the length of the pigeonhole $T^{-\frac{1}{2}}$.
Therefore, the total discrepancy for the first $(T-T^{\frac{1}{2}})$ subjects that are matched in this way is no more than $T^{-\frac{1}{2}}(T-T^{\frac{1}{2}})=T^{\frac{1}{2}}-1$.
However, among the first $(T-T^{\frac{1}{2}})$ subjects, there are at most $T^{\frac{1}{2}}$ of them that remain unmatched.
This is because there are $T^{\frac{1}{2}}$ pigeonholes and there is at most one subject in each pigeonhole that remains unmatched.
Next, observe that there are $T^{\frac{1}{2}}$ subjects remaining.
No matter what their covariates are, they can be matched with the $T^{\frac{1}{2}}$ unmatched subjects, with each pair generating at most $1$ discrepancy.
The total discrepancy generated from the last $T^{\frac{1}{2}}$ subjects is at most $T^{\frac{1}{2}}$.
Therefore, the total discrepancy is at most 
\begin{align*}
T^{\frac{1}{2}}-1+T^{\frac{1}{2}}=2T^{\frac{1}{2}}-1.
\end{align*}
The pigeonhole design outlined above has a $O(T^{\frac{1}{2}})$ discrepancy.
It turns out that we can improve the above design to obtain an even smaller discrepancy.
Before diving into details, we introduce what to improve in the above pigeonhole design.

First, for the first $T-T^{\frac{1}{2}}$ arriving subjects, we have shown that each pair of subjects matched within the same pigeonhole incurs a discrepancy equal to the length of the pigeonhole.
If we take a closer look at these subjects, due to the random nature of the design, the expected discrepancy that each pair incurs within each pigeonhole is actually equal to the square root of its length.
Second, out of the first $T-T^{\frac{1}{2}}$ arriving subjects, we have shown that there will be at most $T^{\frac{1}{2}}$ subjects that remain unmatched, one from each pigeonhole.
If we take a closer look at these subjects, we actually randomly assign them into either the control or the treated group with half probability each.
Therefore, there exist certain probabilities that two adjacent subjects are assigned to different groups, and we can match them in a pair.

In Lemma~\ref{lem:Pigeonhole:p=1} and the discussions below, we describe in details how to leverage the above two sources of randomness in the assignment of subjects, which leads to a pigeonhole design with significantly smaller discrepancy.

\begin{lemma}
\label{lem:Pigeonhole:p=1}
Fix any $\eta \in (0,1)$.
When $p=1, q=0$, the pigeonhole design using uniform pigeonholes $\cP(\cS) = \left\{[0, T^{-\eta}), [T^{-\eta},2T^{-\eta}), \ldots, [1-T^{-\eta},1] \right\}$ has an expected discrepancy on the order of $O\left( T^{\frac{1}{2}-\frac{\eta}{2}} + T^{\frac{\eta}{2}} (\log{T})^\frac{1}{2}\right)$.
\end{lemma}

\begin{corollary}
\label{coro:Pigeonhole:p=1}
When $\eta = \frac{1}{2}$, the pigeonhole design as parameterized in Lemma~\ref{lem:Pigeonhole:p=1} has an expected discrepancy on the order of $O\big(T^{\frac{1}{4}}(\log{T})^\frac{1}{2}\big)$.
\end{corollary}

Lemma~\ref{lem:Pigeonhole:p=1} and Corollary~\ref{coro:Pigeonhole:p=1} suggest that, through properly splitting the pigeonholes, the pigeonhole design can achieve an expected discrepancy on the order of $O(T^\frac{1}{4} (\log{T})^\frac{1}{2})$.
This is a significant reduction from the $\Theta(T^{\frac{1}{2}})$ expected discrepancy obtained by the completely randomized design.
Next, we show that an $O(T^\frac{1}{4})$ expected discrepancy is best-possible.

\begin{lemma}
\label{lem:Pigeonhole:p=1:LB}
When $p=1, q=0$, any pigeonhole design must suffer from an expected discrepancy on the order of $\Omega(T^{\frac{1}{4}})$.
\end{lemma}

\subsubsection*{Sketch proofs.}

To conclude this section, we sketch the main ideas behind the analysis of the pigeonhole design.
First, we explain the main idea of the proof of Lemma~\ref{lem:Pigeonhole:p=1} here in an unrigorous way, and defer the detailed proof to Appendix~\ref{sec:Proof:Pigeonhole:p=1}.
We also provide the pseudo-codes for the pigeonhole design, under the parameters in Lemma~\ref{lem:Pigeonhole:p=1}, in Algorithm~\ref{alg:Pigeonhole:p=1} in Appendix~\ref{sec:Pseudo:p=1}.

\proof{Sketch Proof of Lemma~\ref{lem:Pigeonhole:p=1}.}
Now that we have built the intuition that uniform pigeonholes are preferable in hedging against adversarial arrival sequences, we only need to decide how many pigeonholes to have.
Suppose the length of each pigeonhole is $T^{-\eta}$; then the number of pigeonholes is $T^\eta$.
Consider the moment when either the control or the treated group reaches $T/2$ in size, and denote it as $\tau$.
With high probability, $\tau$ is close to the end of the horizon, i.e., $T - \tau \leq T^\frac{\eta}{2}.$

We analyze the discrepancy generated from using $T^\eta$ many pigeonholes, by considering the following two perspectives.
On one hand, consider the first $\tau$ subjects. 
We match every two subjects in a pair when they arrive at the same pigeonhole.
If the adversary makes an even number of subjects arrive at each pigeonhole and there are no unmatched subjects, the discrepancy of the first $\tau$ subjects is on the order of $O\big(T^{\eta} \cdot (\frac{\tau}{T^\eta})^{\frac{1}{2}} \cdot T^{-\eta}\big)$, which is equal to $O(T^{\frac{1}{2}-\frac{\eta}{2}})$. 
In the first expression, $T^\eta$ stands for the number of pigeonholes, $(\frac{\tau}{T^\eta})^{\frac{1}{2}}$ stands for the discrepancy from each pigeonhole if the arriving subjects are uniformly distributed over all pigeonholes, and $T^{-\eta}$ stands for the maximum discrepancy generated from two subjects in the same pigeonhole.
The above analysis shows that the length of each pigeonhole should be as small as possible to reduce the discrepancy generated from the first phase. We formalize this result in Proposition~\ref{prop:OnePigeonhole} in Appendix \ref{sec:Proof:prop:OnePigeonhole}. 

On the other hand, the adversary could make an odd number of subjects arrive at each pigeonhole, leaving a total number of $T^\eta$ unmatched subjects.
Since the pigeonhole design is a randomized design of experiment, the unmatched subject has a half-half chance of being assigned to the treated and control group, respectively.
Using a coupling technique, we show that to matching these unmatched subjects across different pigeonholes will incur a discrepancy on the order of $O(T^\frac{\eta}{2})$.
Finally, matching the remaining unmatched subjects with the remaining subjects after $\tau$ will incur a discrepancy on the order of $O(T^\frac{\eta}{2} (\log{T})^\frac{1}{2})$.
Therefore, the above analysis shows that the number of pigeonholes should be as small as possible to reduce such discrepancy, which is equivalent to saying that the length of each pigeonhole should be large.

The above two perspectives suggest a trade-off between having small and large pigeonholes.
We balance these two perspectives by selecting $\eta=\frac{1}{2}$, which leads to $O(T^{\frac{1}{4}}(\log{T})^\frac{1}{2})$ discrepancy.
\Halmos
\endproof

Next, we explain the main idea of Lemma~\ref{lem:Pigeonhole:p=1:LB} in an unrigorous way.
We defer the rigorous proof of Lemma~\ref{lem:Pigeonhole:p=1:LB} to Appendix~\ref{sec:Proof:Pigeonhole:p=1:LB}. 

\proof{Sketch Proof of Lemma~\ref{lem:Pigeonhole:p=1:LB}.}
Suppose that there are $K$ pigeonholes.
For each $k \in [K]$, suppose the $k$-th pigeonhole has length $L_k$.
Then consider an adversarial sequence which, for each $k \in [K]$, places $n_k = \frac{L_k^2}{\sum_{i=1}^K L_i^2} T$ subjects in each pigeonhole. 
Inside each pigeonhole $k$, the adversarial sequence places $\frac{n_k}{2}$ subjects on the left end of this pigeonhole and $\frac{n_k}{2}$ subjects on the right end of the pigeonhole.
The adversarial sequence choose the sequence of these subjects such that small and large values alternate.
In other words, for the subjects that arrive at the same pigeonhole, the subjects on the left end and the subjects on the right end alternate to arrive.
To focus on the subjects on the left end, the number of such subjects who are assigned control and treated follow the binomial distribution, which we denote as $\mathrm{Bin}(\frac{n_k}{2},\frac{1}{2})$.
Due to Lemma~\ref{lem:lowerfluid}, these subjects yield an expected discrepancy of approximately $c_k \cdot L_k \cdot (n_k)^{\frac{1}{2}}$ for some constant $c_k$. 
We denote $\underline{c} = \min_k c_k$. 

Due to Cauchy-Schwarz inequality, the discrepancy from all the pigeonholes is at least 
\begin{align*}
\underline{c} \cdot \sum_{k=1}^K \cdot L_k \cdot (n_k)^\frac{1}{2} = \underline{c} \cdot \sqrt{\Big(\sum_{k=1}^K L_k^2\Big) \Big(\sum_{k=1}^K n_k\Big)} = \underline{c} \sqrt{T \sum_{k=1}^K L_k^2}.
\end{align*}

Next, since $L_k^2$ is convex in $L_k$, by Jensen's inequality, 
\begin{align*}
\sqrt{T \sum_{k=1}^K L_k^2} \geq \sqrt{T} \sqrt{K (\frac{1}{K})^2} = \sqrt{\frac{T}{K}},
\end{align*}
where the inequality takes equality if and only if $L_k = \frac{1}{K}$ for any $k \in [K]$. 
If there are at most $K=O(T^{\frac{1}{2}})$ pigeonholes, the above expression yields a discrepancy on the order of $\underline{c} \cdot \sqrt{\frac{T}{K}} = \Omega(T^{\frac{1}{4}})$.

Second, if there are at least $\Omega(T^{\frac{1}{2}})$ pigeonholes, the adversarial sequence places an odd number of subjects in each pigeonhole.
Under this sequence, there is one subject unmatched in each pigeonhole. 
Due to Lemma~\ref{lem:Pigeonhole:p=1}, these $\Omega(T^{\frac{1}{2}})$ unmatched subjects lead to a discrepancy of $\Omega(T^{\frac{1}{4}})$.
\Halmos
\endproof

\section{Performance Analyses of Different Designs in the General Cases}
\label{sec:MultiDimension}

So far we have seen that, in the case where there is one single continuous dimension, the pigeonhole design has a smaller discrepancy than the completely randomized design;
yet it has a larger discrepancy than the matched-pair design.
In this section, we will analyze the discrepancies of the three aforementioned designs in the general multi-dimensional cases.
Note that Definitions~\ref{defn:MatchedPairDesign} --~\ref{defn:PigeonholeDesign} of the three designs are general.
Our definitions of three designs all hold in the general cases.
It is their theoretical performances and the associated analyses that will be different.

In the general multi-dimensional cases, the theoretical performances of the aforementioned designs critically depend on $p$ the number of continuous covariates.
We distinguish three cases.
First, all the covariates are discrete, i.e., $p = 0, q \geq 1$.
In this case, the matched-pair design has a discrepancy on the order of $\Theta(1)$; the completely randomized design has an expected discrepancy on the order of $\Theta(T^{\frac{1}{2}})$; the pigeonhole design has an expected discrepancy on the order of $\Theta(1)$.
This case is more common than it seems to be, as many online platforms keep their user demographics using categorical, or even binary, records.
We also conduct simulations in Section~\ref{sec:Simulations:ATE} using this data.

Second, there is one continuous covariate and possibly many discrete covariates, i.e., $p=1, q \geq 0$.
In this case, the performances are similar to Lemmas~\ref{lem:MatchedPair:p=1}--~\ref{lem:Pigeonhole:p=1}.
The matched-pair design has a discrepancy on the order of $\Theta(1)$; the completely randomized design has an expected discrepancy on the order of $\Theta(T^{\frac{1}{2}})$; the pigeonhole design has an expected discrepancy on the order of $\Theta\big(T^{\frac{1}{4}}\big)$ (ignoring logarithmic factors).
The first and the second cases are where the pigeonhole design demonstrates the most values.

Third, there are more than one continuous covariates and possibly many discrete covariates, i.e., $p \geq 2, q \geq 0$.
In this case, even the matched-pair design has a non-negligible discrepancy on the order of $\Theta\big(T^\frac{p-1}{p}\big)$.
The completely randomized design has an expected discrepancy at least $\Omega\big(T^\frac{p-1}{p}\big)$, as the discrepancy should be at least as big as the matched-pair design.
But we are unable to provide an upper bound on the discrepancy, as it still remains an open question in the optimal transport literature \citep{fournier2015rate}.
The pigeonhole design has an expected discrepancy of $O(T^\frac{p-1}{p} (\log{T})^\frac{3}{2})$ when $p=2$, and $O(T^\frac{p-1}{p})$ when $p\geq 3$.
The performance of the pigeonhole design closely matches that of the matched-pair design.

\subsection{Performance Analyses with Many Discrete Covariates}

In the first case, we introduce some more notations.
Recall that there are $q$ discrete dimensions, each having a finite number of supports.
For any $i \in [q]$, denote $S_i$ to be the finite supports along the $i$-th dimension.
Denote the number of supports in this dimension to be $\vert S_i \vert = m_i$, which does not depend on $T$.
Using such notations, there are a total of $\prod_{i\in[q]} m_i$ many supports in total.

\subsubsection{The matched-pair design.}

When there are only discrete covariates, the matched-pair design has $\Theta(1)$ discrepancy.
To see this, note that overall there are finitely many supports.
One feasible matching is to match every two subjects at the same support in pairs, and then to match the remaining subjects between different supports.
Since there is at most one unmatched subject from each support, there are at most $\prod_{i\in[q]} m_i$ many unmatched subjects in total.
Matching them in pairs incurs $\Theta(1)$ discrepancy.
Since this is one feasible matching, the minimum weight matching must incur discrepancy at most this much.
We formally state it as Theorem~\ref{thm:MatchedPair:p=0} below, and defer the proof to Appendix~\ref{sec:Proof:Thm123}.

\begin{theorem}
\label{thm:MatchedPair:p=0}
When $p = 0, q \geq 1$, the matched-pair design has a discrepancy on the order of $\Theta(1)$.
\end{theorem}

\subsubsection{The completely randomized design.}

In contrast to the matched-pair design, the completely randomized design incurs a significantly larger discrepancy when dealing with discrete covariates. 
To see this, consider a special case where there are only two distinct supports, each containing an even number of subjects.
In this special case, since there is an even number of subjects in each support, the matched-pair design assigns them into control and treated groups evenly, resulting in zero discrepancy.
The completely design, however, randomly assigns subjects to control and treated groups without considering the covariate information.
The discrepancy critically depends on how unevenly are the subjects assigned, which, under random assignment, has an expectation on the order of $\Theta(T^\frac{1}{2})$.
More generally, the above intuition also holds true when there are many discrete covariates.
We formally state it as Theorem~\ref{thm:CompletelyRandomized:p=0} below.

\begin{theorem}
\label{thm:CompletelyRandomized:p=0}
When $p = 0, q \geq 1$, the completely randomized design has an expected discrepancy on the order of $\Theta(T^\frac{1}{2})$.
\end{theorem}

The proof is in Appendix~\ref{sec:Proof:Thm123}.
Theorem~\ref{thm:CompletelyRandomized:p=0} offers a precise characterization of the expected discrepancy, which is shown to be on the order of $\Theta(T^\frac{1}{2})$.
This theorem serves as both an upper bound and a lower bound.
The proof of this result requires projecting the original multi-dimensional problem onto lower dimensions, and analyzing the performance on the lower dimensions separately.

\subsubsection{The pigeonhole design.}

When there are many discrete covariates but no continuous covariate, a pigeonhole design simply uses the finitely many supports as natural pigeonholes.
In each pigeonhole, the pigeonhole design sequentially matches two subjects into pairs, leaving at most one unmatched subject in each pigeonhole.
The subjects matched within each pigeonhole has zero discrepancy.
For the subjects that are not matched within each pigeonhole, matching these unmatched subjects across pigeonholes incurs $\Theta(1)$ discrepancy.
We formally state the above as Theorem~\ref{thm:Pigeonhole:p=0} below.

\begin{theorem}
\label{thm:Pigeonhole:p=0}
When $p = 0, q \geq 1$, the pigeonhole design using the natural pigeonholes $\cP(\cS) = S_1 \times S_2 \times ... \times S_q$ has an expected discrepancy on the order of $\Theta(1)$.
\end{theorem}

The proof can be found in Appendix~\ref{sec:Proof:Thm123}.
By comparing Theorem~\ref{thm:Pigeonhole:p=0} with Theorems~\ref{thm:MatchedPair:p=0} and~\ref{thm:CompletelyRandomized:p=0}, we see that the performance of the pigeonhole design closely matches that of the matched-pair design, and both are much smaller than that of the completely randomized design.
This setting with all discrete covariates highlights the value that the pigeonhole design brings as an alternative to the traditional completely randomized design.
Moreover, this discrete covariate setting is more common that what it seems to be.
Many online platforms collect user demographics using categorical and binary records, such as gender, browsing type, device type, and other behavior-targeting covariates.
Notably, the Yahoo! front page user click log data is even entirely stored as a binary matrix, i.e., all the covariates in the data are binary; see Section~\ref{sec:Simulations:ATE} for more details.

\subsection{Performance Analyses with One Continuous and Many Discrete Covariates}

\subsubsection{The matched-pair design.}

When there is one continuous covariate and many discrete covariates, the matched-pair design has $\Theta(1)$ discrepancy.
To see this, suppose that the first $q$ dimensions are discrete and the last $p=1$ dimension is continuous.
For each support in the first $q$ dimensions, we focus on those subjects whose first $q$ dimensions fall into this support.
These subjects only differ in the values in the last dimension, which is a continuous dimension between $[0,1]$.
Match all such subjects together, due to Lemma~\ref{lem:MatchedPair:p=1}, incurs at most $\Theta(1)$ discrepancy.
We repeat the same procedure for all the finitely many supports in the first $q$ discrete dimensions to see that the overall discrepancy is $\Theta(1)$.
We formalize the above statement as Theorem~\ref{thm:MatchedPair:p=1} below, and defer its proof to Appendix~\ref{sec:Proof:Thm456}.

\begin{theorem}
\label{thm:MatchedPair:p=1}
When $p = 1, q \geq 0$, the matched-pair design has a discrepancy on the order of $\Theta(1)$.
\end{theorem}

\subsubsection{The completely randomized design.}

Then completely randomized design again incurs a significantly larger discrepancy than the matched-pair design.
Informally, we see that on the $q$ discrete covariates, the completely randomized design incurs an expected discrepancy on the order of $\Theta(T^\frac{1}{2})$;
on the only continuous covariate, the completely randomized design incurs an expected discrepancy on the order of $\Theta(T^\frac{1}{2})$, as well.
Combining both parts, the completely randomized design should incur an overall expected discrepancy that is on the same order.
We formally state it as Theorem~\ref{thm:CompletelyRandomized:p=1} below, and defer its proof to Appendix~\ref{sec:Proof:Thm456}.

\begin{theorem}
\label{thm:CompletelyRandomized:p=1}
When $p = 1, q \geq 0$, the completely randomized design has an expected discrepancy on the order of $\Theta(T^\frac{1}{2})$.
\end{theorem}

\subsubsection{The pigeonhole design.}

When there are both continuous and discrete covariates, a pigeonhole design combines both the uniform pigeonholes in the continuous dimension and the natural pigeonholes in the discrete dimensions.
Informally, we see that on the $q$ discrete covariates, the pigeonhole design incurs an expected discrepancy on the order of $\Theta(1)$;
on the only continuous covariate, the completely randomized design incurs an expected discrepancy on the order of $\Theta\big(T^\frac{1}{4} (\log{T})^\frac{1}{2}\big)$.
Combining both parts, the completely randomized design should incur an overall expected discrepancy that is on the order of $\Theta\big(T^\frac{1}{4} (\log{T})^\frac{1}{2}\big)$.
We formally state it as Theorem~\ref{thm:Pigeonhole:p=1} below, and defer its proof to Appendix~\ref{sec:Proof:Thm456}.

\begin{theorem}
\label{thm:Pigeonhole:p=1}
When $p = 1, q \geq 0$, the pigeonhole design using pigeonholes $\cP(\cS) = S_1 \times S_2 \times ... \times S_q \times \left\{[0, T^{-\frac{1}{2}}), [T^{-\frac{1}{2}},2T^{-\frac{1}{2}}), \ldots, [1-T^{-\frac{1}{2}},1] \right\}$ has an expected discrepancy on the order of $O\big(T^\frac{1}{4} (\log{T})^\frac{1}{2}\big)$. In contrast, any pigeonhole design must suffer from an expected discrepancy on the order of $\Omega(T^\frac{1}{4})$.
\end{theorem}

\subsection{Performance Analyses with At Least Two Continuous and Many Discrete Covariates}

\subsubsection{The matched-pair design.}

Unlike in the previous two cases (with at most one continuous covariate and many discrete covariates) where the minimum weight perfect matching can be found easily, the matched-pair design when there are at least two continuous covariates needs to solve problem \eqref{eqn:SmallestDiscrepancy}.
Finding an optimal solution to \eqref{eqn:SmallestDiscrepancy} could be computationally more challenging, though still polynomial time solvable \citep{derigs1988solving, lu2011optimal, oncan2013minimum}.
Not only is the matched-pair design more challenging to solve, but the discrepancy of the design is also much larger.
Theorem~\ref{thm:MatchedPair:p>=2} below characterizes the hardness of the matched-pair design.

\begin{theorem}
\label{thm:MatchedPair:p>=2}
When $p \geq 2$, the matched-pair design has a discrepancy on the order of $\Theta\big(T^{\frac{p-1}{p}}\big)$.
\end{theorem}

We explain the main idea of the proof of Theorem~\ref{thm:MatchedPair:p>=2} here in an unrigorous way, and defer the detailed proof to Appendix~\ref{sec:Proof:MatchedPair:p>=2}.

\proof{Sketch proof of Theorem~\ref{thm:MatchedPair:p>=2}.}
We prove the $\Omega\big(T^{\frac{p-1}{p}}\big)$ part, as the $O\big(T^{\frac{p-1}{p}}\big)$ part is suggested in \citet{bai2021inference}.
We construct an arrival sequence such that the discrepancy on this sequence is $\Omega\big(T^{\frac{p-1}{p}}\big)$. 
All subjects in this sequence takes one single value in the $q$ discrete dimensions.
In the $p$ continuous dimensions, we evenly split the covariate space $[0,1]^p$ into $T$ smaller hypercubes, such that each hypercube has edge length $T^{-\frac{1}{p}}$.
Then, let the covariate of each subject be at the center of each smaller hypercube.
In this sequence, the distance between the covariates of any two subjects is at least $T^{-\frac{1}{p}}$.
There is a total of $T/2$ pairs of subjects.
Therefore, the discrepancy is at least $T^{-\frac{1}{p}} \cdot T = T^{\frac{p-1}{p}}$.
Combining with Lemma~\ref{lem:MatchedPair:p>=2:UB} we finish the proof.
\Halmos
\endproof

The discrepancy with respect to $T$ increases faster as the dimension $p$ increases.
When $p$ is large, the discrepancy almost increases linearly.
This is the curse of dimensionality of matching estimators in the statistics literature, i.e., the dimension of the covariate information ``plays an important role in the properties of matching estimators'' \citep{abadie2006large}.

\subsubsection{The completely randomized design.}

The performance of the completely randomized design is challenging to analyze.
We only know that the expected discrepancy of the completely randomized design is at least on the order of $\Omega\big(T^{\frac{p-1}{p}}\big)$, as it cannot be smaller than that of the matched-pair design.
We state this claim in Theorem~\ref{thm:CompletelyRandomized:p>=2}, although it is only for a formatting preference.
The proof is trivial and thus omitted.

\begin{theorem}
\label{thm:CompletelyRandomized:p>=2}
When $p \geq 2$, the completely randomized design has an expected discrepancy on the order of $\Omega\big(T^{\frac{p-1}{p}}\big)$.
\end{theorem}

It is challenging to clearly derive the upper bound of the expected discrepancy for the completely randomized design \citep{fournier2015rate}.
Nonetheless, we can still calculate the expected discrepancy under a special family of arrival sequences.
This result is of its own mathematical interests, and will be helpful in analyzing the expected discrepancy of the pigeonhole design.

We construct the family of arrival sequences as follows. 
First, let there be $p$ continuous dimensions and no discrete dimensions.
Then, for the $p$ continuous dimensions, we split the $p$-dimensional unit hypercube into $T$ smaller hypercubes, each with edge-length $T^{-\frac{1}{p}}$.
With all the smaller hypercubes given, let there be exactly one subject in each smaller hypercube.
Mathematically, the family of arrival sequences can be defined as follows.
First, for any set $\bX$, denote ${\displaystyle \bX^{p}}$ to be the $p$-ary Cartesian power of $\bX$, i.e., ${\displaystyle \bX^{p}=\underbrace {\bX \times \bX \times \cdots \times \bX}_{p}=\left\{(x^{1},\ldots ,x^{p})\ |\ x^{i}\in \bX\ {\text{for every}}\ i\in \{1,\ldots ,p\}\right\}.}$
Then, the family of arrival sequences is defined to be
\begin{multline*}
\mathcal{X} = \left\{ (\bm{x}_1, \bm{x}_2, ..., \bm{x}_T) \big| \forall \mathbb{X} \in \left\{[0, T^{-\frac{1}{p}}), [T^{-\frac{1}{p}},2T^{-\frac{1}{p}}), \ldots, [1-T^{-\frac{1}{p}},1] \right\}^p,\right. \\
\left.\exists t, s.t.  \ \bm{x}_t \in \mathbb{X}, \text{ and } \forall t' \ne t, \bm{x}_t' \notin \mathbb{X} \right\}.
\end{multline*}
Under such a family of arrival sequences, the expected discrepancy of the completely randomized design can be upper bounded.

\begin{lemma}
\label{lem:CompletelyRandomized:p>=2}
For any $\bm{x} \in \mathcal{X}$, the expected discrepancy of the completely randomized design under arrival sequence $\bm{x}$ is on the order of $O\big(T^{\frac{p-1}{p}} \log{T}\big)$ when $p = 2$; and on the order of $O\big(T^{\frac{p-1}{p}}\big)$ when $p \geq 3$.
\end{lemma}

The proof of Lemma~\ref{lem:CompletelyRandomized:p>=2} is non-trivial.
We explain the main idea of the proof of Lemma~\ref{lem:CompletelyRandomized:p>=2} here in an unrigorous way, and defer the detailed proof to Appendix~\ref{sec:Proof:lem:CompletelyRandomized:p>=2}.

\proof{Sketch proof of Lemma~\ref{lem:CompletelyRandomized:p>=2}.}
To illustrate, we provide a sketch for $p=2$.
The proof proceeds by induction.
The idea is to show that the discrepancy of the completely randomized design is at most $\sqrt{2}\sqrt{T}\log T\sqrt{\log T}$ for any $T=2^{2i}$ for some $i \in \mathbb{N}$. 
In the induction step, we consider the case where we have doubled the number of subjects and hypercubes, i.e., $T=2^2T_0$, where $T_0$ is the number of subjects and hypercubes from the previous step of the induction. 
To do this, we divide the original unit square into four smaller squares, and we analyze the discrepancy within each of these smaller squares separately.
We use a coupling argument to apply the induction hypothesis to bound the discrepancy within each square.
Finally, we add up their contributions to the total discrepancy, and show that the total discrepancy still satisfies the desired upper bound.
\Halmos
\endproof

Again, note that Lemma~\ref{lem:CompletelyRandomized:p>=2} is only for a special family of arrival sequences, but not for all the arrival sequences.
An upper bound for any arrival sequences still remains unclear.

\subsubsection{The pigeonhole design.}

The performance of the pigeonhole design closely matches that of the matched-pair design.
When $p=2$, the gap is up to logarithmic factors; when $p\geq 3$, there is no gap between the orders.

Recall that for any set $X$, $X^p$ is the $p$-ary Cartesian power of $X$.

\begin{theorem}
\label{thm:Pigeonhole:p>=2}
Fix any $\phi \in (0,1)$, and any $c > 1$. 
The pigeonhole design using pigeonholes $\cP(\cS) = S_1 \times S_2 \times ... \times S_q \times \left\{[0, c^{\frac{1}{p}}T^{-\phi}), [c^{\frac{1}{p}}T^{-\phi},2c^{\frac{1}{p}}T^{-\phi}), \ldots, [1-c^{\frac{1}{p}}T^{-\phi},1] \right\}^p$ has an expected discrepancy:
\begin{itemize}
\item on the order of $O\left(T^{1-\phi} + T^{(p-1)\phi}(\log{T})^\frac{3}{2}\right)$ when $p=2$;
\item on the order of $O\left(T^{1-\phi} + T^{(p-1)\phi}\right)$ when $p\geq 3$.
\end{itemize}
\end{theorem}

\begin{corollary}
\label{coro:Pigeonhole:p>=2}
When $\phi = \frac{1}{p}$, the pigeonhole design using uniform pigeonholes and an associated balancing parameter has an expected discrepancy:
\begin{itemize}
\item on the order of $O\big(T^{\frac{p-1}{p}}(\log{T})^\frac{3}{2}\big)$ when $p=2$;
\item on the order of $O\big(T^{\frac{p-1}{p}}\big)$ when $p\geq 3$.
\end{itemize}
\end{corollary}

We provide the pseudo-codes for the pigeonhole design under the parameters in Theorem~\ref{thm:Pigeonhole:p>=2}, in Algorithm~\ref{alg:Pigeonhole:p>=2} in Appendix~\ref{sec:Pseudo:p=1}.
We explain the main idea of the proof of Theorem~\ref{thm:Pigeonhole:p>=2} here in an unrigorous way, and defer the detailed proof to Appendix~\ref{sec:Proof:Pigeonhole:p>=2}.

\proof{Sketch proof of Theorem~\ref{thm:Pigeonhole:p>=2}.}
To illustrate, we provide a sketch when $q=0$.
We consider the moment when either the control or the treated group reaches $T/2$ in size, and denote this moment as $\tau$.
We consider the following three perspectives.
First, consider the first $\tau$ arriving subjects.
We match every pair of subjects that arrive at the same pigeonhole, with at most one subject unmatched from each pigeonhole.
For the matched subjects, we show that the discrepancy is on the order of $O(T^{1-\phi})$.

Second, there are at most $T^{p\phi}$ many unmatched subjects.
For all the unmatched subjects, they are randomly assigned to either the control or the treated group.
If there are exactly the same number of subjects that are assigned into the control group as into the treated group, we use a coupling technique to show that these $T^{p\phi}$ many unmatched subjects will have an expected discrepancy on the order of $O(T^{(p-1)\phi} (\log{T})^\frac{3}{2})$ when $p = 2$ and $O(T^{(p-1)\phi})$ when $p \geq 3$.

Third, there might not be exactly the same number of subjects that are assigned into the control group as into the treated group.
But due to concentration, the difference between the two groups of subjects is upper bounded by $2T^\frac{p\phi}{2} (\log{T})^\frac{1}{2}$ with high probability.
Together with the last $(T-\tau)$ subjects, they have a discrepancy upper bounded by $O\big(T^\frac{p\phi}{2} (\log{T})^\frac{1}{2}\big)$.
Combining three parts we finish the proof.
\Halmos
\endproof


\subsubsection*{Benefits of pigeonhole design.}

It is worth noting that, when all the covariates are continuous, although Theorem~\ref{thm:Pigeonhole:p>=2} and Corollary~\ref{coro:Pigeonhole:p>=2} do not suggest a better theoretical performance than Theorem~\ref{thm:CompletelyRandomized:p>=2} in the worst case, we can still show that the the pigeonhole design is better than the completely randomized design when the arrival sequence is highly clustered.

\begin{figure}[!tb]
\minipage{0.48\textwidth}
\centering
\includegraphics[width=0.6\textwidth]{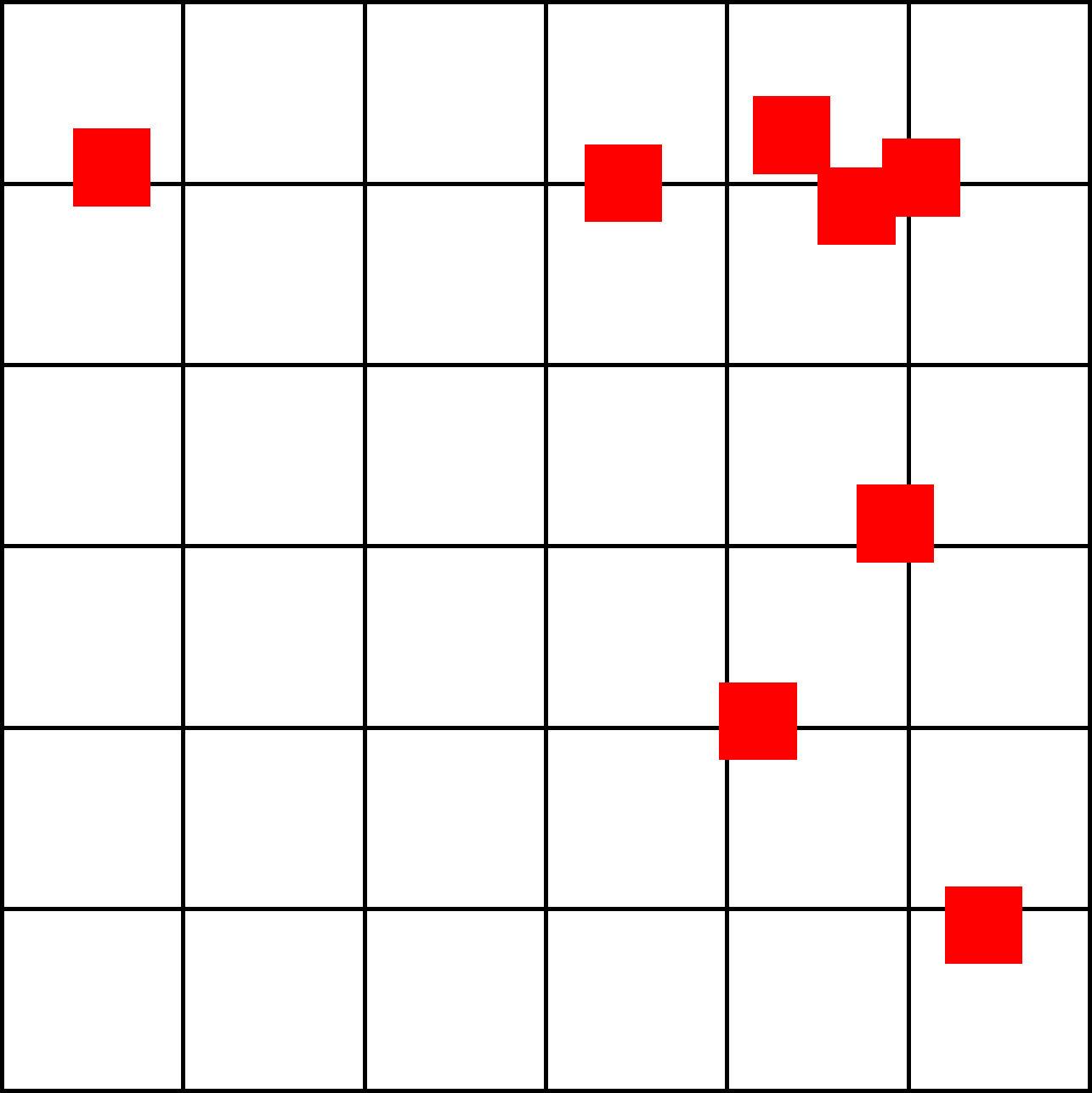}
\endminipage
\hfill
\minipage{0.48\textwidth}
\centering
\includegraphics[width=0.6\textwidth]{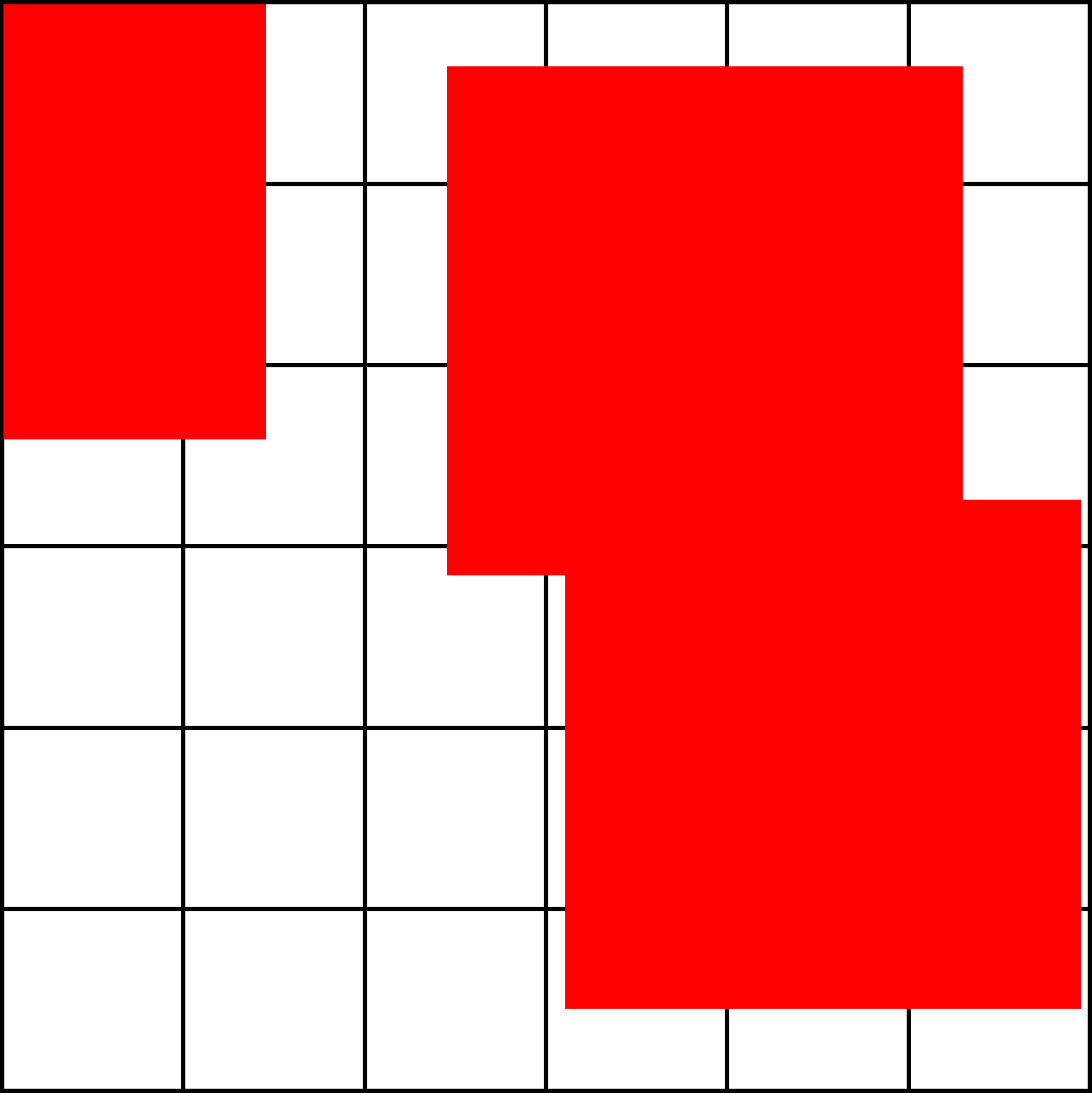}
\endminipage
\vspace{2mm}
\caption{Illustration of a highly clustered arrival sequence (left) and a weakly clustered arrival sequence (right).}
\label{fig:Clustered}
\end{figure}

We define a highly clustered arrival sequence as follows.
For any $\bm{v} \in [0,1]^p$ and $\gamma \geq \underline{\gamma} > 0$, define $\mathcal{C}(\bm{v}, T^{-\gamma}) = \{\bm{u}: \| \bm{u} - \bm{v} \|_{\infty} \leq \frac{1}{2}T^{-\gamma} \}$ to be a cluster centered around $\bm{v}$ with diameter $T^{-\gamma}$.
We say the arrival sequence is ``clustered'' if there exists a finite collection of $N$ clusters $\mathcal{C}(\bm{v}_1, T^{-\gamma})$, $\mathcal{C}(\bm{v}_2, T^{-\gamma})$, $...$, $\mathcal{C}(\bm{v}_N, T^{-\gamma})$, such that all the subjects come from the collection of such clusters $\bigcup_{n=1}^N \mathcal{C}(\bm{v}_n, T^{-\gamma})$.
In addition, we say the arrival sequence is ``highly clustered'' if $\underline{\gamma} > \frac{1}{2}$; we say the arrival sequence is ``weakly clustered'' if $\underline{\gamma} < \frac{1}{2}$.
Note that it is possible that $\gamma > \frac{1}{2} > \underline{\gamma}$, which means that the arrival sequence is highly clustered but we do not know this information; instead, our knowledge only knows $\underline{\gamma} < \frac{1}{2}$.
If this is the case, we say the arrival sequence is weakly clustered.
See Figure~\ref{fig:Clustered} for an illustration.

Now we show that the pigeonhole design has superior performance over the completely randomized design when the the arrival sequence is highly clustered.

\begin{theorem}[Benefits of Pigeonhole Design]
\label{thm:Pigeonhole:clustered}
Let there be $p$ continuous dimensions.
Suppose $\gamma \geq \underline{\gamma} > 0$.
Suppose all the subjects come from a finite collection of $N$ clusters $\bigcup_{n=1}^N \mathcal{C}(\bm{v}_n, T^{-\gamma})$.
\begin{enumerate}
\item If $\gamma$ is known, but the centers $\bm{v}_n, \forall n \in [N]$ are unknown, then fix $c > 1$ and $\zeta = \frac{1}{p} + \frac{p-1}{p} \gamma$, the pigeonhole design using pigeonholes $\cP(\cS) = \left\{[0, c^{\zeta}T^{-\zeta}), [c^{\zeta}T^{-\zeta},2c^{\zeta}T^{-\zeta}), \ldots, [1-c^{\zeta}T^{-\zeta},1] \right\}^p$ has an expected discrepancy:
\begin{itemize}
\item on the order of $O\big(T^{\frac{p-1}{p}(1-\gamma)} \big(\log{T}\big)^\frac{3}{2}\big)$ when $p = 2$;
\item on the order of $O\big(T^{\frac{p-1}{p}(1-\gamma)}\big)$ when $p \geq 3$.
\end{itemize}
\item If both $\gamma$ and the centers $\bm{v}_n, \forall n \in [N]$ are unknown, then fix $c > 1$ and $\zeta = \frac{1}{p} + \frac{p-1}{p} \underline{\gamma}$, the pigeonhole design using pigeonholes $\cP(\cS) = \left\{[0, c^{\zeta}T^{-\zeta}), [c^{\zeta}T^{-\zeta},2c^{\zeta}T^{-\zeta}), \ldots, [1-c^{\zeta}T^{-\zeta},1] \right\}^p$ has an expected discrepancy:
\begin{itemize}
\item on the order of $O\big(T^{\frac{p-1}{p}(1-\underline{\gamma})} \big(\log{T}\big)^\frac{3}{2}\big)$ when $p = 2$;
\item on the order of $O\big(T^{\frac{p-1}{p}(1-\underline{\gamma})}\big)$ when $p \geq 3$.
\end{itemize}
\end{enumerate}
No matter if $\gamma$ is known or not, the matched pair design has a discrepancy on the order of $\Theta\big(T^{\max\{1-\frac{1}{p}-\gamma, 0\}}\big)$; the completely randomized design has an expected discrepancy on the order of at least $\Omega\big(T^{\max\{1-\frac{1}{p}-\gamma, \frac{1}{2}\}}\big)$.
\end{theorem}

Note that, for Theorem~\ref{thm:Pigeonhole:clustered}, the pigeonhole design does not need to know the information of where the clusters are located.
It only needs to know that the arrival sequence is clustered.
Depending on whether the diameter is known or not, we have different performance guarantees.
And when the arrival sequence is highly clustered, i.e., $\underline{\gamma} > \frac{1}{2}$, regardless of whether the diameter is known or not, the pigeonhole design outperforms the completely randomized design.
We explain the main idea of the proof of Theorem~\ref{thm:Pigeonhole:clustered} here in an unrigorous way, and defer the detailed proof to Appendix~\ref{sec:Proof:thm:Pigeonhole:clustered}.

\proof{Sketch proof of Theorem~\ref{thm:Pigeonhole:clustered}.}
If the edge length of pigeonholes is bigger than $T^{-\gamma}$ the diameter of the clusters, the worst-case scenario emerges when two clusters are situated at two corners of the pigeonhole. 
This situation generates a discrepancy of $T \cdot c^{\zeta}T^{-\zeta} = O(T^{1-\zeta})$. 
On the other hand, if the edge length of pigeonholes is smaller than the diameter of the clusters, a single cluster overlaps with at most $O(T^{\zeta-\gamma})$ pigeonholes. 
Due to Theorem~\ref{thm:Pigeonhole:p>=2}, they will contribute $\sqrt{p} \cdot O\big(T^{\max\{(p-1)(\zeta-\gamma), \frac{p}{2}(\zeta-\gamma)\}}\big) = O\big(T^{(p-1)(\zeta-\gamma)}\big)$ to the total discrepancy.

If the value of $\gamma$ is known, the ideal edge length of pigeonholes is to set $\zeta$ to balance the two scenarios discussed above, which obtains a discrepancy on the order of $O\big(T^{\frac{p-1}{p} (1-\gamma)}\big)$.
If the value of $\gamma$ is unknown, the only thing to do is to pretend $\gamma = \underline{\gamma}$ and use the same design as in the known case but replacing $\gamma$ by $\underline{\gamma}$.
This obtains a discrepancy on the order of $O\big(T^{\frac{p-1}{p} (1-\underline{\gamma})}\big)$.
\Halmos
\endproof

\section{Simulation Study: Average Treatment Effect Estimation Using Yahoo! Data}
\label{sec:Simulations:ATE}

In this section, we conduct simulations using user click log data from Yahoo! front page\footnote{This data set can be downloaded as ``R6B - Yahoo! Front Page Today Module User Click Log Dataset, version 2.0'' at \href{http://webscope.sandbox.yahoo.com/catalog.php?datatype=r}{http://webscope.sandbox.yahoo.com/catalog.php?datatype=r}.} to illustrate the values of the pigeonhole design.
We combine the Yahoo! data with a synthetic data generating process.
By fixing one realization of the data generating process, we simulate the average treatment effect estimator under the pigeonhole design, and compare it against the same estimator under the completely randomized design.
The simulations suggest a $10.2\%$ reduction in the variance.
We describe the details below.

\subsubsection*{Raw data and pre-processing.}

The dataset stores user click data when they visited the ``Yahoo! front page today module'' between October 2 and October 16, 2011.
Each record is stored as a row in the data, and there are two important components associated with each record.
The first is a binary response, $0$ for no-click and $1$ for click;
the other is a 136-dimensional binary covariate vector such as gender, browsing type, device type, and other behavior-targeting covariates.
Since the dataset is huge, we only focus on the first $T = 100000$ records of user visits, which all happened on October 2, 2011.

We pre-process the data as follows.
First, we discard the covariates that are constant throughout the data.
We discard those covariates such that less than $1\%$ of data is zero, and those covariates such that less than $1\%$ of data is one.
After discarding such constant covariates, there are only $39$ covariates remaining. 
Second, we use the stepwise selection method to find the most important covariates among these $39$ covariates.
Stepwise selection yields $16$ covariates in total.
See Table~\ref{tbl:SampleData} for a sample of data after stepwise selection.
Our two-step procedure is in contrast to the feature selection procedure in \citet{bhat2020near}, where they first removed the duplicate and co-linear covariates and then selected the features at random until up to $40$ features were collected.

\begin{table}[!tb]
\centering
\TABLE{Data sample after stepwise selection.
\label{tbl:SampleData}}
{\small
\begin{tabular}{c|cccccccccccccccc}
Click & X3 & X4 & X5 & X7 & X10 & X13 & X17 & X24 & X26 & X27 & X30 & X32 & X33 & X35 & X37 & X38 \\ \hline
0     & 0  & 1  & 1  & 1  & 1   & 1   & 1   & 0   & 0   & 0   & 0   & 1   & 0   & 0   & 0   & 0   \\
0     & 0  & 1  & 1  & 1  & 1   & 0   & 0   & 0   & 0   & 0   & 0   & 0   & 0   & 0   & 0   & 0   \\
1     & 1  & 0  & 1  & 1  & 1   & 1   & 1   & 0   & 1   & 1   & 0   & 0   & 0   & 0   & 1   & 0   \\
...   & \multicolumn{16}{c}{...}
\end{tabular}
}
{This table presents a sample of the first $3$ rows of data after stepwise selection. There are $16$ binary covariates, whose names are censored and replaced by their respective column numbers. The outcome is also binary, with $1$ standing for a click and $0$ standing for a no-click.}
\end{table}

\subsubsection*{Data generating process.}

We combine the Yahoo! data with a synthetic data generating process.
As there is no treatment intervention in the dataset, we will need to generate treatment outcomes for the simulation.
To do so, we first build a linear regression model based on the dataset to predict the click probabilities.
To reflect the practice that some covariates might be more important than other covariates, we increase the magnitude of the $5$ largest coefficients in the linear regression model.
See Appendix~\ref{sec:simu:Robustness} for robustness checks of this data generating process.
We then use this linear model to generate the click probabilities under ``control,'' which yields a $5.08\%$ probability on average.
We then add $U[0, 5.08\%]$ independent and identically distributed uniform noises to the above probabilities, and generate the click probabilities under ``treatment,'' which yields a $7.62\%$ probability on average.
If a probability is either negative or greater than $1$, we trim the probability number to fall between $[0,1]$.

Finally, we generate independent Bernoulli variables using such probabilities for both control and treatment.
When a subject is assigned to the control group, i.e., $W_t = 0$, we observe the binary variable $Y_t(0)$ obtained from its click probability under control; when it is assigned to the treated group instead, i.e., $W_t = 1$, we observe the other binary variable $Y_t(1)$ obtained from its click probability under treatment.
For this simulation, we fix one realization of the above data generating process.
Since we generate both outcomes under control and treatment, we can calculate the average treatment effect as
\begin{align*}
\tau = \frac{1}{T} \sum_{t=1}^T Y_t(1) - \frac{1}{T} \sum_{t=1}^T Y_t(0) = 7.64\% - 5.07\% = 2.57\%.
\end{align*}

\subsubsection*{Experimental design and difference-in-means estimator.}

In this simulation, we use both the pigeonhole design and the completely randomized design to determine the treatment assignment $W_t$ for each subject $t$.
Once the treatment assignment is determined, the corresponding outcome $Y_t(W_t)$ is observed.
We then use the following difference-in-means estimator to estimate the average treatment effect,
\begin{align*}
\widehat{\tau} = \frac{2}{T} \sum_{t=1}^T \bI\{W_t = 1\} Y_t - \frac{2}{T} \sum_{t=1}^T \bI\{W_t = 0\} Y_t.
\end{align*}
Denote $\widehat{\tau}^\mathrm{PhD}$ and $\widehat{\tau}^\mathrm{CRD}$ as the difference-in-means estimators obtained when the treatment assignments $W_t$ are determined by the pigeonhole design (PhD) and the completely randomized design (CRD).
Note that, as we consider random experiments, these two estimators are random in nature.
We simulate $10000$ samples of both estimators to obtain their respective empirical distributions.

\subsubsection*{Results.}

\begin{figure}[!tb]
\center
\includegraphics[width=0.7\textwidth]{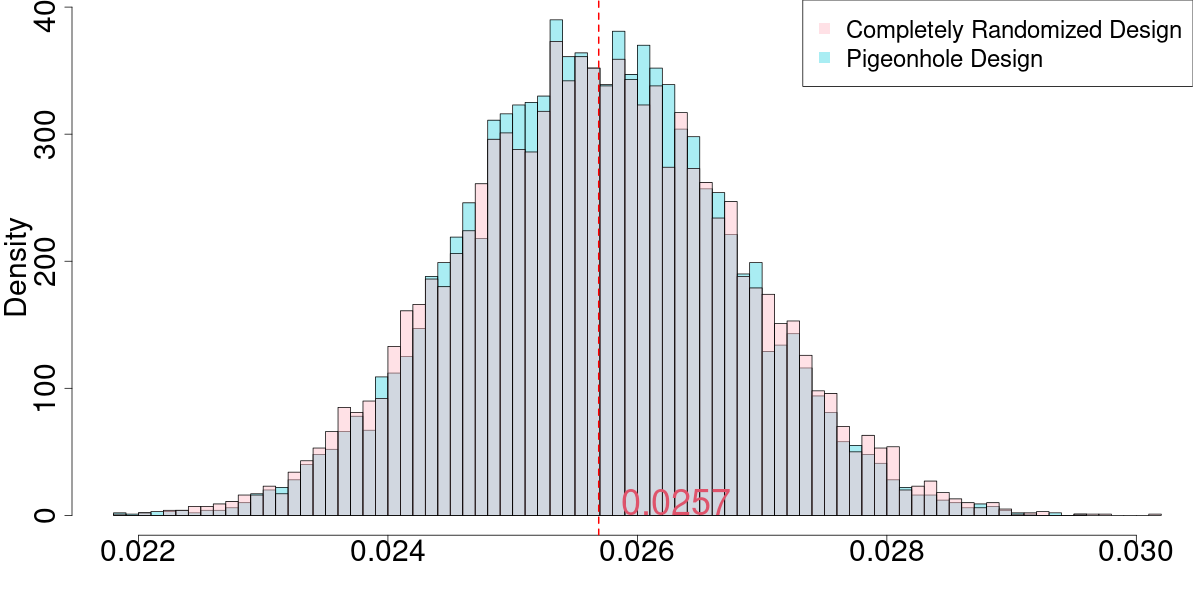}
\caption{Empirical distributions of the estimators under two different designs.}
\label{fig:Histogram}
\end{figure}

Now we compare the performances of $\widehat{\tau}^\mathrm{PhD}$ and $\widehat{\tau}^\mathrm{CRD}$.
We present their empirical distributions in Figure~\ref{fig:Histogram}.
In Figure~\ref{fig:Histogram}, the pink histogram stands for the distribution of $\widehat{\tau}^\mathrm{CRD}$ while the blue one stands for the distribution of $\widehat{\tau}^\mathrm{PhD}$.
The red vertical line indicates the average treatment effect $\tau$.
Figure~\ref{fig:Histogram} shows that the pigeonhole design yields an estimator that is more concentrated around $\tau$ than the completely randomized design.
Quantitatively, $\bE[\widehat{\tau}^\mathrm{CRD}] = \bE[\widehat{\tau}^\mathrm{PhD}] = 2.57\% = \tau$. 
So both estimators are unbiased.
Moreover, $\Var[\widehat{\tau}^\mathrm{CRD}] = 1.26 \cdot 10^{-6}$ and $\Var[\widehat{\tau}^\mathrm{PhD}] = 1.13 \cdot 10^{-6}$.
The pigeonhole design reduces $10.2\%$ variance compared to the completely randomized design.

\begin{figure}[!tb]
\center
\includegraphics[width=0.7\textwidth]{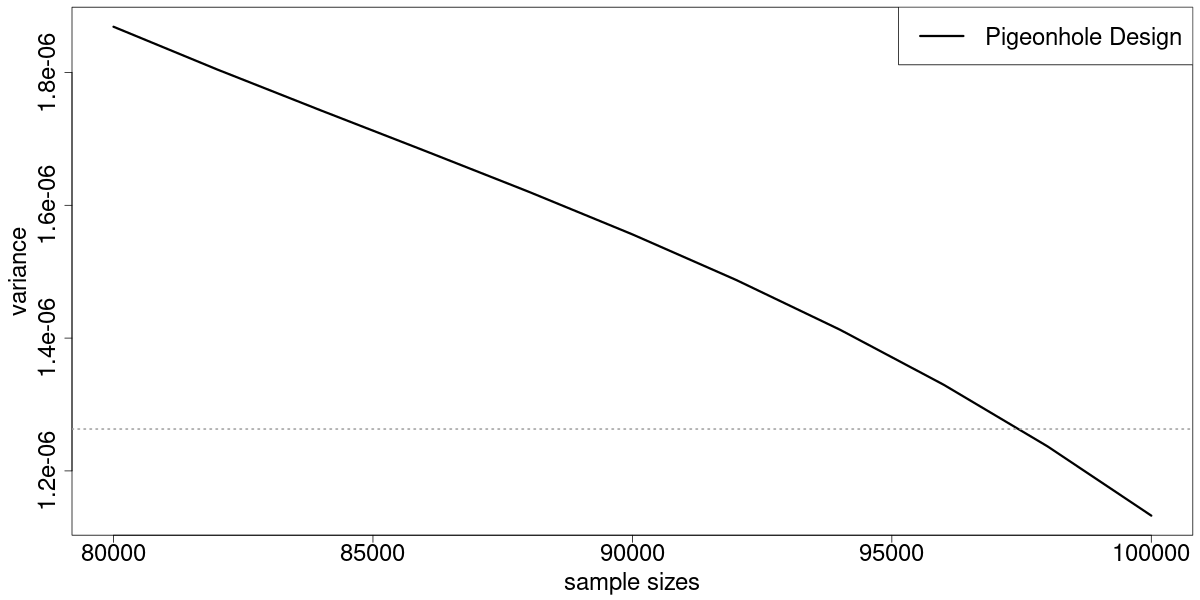}
\caption{Simulated variances of the estimator under pigeonhole design under different sample sizes.}
\label{fig:SampleSizes}
\end{figure}

For practitioners, variance reduction typically means that an experimenter can reduce the sample size required when conducting an experiment.
In Figure~\ref{fig:SampleSizes}, we simulate how much sample would be required to achieve the same level of variance.
To see this, we first fix the completely randomized design to have $T = 100000$ samples, and use the dotted horizontal line to denote its expected variance $\Var[\widehat{\tau}^\mathrm{CRD}]$.
We then conduct experiments using the pigeonhole design and change the sample size to come from $\{80000, 82000, ..., 100000\}$.
We use spline fitting to smooth out the curve.
Figure~\ref{fig:SampleSizes} suggests an approximate reduction of $3\%$ in sample size, if a pigeonhole design is used to achieve the same level of variance.

\section{Practical Suggestions, Limitations, and Future Research Directions}
\label{sec:Conclusion}

In this paper, we consider the problem of designing online experiments, when experimental subjects with heterogeneous covariate information arrive sequentially.
Each subject must be immediately assigned into either the control or the treated group, with an objective of minimizing the total discrepancy.
To solve this problem, we propose the pigeonhole design method.
In a pigeonhole design, we first partition the covariate space into smaller spaces, and then balance the number of control and treated subjects when they arrive at the same pigeonhole.
In both single-dimensional and multi-dimensional cases, we have shown strong theoretical performance of the pigeonhole design, compared to the completely randomized design as a benchmark.
In the multi-dimensional case, the theoretical performance of the pigeonhole design even closely matches that of the match-pair design, which is an offline design that requires clairvoyant covariate information.
We conduct simulations using Yahoo! data and demonstrate the effectiveness of the pigeonhole design.

\subsubsection*{Practical suggestions.}

When a firm decide to use a pigeonhole design, they have to make multiple decisions to ensure that the results are reliable.
First, the firm must determine the possible subject covariates using substantive domain knowledge, which adequately captures the impact on the objective of evaluation; see \cite{kohavi2020trustworthy} for a discussion of metric definition strategies.

Second, after the metrics are defined, the firm need to select the proper covariates.
It is fine to select more covariate information than necessary, as \citet{greevy2004optimal} explicitly comments that ``blocking on irrelevant covariates wastes computer time but does not harm statistical efficiency or power.''
This has also been discussed in, e.g., \citealp[Section~4.26]{cochran1957experimental}, as well as \citet{chase1968efficiency}.
Yet it is still recommended that the firm select the more relevant covariates to keep the number of covariates reasonably small.
This is because the theoretical results derived in this paper may get worse, as the constant of the bounds may grow large when the dimension of discrete covariates $q$ grows.

Third, after the covariates are determined, the firm must use in-depth knowledge of the covariate information to describe the covariates as binary, categorical, or continuous.
Categorical covariates can be represented using multiple binary covariates when the number of categories is small; and they can be treated as continuous covariates when the categories are granular enough. 
Recall that, in Theorem~\ref{thm:Pigeonhole:p>=2}, the number of binary dimensions do not play a role in the discrepancy.
This suggests that the existence of binary covariates does not hurt the performance of the pigeonhole design.
We recommend that whenever possible, the experimenter distinguishes binary covariates and continuous covariates.

\subsubsection*{Limitations and future research directions.}

We point out two limitations of our paper, which could each lead to a future research direction.
First, this paper models the covariates of all the experimental subjects to be adversarially chosen.
Under the adversarial model, further improving the $O(T^\frac{1}{4})$ expected discrepancy in Lemma~\ref{lem:Pigeonhole:p=1} is difficult, as the number of subjects in any pigeonhole could be adversarially chosen.
When the covariates are drawn from a known stochastic distribution, it remains unknown if there will be better designs of experiments, such as allowing the pigeonholes to be adaptively partitioned.
It is an interesting question to study the generic re-solve techniques or semi-adaptive policies under the stochastic setting, which have proved to be useful in the revenue management and stochastic control literature \citep{arlotto2019uniformly, bumpensanti2018re, chen2016real, chen2023real, jasin2012re, jasin2014reoptimization, lei2020real, vera2019bayesian}.

Second, we make the control group and treated group to have the same number of subjects.
Although this is a widely adopted assumption to make, especially in the literature of matching without replacement, this is not desirable in many applications when treatment assignments are costly.
It would require a new objective function, e.g., minimizing the total discrepancy plus variance \citep{bertsimas2015power}, or some modeling assumptions on the distributions of the covariate information \citep{bhat2020near}.
We leave this as a future research direction.

\ACKNOWLEDGMENT{
The authors would like to thank Dean Eckles, Rui Gao, David Hughes, Nian Si, John Silberholz, Chung-Piaw Teo, and participants at Harvard University, Nanyang Technological University, University of Chicago, and University of Toronto whose comments have significantly improved the manuscript.
The authors would also like to thank the department editor George Shanthikumar, the anonymous associate editor, and the anonymous referees whose comments significantly improved the manuscript throughout the reviewing process.
In particular, the proof of Lemma~\ref{lem:Pigeonhole:p=1} borrowed ideas from the reports of two anonymous referees.
}

\bibliographystyle{informs2014}
\bibliography{bibliography}

\ECSwitch


\ECHead{Online Appendix}

\vspace{8mm}
\begin{center}
{\large \textbf{Pigeonhole Design: Balancing Sequential Experiments}}\\[-0.9ex]
{\large \textbf{from an Online Matching Perspective}}\\[0.3ex]
Jinglong Zhao and Zijie Zhou\\[-0.9ex]
\today
\end{center}
\vspace{5mm}

\section{Pseudo-codes of the Pigeonhole Design}
\label{sec:Pseudo:p=1}

We provide the pseudo-codes for the pigeonhole design under the parameters in Lemma~\ref{lem:Pigeonhole:p=1}.

\begin{algorithm}[!h]
\caption{The Pigeonhole Design when $p=1, q = 0$}
\label{alg:Pigeonhole:p=1}
\begin{algorithmic}[1]
\STATE \textbf{Initialize:} $\eta \gets \frac{1}{2}$, $\cP(\cS) \gets \left\{[0, T^{-\eta}), [T^{-\eta},2T^{-\eta}), \ldots, [1-T^{-\eta},1] \right\}$.
\FOR{$t =1, 2, \ldots, T$}
\IF{One of the control or treated group has $\frac{T}{2}$ subjects}
\STATE{Assign $x_t$ to the other group.}
\ELSE
\STATE{Observe $x_t$ and find $I_k \in \cP(\cS)$ such that $x_t \in I_k$.}
\IF{There is even number of subjects in $I_k$}
\STATE{Assign $x_t$ to control or treated group with half probability each.}
\ELSE
\STATE{Assign $x_t$ to the group which has less subjects in interval $I_k$.}
\ENDIF
\ENDIF
\ENDFOR
\end{algorithmic}
\end{algorithm}

\clearpage

We then provide the pseudo-codes for the pigeonhole design under the parameters in Theorem~\ref{thm:Pigeonhole:p>=2}.

\begin{algorithm}[h]
\caption{The Pigeonhole Design when $p \geq 2, q \geq 0$}
\label{alg:Pigeonhole:p>=2}
\begin{algorithmic}[1]
\STATE \textbf{Input:} $c>1$.
\STATE \textbf{Initialize:} $\phi \gets \frac{1}{p}$, \\
$\cP(\cS) = S_1 \times S_2 \times ... \times S_q \times \left\{[0, c^{\frac{1}{p}}T^{-\phi}), [c^{\frac{1}{p}}T^{-\phi},2c^{\frac{1}{p}}T^{-\phi}), \ldots, [1-c^{\frac{1}{p}}T^{-\phi},1] \right\}^p$. 
\FOR{$t =1, 2, \ldots, T$}
\IF{One of the control or treated group has $\frac{T}{2}$ subjects}
\STATE{Assign $\bm{x}_t$ to the other group.}
\ELSE
\STATE{Observe $\bm{x}_t$ and find the unique $\mathbb{X}_k \in \cP(\cS)$ such that $\bm{x}_t \in \mathbb{X}_k$.}
\IF{There is even number of subjects in $\mathbb{X}_k$}
\STATE{Assign $\bm{x}_t$ to control or treated group with half probability each.}
\ELSE
\STATE{Assign $\bm{x}_t$ to the group that has less subjects in $\mathbb{X}_k$.}
\ENDIF
\ENDIF
\ENDFOR
\end{algorithmic}
\end{algorithm}

\clearpage

\section{Useful Lemmas}

To make this paper self-contained, we review in this section three existing results that we have referred to in the paper.
The first two Lemmas are useful concentration inequalities.
\begin{lemma}[Hoeffding's Inequality, \citet{hoeffding1994probability}]
\label{lem:hoeffding}
Let $H(n)$ be a binomial random variable with probability $p$ and trial $n$, then
$$P(\vert H(n)-pn \vert \leq \sqrt{n \log n}) \geq 1-\frac{2}{n^2}.$$
\end{lemma}

\begin{lemma}[Binomial Mean Absolute Deviation, Equation~(2.4) in \citet{blyth1980expected}]
\label{lem:blyth}
Let $X$ be a Binomial distribution with $n$ trials and success probability $p$, then we have
$$E\left[|X-np|\right]= \sqrt{\frac{2p(1-p)}{\pi}n}+O(n^{-\frac{1}{2}}).$$
\end{lemma}

Next, we describe Theorem~4.2 in \citet{bai2021inference}, which suggests an upper bound on the discrepancy of the matched-pair design when there are at least two continuous covariates.
\begin{lemma}
[Theorem~4.2, \citep{bai2021inference}]
\label{lem:MatchedPair:p>=2:UB}
When $p \geq 2$, the matched-pair design has a discrepancy on the order of $O\big(T^{\frac{p-1}{p}}\big)$.
\end{lemma}

\section{Proof of Lemma~\ref{lem:MatchedPair:p=1}}
\label{sec:Proof:MatchedPair:p=1}
\proof{Proof of Lemma~\ref{lem:MatchedPair:p=1}.}
Suppose the adversary chooses input $(x_1,x_2,...,x_T)$. 
As the matched-pair design is endowed with clairvoyant information, the decision maker knows $(x_1,x_2,...,x_T)$ at the beginning. 
Let $(x_1',x_2',...,x_T')$ be smallest-to-largest rearrangement of $(x_1,x_2,...,x_T)$ such that 
\[
x_1' \leq x_2' \leq ... \leq x_T'.
\]
The matched-pair design matches $(x_t',x_{t+1}')$ into a size-two pair for any odd $t$. Therefore, the discrepancy under the input $(x_1,x_2,...,x_T)$ is
\[
\sum_{\tau=1}^{\frac{T}{2}}(x_{2\tau}-x_{2\tau-1}) \leq x_T-x_1 \leq 1.
\]
So the total discrepancy will be no more than $1$.
\Halmos
\endproof

\section{Proof of Lemma~\ref{lem:CompletelyRandomized:p=1}}
\label{sec:Proof:CompeltelyRandomzied:p=1}

For any arrival sequence $\bm{x}$, let $\ALG^\sC(\bm{x})$ be the expected discrepancy incurred by a completely randomized design when the arrival sequence is $\bm{x}$.

Let $V$ be the (random) set of indices that the completely randomized design selects to assign into the control group.
As a result, $([T] \setminus V)$ is the set of indices that the completely randomized design selects to assign into the treated group.
We denote $\mathcal{V}$ to be the set that contains all possible $V$.
We know that the size of $\mathcal{V}$ is $|\mathcal{V}|={T \choose T/2}$.
In addition, since we select $V$ randomly, the probability for any $V \in \mathcal{V}$ to be selected is $\frac{1}{{T \choose T/2}}$.

For any arrival sequence $\bm{x}$ and any set $V \in \mathcal{V}$, denote $\Dis(\bm{x},V)$ to be the (deterministic) discrepancy incurred by a completely randomized design when the arrival sequence is $\bm{x}$ and when the realized set of indices is $V$.
We can take expectation over $V$ and have the following, $$\ALG^\sC(\bm{x}) = E_{V}[\Dis^\sC(\bm{x}, V)].$$

We first introduce the following three Lemmas that help us prove Lemma~\ref{lem:CompletelyRandomized:p=1}.

\begin{lemma}
\label{lem:lowerfluid}
Let the arrival instance $\bm{x}^*$ be $(0,0,...,0,1,1,...,1)$, which contains $\frac{T}{2}$ many zeros and $\frac{T}{2}$ many ones.
The expected discrepancy on this arrival sequence $\bm{x}^*$ is $\ALG^\sC(\bm{x}^*) = \frac{1}{\sqrt{\pi}}\sqrt{T}+O(T^{-\frac{1}{2}})$.
\end{lemma}

\proof{Proof of Lemma~\ref{lem:lowerfluid}.}
The completely randomized design select $\frac{T}{2}$ subjects from $T$ subjects randomly.
Given that the $T$ subjects consists of $\frac{T}{2}$ many zeros and $\frac{T}{2}$ many ones, if we select $H$ many zeros to the control group, then we automatically select $\frac{T}{2}-H$ many ones to the control group. Therefore, the total discrepancy can be calculated by
\begin{align*}
\bE\left[\left\vert H-(\frac{T}{2}-H) \right\vert \right]= 2 \bE\left[\left\vert H-\frac{T}{4} \right\vert \right].
\end{align*}

Observe that $H$ follows a binomial random variable with $\frac{T}{2}$ trials and success probability $\frac{1}{2}$.
In addition, $\bE[H]=\frac{T}{4}$.
By equation 2.4 in \cite{blyth1980expected} (re-stated using our notations in Lemma~\ref{lem:blyth}), we have $E\left[|H-\frac{T}{4}|\right]= \frac{1}{\sqrt{\pi}}\sqrt{T}+O(T^{-\frac{1}{2}})$.
\Halmos
\endproof




\begin{lemma} \label{lem:permutation}
Consider two arriving sequences $\bm{x}'$ and $\bm{x}''$.
If there exists an one-to-one correspondence $\sigma: [T] \to [T]$, such that $\forall \ t \in [T]$, $x'_t = x''_{\sigma(t)}$, then the expected discrepancy of the completely randomized design will be the same on these two arrival sequences $\bm{x}'$ and $\bm{x}''$, i.e., $$\ALG^\sC(\bm{x}')=\ALG^\sC(\bm{x}'').$$
\end{lemma}

\proof{Proof of Lemma~\ref{lem:permutation}.}
Recall that, we use $V$ for the (random) set of indices that the completely randomized design selects to assign into the control group.
And recall that we denote $\Dis(\bm{x},V)$ to be the (deterministic) discrepancy incurred by a completely randomized design when the arrival sequence is $\bm{x}$ and when the realized set of indices is $V$.
Then we have,
\[
\ALG^\sC(\bm{x}') = E_V[\Dis(\bm{x}',V)] =\frac{1}{{T \choose T/2}}\sum_{V \in \mathcal{V}}\Dis(\bm{x}',V),
\]
\[
\ALG^\sC(\bm{x}'') = E_V[\Dis(\bm{x}'',V)] = \frac{1}{{T \choose T/2}}\sum_{V \in \mathcal{V}}\Dis(\bm{x}'',V).
\]

Now denote $\sigma(V)=\{u \in[T] \mid u=\sigma(t), \forall t \in V\}$. 
In words, $\sigma(V)$ is the set of indices to receive control after permutation $\sigma(\cdot)$.
Since $\sigma(\cdot)$ is an one-to-one correspondence, we know that $\mathcal{V}$ is not only the set that contains all possible $V$, but also the set that contains all possible $\sigma(V)$.
In addition, since we select $V$ randomly, the probability for any $\sigma(V) \in \mathcal{V}$ to be selected is $\frac{1}{{T \choose T/2}}$.

Therefore, for any $V \in \mathcal{V}$, $$\Dis(\bm{x}',V) = \Dis(\bm{x}'', \sigma(V)).$$
If we add them all together,
\[
\sum_{V \in \mathcal{V}}\Dis(\bm{x}',V)=\sum_{\sigma(V) \in \mathcal{V}}\Dis(\bm{x}'',\sigma(V))=\sum_{V \in \mathcal{V}}\Dis(\bm{x}'',V).
\]
This suggests that
\[
\ALG^\sC(\bm{x}') = \ALG^\sC(\bm{x}''),
\]
which completes the proof.
\Halmos
\endproof

\begin{lemma} \label{lem:upperfluid}
For all possible arrival instances $\bm{x} \in [0,1]^{T}$, we have
\begin{align*}
\max_{\bm{x} \in [0,1]^{T}} \ALG^\sC(\bm{x}) = \ALG^\sC(\bm{x}^*),
\end{align*}
where $\bm{x}^{*} = (0,0,...,0,1,1,...,1)$, which contains $\frac{T}{2}$ many zeros and $\frac{T}{2}$ many ones.
\end{lemma}

\proof{Proof of Lemma~\ref{lem:upperfluid}.}
Consider any arrival sequence $\bm{x}=(x_1,x_2,\ldots,x_T)$.
Due to Lemma~\ref{lem:permutation}, we have that the discrepancy of $\ALG^\sC$ does not change under any permutation $\sigma(\cdot)$.
Therefore, we can sort the subjects in $\bm{x}$ from the smallest to the largest, i.e., $x_1 \leq x_2 \leq \ldots \leq x_T$.

Next, we select $\frac{T}{2}$ subjects from $(x_1,x_2,\ldots,x_T)$ randomly.
There are $|\mathcal{V}| = {T \choose T/2}$ possible cases.
We have
\begin{align*}
\ALG^\sC(\bm{x})=\frac{1}{{T \choose T/2}} \sum_{V \in \mathcal{V}}\Dis(\bm{x},V).
\end{align*}

For each realization $V \in \mathcal{V}$, we let the ordered subjects in the control group and treated group be $C(V)=\{y_1,y_2,\ldots,y_{\frac{T}{2}}\}$ and $T(V)=\{z_1,z_2,\ldots,z_{\frac{T}{2}}\}$ respectively. We have $\Dis(\bm{x},V)=\sum_{i=1}^{\frac{T}{2}}|y_i-z_i|$.

If we sum up the discrepancy for all realizations $V \in \mathcal{V}$, we have the total discrepancy is in the following form:
\begin{align*}
\sum_{V \in \mathcal{V}}\Dis(\bm{x},V) = \sum_{t=1}^T \xi_t x_t,
\end{align*}
where $\xi_t \in \mathbb{Z}$ is an abstract notation that counts how many times (out of a total of ${T \choose T/2}$ many selections of $V$) subject $t$ is matched to another subject with a larger index (negative number stands for the number of time matched to another subject with a smaller index).

Next, since each $x_t$ appears in a number of additions as well as a number of subtractions, we can merge the terms and focus on the coefficients of each $x_t$ term. 
The sum of discrepancies can be calculated by examining the coefficients of each $x_t$ term. For example, there are $\binom{T}{\frac{T}{2}}$ instances of $-x_1$ and the same number of $+x_T$ in the sum of discrepancies over all realizations.
This is because $x_1$ is the smallest and $x_T$ is the largest. For any other $t \in \{2,3,\ldots,\frac{T}{2}\}$, the sum of discrepancies over all realizations includes both a number of $+x_t$ and a number of $-x_t$.
As we examine all possible realizations, the number of $+x_t$ is fewer than that of $-x_t$, given $t \in \{2,3,\ldots,\frac{T}{2}\}$ and $x_t$'s relatively smaller size among T numbers, causing it to more frequently be negative. 
Thus, the coefficient of $x_t$ is negative, which we denote as $-\xi_t$. 
By symmetry, there exists a bijective function which maps each instance that $x_t$ is positive in discrepancy calculation and each instance that $x_{T-t+1}$ is negative in discrepancy calculation. 
Therefore, the coefficient of $x_{T-t+1}$ is the absolute value of that of $x_t$, which is equal to $\xi_t$.


Therefore, we have
\begin{align*}
\ALG^\sC(\bm{x}) & = \frac{1}{{T \choose T/2}} \sum_{t=1}^T \xi_t x_t \\
& = \frac{1}{{T \choose T/2}} \sum_{t=1}^{\frac{T}{2}}\vert \xi_t \vert (x_{T-t+1} - x_t) \\
& \leq \frac{1}{{T \choose T/2}} \sum_{t=1}^{\frac{T}{2}}\vert \xi_t \vert(1-0) \\
& = \frac{1}{{T \choose T/2}} \sum_{t=1}^{\frac{T}{2}}\vert \xi_t \vert.
\end{align*}
The equality sign holds if and only if $x_t=1$ for all $t > \frac{T}{2}$ and $x_t = 0$ for all $t \leq \frac{T}{2}$.
Under this case, $(x_1,x_2,\ldots,x_T)=\bm{x}^{*}$, and $\max_{\bm{x} \in [0,1]^{T}} \ALG^\sC(\bm{x}) = \ALG^\sC(\bm{x}^*)$.
\Halmos
\endproof

\proof{Proof of Lemma~\ref{lem:CompletelyRandomized:p=1}}
For any instance of arrival sequence $\bm{x}$, denote $\ALG^\sC(\bm{x})$ to be the expected discrepancy of the completely randomized design on sequence $\bm{x}$. 
Let $[0,1]^T$ be the set that contains all possible arrival inputs.
By Lemma~\ref{lem:lowerfluid}, we find an arrival instance $\bm{x}^{*}$ such that $\ALG^\sC(\bm{x}^{*}) = \frac{1}{\sqrt{\pi}}\sqrt{T}+O(T^{-\frac{1}{2}})$.
This suggests that the discrepancy of the completely randomized design is at least $\frac{1}{\sqrt{\pi}}\sqrt{T}$.

Next, we show that the upper bound of the discrepancy is also on the same order. 
First, by Lemma~\ref{lem:permutation}, we have the completely randomized design under $p=1$ is permutation-invariant to the arriving sequence.
Second, Lemma~\ref{lem:upperfluid} shows that among all possible arrival instances in $[0,1]^T$, $\bm{x}^{*}$ has the largest discrepancy.
Therefore, combining this result with Lemma~\ref{lem:lowerfluid}, we know that the discrepancy of any completely randomized design is upper bounded by
\begin{align*}
\max_{\bm{x} \in [0,1]^T}\ALG^\sC(\bm{x}) = \ALG^\sC(\bm{x}^{*}) = \frac{1}{\sqrt{\pi}}\sqrt{T}+O(T^{-\frac{1}{2}}).
\end{align*}
Combining the lower bound and the upper bound we show that the completely randomized design has an expected discrepancy on the order of $\Theta(T^{\frac{1}{2}})$.
\Halmos
\endproof

\section{Proof of Lemma~\ref{lem:Pigeonhole:p=1}}
\label{sec:Proof:Pigeonhole:p=1}

\proof{Proof of Lemma~\ref{lem:Pigeonhole:p=1}}
Consider the case when the length of each pigeonhole is $T^{-\eta}$ and there are $T^\eta$ many pigeonholes, i.e., $\cP(\cS) = \left\{[0, T^{-\eta}), [T^{-\eta},2T^{-\eta}), \ldots, [1-T^{-\eta},1] \right\}$.
Following an execution of Algorithm~\ref{alg:Pigeonhole:p=1}, there must be a moment that either the control or the treated group reaches $\frac{T}{2}$ in size.
We define a stopping time based on the above.

First, for any subject $t \in [T]$, let $n^0_t$ and $n^1_t$ be the total number of control and treated subjects after subject $t$ is assigned, respectively.
Denote $n^0_t = n^1_t = 0$ to reflect that no subject was assigned yet at the beginning of the entire horizon.
With the above definitions, define
\begin{align*}
\tau = \min \Bigg\{ t \in [T] \bigg\vert n^0_t = \frac{T}{2} \ \text{or} \ n^1_t = \frac{T}{2} \Bigg\}.
\end{align*}

So far until subject $\tau$, we have not entered the balancing periods, i.e., the ``if'' condition in Line~3 of Algorithm~\ref{alg:Pigeonhole:p=1} has not been triggered.
As a result, each pigeonhole must have at most one unmatched subject.
We analyze the expected discrepancy that comes from the following two components.

\noindent \textbf{Component 1}:
Consider the first $\tau$ subjects.
Denote the set of time indices for the matched subjects to be $\mathcal{T}_M$; denote the set of time indices for the unmatched subjects to be $\mathcal{T}_U$.
Component 1 provides an upper bound on the discrepancy generated from $\mathcal{T}_M$ the matched subjects.

Out of the matched subjects $\mathcal{T}_M$, and for any $k \in [T^\eta]$, let the number of subjects from the $k$-th pigeonhole $\big[(k-1)T^{-\eta}, kT^{-\eta}\big)$ be denoted as $n_{k,\tau}$.
Then, due to Proposition~\ref{prop:OnePigeonhole}, there exists an absolute constant $c$ such that the discrepancy generated from the $k$-th pigeonhole can be upper bounded by $T^{-\eta} \cdot c \cdot (n_{k,\tau})^{\frac{1}{2}}$.
Across all the pigeonholes, the total discrepancy can be upper bounded by
\begin{align*}
\sum_{k=1}^{T^{\eta}} T^{-\eta} \cdot c \cdot (n_{k,\tau})^{\frac{1}{2}} & \leq c T^{-\eta} \Big(\sum_{k=1}^{T^{\eta}} n_{k,\tau}\Big)^\frac{1}{2} \Big(\sum_{k=1}^{T^{\eta}} 1\Big)^\frac{1}{2} \\
& \leq c T^{-\eta} \Big( T^\eta \tau \Big)^\frac{1}{2} \\
& \leq c T^{\frac{1}{2} - \frac{\eta}{2}},
\end{align*}
where the first inequality is due to Cauchy-Schwarz, and the second inequality is because the number of matched subjects can be upper bounded by $\tau$, which leads to $\sum_{k=1}^{T^{\eta}} n_{k,\tau} \leq \tau$.
To conclude Component 1, the discrepancy generated from $\mathcal{T}_M$ is on the order of $O(T^{\frac{1}{2}-\frac{\eta}{2}})$.

\noindent \textbf{Component 2}:
Next, we resume our analysis and focus on $\mathcal{T}_U$, the unmatched subjects, together with the last $(T-\tau)$ subjects.

Recall that each pigeonhole must have at most one unmatched subject.
Now focus on the pigeonholes such that there is exactly one unmatched subject in that pigeonhole.
Let $H^0_\tau$ and $H^1_\tau$ be the number of pigeonholes such that there is exactly one unmatched subject in the control group and in the treated group, respectively.
Denote $H = H^0_\tau + H^1_\tau$.
It is obvious to see that $H$ is upper bounded by the number of pigeonholes, i.e., $H \leq T^\eta$.

Next, note that 
\begin{align*}
\vert H^0_\tau - H^1_\tau \vert = T - \tau,
\end{align*}
as the difference between the number of control subjects and the number of treated subjects is exactly equal to the number of pigeonholes with one unmatched control subject and the number of pigeonholes with one unmatched treated subject.

To understand the distribution of $(T - \tau)$, we characterize the distributions of $H^0_\tau$ and $H^1_\tau$, respectively.
For each pigeonhole that has exactly one unmatched subject, the assignment of this unmatched subject is control with probability $\frac{1}{2}$, and treated with probability $\frac{1}{2}$.
Since the adversary is non-adaptive (also referred to as oblivious adversary), the assignment of control or treated of each unmatched subject is independent of others, and independent of the total number of pigeonholes which has exactly one unmatched subject.
As a result, conditional on $H = H^0_\tau + H^1_\tau$, we know that
\begin{align*}
H^0_\tau \sim \textrm{Bin}\big(H, \frac{1}{2}\big), && H^1_\tau = H - H^0_\tau.
\end{align*}
Then, $H^0_\tau - H^1_\tau = 2 H^0_\tau - H$, and we know that $\bE[H^0_\tau - H^1_\tau] = 0$.

Define the following event conditional on any $1 \leq H \leq T^\eta$,
\begin{align*}
\mathcal{E}_0 = \Bigg\{ \big\vert H^0_\tau - H^1_\tau \big\vert \leq 2 \sqrt{H \log{H}} \Bigg\}.
\end{align*}
Then, due to Lemma~\ref{lem:hoeffding},
\begin{align*}
\Pr(\bar{\mathcal{E}}_0) = \Pr\Big( \big\vert H^0_\tau - H^1_\tau \big\vert > 2 \sqrt{H \log{H}} \Big) \leq \frac{2}{H^2}.
\end{align*}
We next distinguish the low probability event case and the high probability event case.

\noindent \textbf{Case 1} (low probability): For any $H$ and conditional on the low probability event $\bar{\mathcal{E}}_0$, we have that the discrepancy generated from $\mathcal{T}_U$ the unmatched subjects together with the last $(T-\tau)$ subjects can be upper bounded by
\begin{align*}
(T-\tau) \cdot 1 = \big\vert H^0_\tau - H^1_\tau \big\vert \leq H^0_\tau + H^1_\tau = H.
\end{align*}
So this low probability event contributes at most $\frac{2}{H^2} \cdot H \leq 2$ discrepancy in expectation.

\noindent \textbf{Case 2} (high probability): In this case, we can no longer crudely upper bound the discrepancy by the total number of unmatched subjects because we can no longer take advantage of the small enough probability.
However, we can take advantage of the fact that there is only a small difference between the number of unmatched subjects in the control and treated groups.

For any $H$ and conditional on the high probability event $\mathcal{E}_0$, we define a coupling of two stochastic processes.
First, we define a stochastic process $Z$ as follows.
For every $i \in \{1,2,...,H\}$, $Z_i=1$ with probability $\frac{1}{2}$, and $Z_i=0$ with probability $\frac{1}{2}$.
The assignment of all the unmatched subjects follows $Z$.
Second, we define a stochastic process $Y$ as follows.
Conditional on $H^0_\tau$ and $H^1_\tau$, we randomly ``select'' $\big(2\min\{H^0_\tau, H^1_\tau\}\big)$ indices from $\{1,2,...,H\}$.
On the selected indices, $Y$ follows a completely randomized design, i.e., we randomly pick $\min\{H^0_\tau, H^1_\tau\}$ indices and assign $Y_i=1$, and the other $\min\{H^0_\tau, H^1_\tau\}$ indices are assigned $Y_i=0$.
On the remaining $\vert H^0_\tau - H^1_\tau \vert$ many indices that are not ``selected,'' we set $Y_i=1$ with probability $\frac{1}{2}$, and $Y_i=0$ with probability $\frac{1}{2}$.
Conditional on any pair of $H^0_\tau$ and $H^1_\tau$, $Y$ is a coupling process of $Z$.
This is because a completely randomized design is a Bernoulli design conditional on the numbers of control and treated subjects being fixed.

Now that we have the coupling, we can view the assignments of the $H$ many unmatched subjects as if they were generated from the process $Y$ as we have defined above.
There are two components of subjects that we analyze separately.
First, in process $Y$, there are $\big( 2 \min\{H^0_\tau, H^1_\tau\} \big)$ many subjects that are assigned based on the completely randomized design.
Due to Lemma~\ref{lem:CompletelyRandomized:p=1}, we know that the total discrepancy between these subjects can be upper bounded by 
\begin{align*}
\frac{1}{2\sqrt{\pi}}\Big(\min\{H^0_\tau, H^1_\tau\}\Big)^{\frac{1}{2}} \leq \frac{1}{2\sqrt{\pi}} (\frac{1}{2}H)^{\frac{1}{2}} \leq \frac{1}{2\sqrt{\pi}} (\frac{1}{2}T^\eta)^{\frac{1}{2}},
\end{align*} 
which is on the order of $O\left(T^{\frac{\eta}{2}}\right)$.

Second, in process $Y$, there are $\vert H^0_\tau - H^1_\tau\vert$ many subjects that are assigned into either the control group or the treatment group with probability $\frac{1}{2}$ each.
On these subjects, together with the last $(T-\tau)$ subjects, we match them in any arbitrary fashion.
The discrepancy between each pair is no more than $1$.
Therefore, the total discrepancy can be upper bounded by 
\begin{align*}
\frac{1}{2}(\vert H^0_\tau - H^1_\tau\vert + (T-\tau)) \leq 2 \sqrt{H \log{H}} \leq 2 T^\frac{\eta}{2} (\log{T^\frac{\eta}{2}})^\frac{1}{2} = (2\eta)^\frac{1}{2} T^\frac{\eta}{2} (\log{T})^\frac{1}{2},
\end{align*}
where the first inequality is because we are conditional on the high probability event. 
The above term is on the order of $O(T^\frac{\eta}{2} (\log{T})^\frac{1}{2})$.

So this high probability event contributes an expected discrepancy on the order of $O(T^\frac{\eta}{2} (\log{T})^\frac{1}{2})$.
Combining the low probability event and the high probability event, the expected discrepancy is on the order of $O\big(T^\frac{\eta}{2} (\log{T})^\frac{1}{2}\big)$ for Component 2.

Finally, combining both Component 1 and Component 2, the total expected discrepancy is on the order of $O\left( T^{\frac{1}{2}-\frac{\eta}{2}} + T^{\frac{\eta}{2}} (\log{T})^\frac{1}{2}\right)$.
\Halmos
\endproof

\proof{Proof of Corollary~\ref{coro:Pigeonhole:p=1}.}
From Lemma~\ref{lem:Pigeonhole:p=1}, the total expected discrepancy is on the order of $O\left( T^{\frac{1}{2}-\eta} + T^{\frac{\eta}{2}} (\log{T})^\frac{1}{2}\right)$.
If we choose $\eta$ such that $$T^{\frac{1}{2}-\frac{\eta}{2}}=T^{\frac{\eta}{2}},$$
which implies $\eta=\frac{1}{2}$, the pigeonhole design has an expected discrepancy on the order of $O\big(T^{\frac{1}{4}} (\log{T})^\frac{1}{2}\big)$.
\Halmos
\endproof

\section{Proof of Proposition~\ref{prop:OnePigeonhole}} \label{sec:Proof:prop:OnePigeonhole}

In this section, we first introduce Proposition~\ref{prop:OnePigeonhole} and Lemma~\ref{lem:adjustment}, and then use Lemma~\ref{lem:adjustment} to prove Proposition~\ref{prop:OnePigeonhole}.

\begin{proposition}
\label{prop:OnePigeonhole}
When $p=1, q=0$, the pigeonhole design using a single pigeonhole $\cP(\cS) = \left\{[0,1] \right\}$ has an expected discrepancy on the order of $O(T^\frac{1}{2})$.
\end{proposition}

\begin{lemma} 
\label{lem:adjustment}
For any arrival instance $(x_1,\ldots,x_T)$, let $x_1',x_2',...,x_T'$ be the smallest-to-largest rearrangement such that $x_1' \leq x_2' \leq ... \leq x_T'$. If there exists an odd index $i$ such that $x_i, x_{i+1}\in \{x_1', ..., x_{T/2}'\}$ or $x_i, x_{i+1}\in \{x_{T/2+1}', ..., x_T'\}$, then there must exist a rearrangement of the arrival instance $(x_{\sigma(1)},\ldots,x_{\sigma(T)})$ such that the pigeonhole design described in Proposition~\ref{prop:OnePigeonhole} incurs a larger discrepancy on the rearranged arrival instance.
\end{lemma}

\subsection{Proof of Lemma~\ref{lem:adjustment}}

\proof{Proof of Lemma~\ref{lem:adjustment}.}
For any arrival instance $(x_1,\ldots,x_T)$, if there exists an odd index $i$ such that $x_i, x_{i+1}\in \{x_1', ..., x_{T/2}'\}$, then
due to the pigeonhole principle, we see that there also exists an odd index $j$ such that $x_j \in \{x_{T/2+1}', ..., x_T'\}$ and $x_{j+1} \in \{x_{T/2+1}', ..., x_T'\}$. 

Without loss of generality, we can assume that $x_i \leq x_{i+1}$ and $x_j \leq x_{j+1}$.
We will show that for the rearranged arrival sequence by swapping only $i+1$ and $j+1$ leads to a larger discrepancy.
Mathematically, we show that the rearranged sequence $(x_{\sigma(1)},\ldots,x_{\sigma(T)})$, where $\sigma(k)=k$ for $k \neq i+1, j+1$, and $\sigma(i+1)=j+1$ and $\sigma(j+1)=i+1$, the pigeonhole design as described in Proposition~\ref{prop:OnePigeonhole} incurs a larger discrepancy on $(x_{\sigma(1)},\ldots,x_{\sigma(T)})$ compared to the original sequence $(x_1,\ldots,x_T)$.

Recall that we denote the control group as group A and treated group as group B. 
Recall that according to the pigeonhole design, for the arriving subjects on the odd indices, we randomly (with probability a half) assign the subjects to either $A$ or $B$. 
For the arriving subjects on the even indices, we assign the subjects to the opposite group to the subject who arrived earlier. 
Now consider the four subjects and their treatment assignments $w_i,w_{i+1},w_j,w_{j+1}$. 
Under the pigeonhole design, the four possible treatment assignments are $(A,B,A,B)$, $(A,B,B,A)$, $(B,A,A,B)$, $(B,A,B,A)$, where each realization happens with probability $\frac{1}{4}$. 
Let $\bm{w}_{-i,-j}$ be the realized treatment assignments of all other subjects except $x_i,x_{i+1},x_j,x_{j+1}$. 
Due to the definition of a pigeonhole design, the treatment assignments $\bm{w}_{-i,-j}$ and the treatment assignments $w_i, w_{i+1}, w_j, w_{j+1}$ are generated independently. 

Denote $\Dis(\bm{x},\bm{w})$ to be the (deterministic) discrepancy incurred by a pigeonhole design when the arrival sequence is $\bm{x}$ and when the realized treatment assignments are $\bm{w}$. 
Denote $\bm{w}=(\bm{w}_{-i,-j},(w_i, w_{i+1}, w_j, w_{j+1}))$ to highlight the four special indices.
Further, with some abuse of notations, let $|\bm{w}_{-i,-j}| = 2^{T/2-2}$ denote the number of supports that $\bm{w}_{-i,-j}$ can take values from.
Then, the total expected discrepancy under $(x_1,\ldots,x_T)$ is
\begin{align} 
\label{eq:disoriginal}
\bE_{\bm{w}}[\Dis((x_1,\ldots,x_T),\bm{w})] = \frac{1}{|\bm{w}_{-i,-j}|}\sum_{\bm{w}_{-i,-j}}& \nonumber \Big( \frac{1}{4}\Dis((x_1,\ldots,x_T),(\bm{w}_{-i,-j},(A,B,A,B))) \\& \nonumber +\frac{1}{4}\Dis((x_1,\ldots,x_T),(\bm{w}_{-i,-j},(A,B,B,A))) \\& \nonumber +\frac{1}{4}\Dis((x_1,\ldots,x_T),(\bm{w}_{-i,-j},(B,A,A,B))) \\&+ \frac{1}{4}\Dis((x_1,\ldots,x_T),(\bm{w}_{-i,-j},(B,A,B,A))) \Big).
\end{align}

On the other hand, consider the rearranged arrival instance $(x_{\sigma(1)},\ldots,x_{\sigma(T)})$.
For this rearranged arrival sequence, and under the pigeonhole design, the four possible treatment assignments $w_i,w_{i+1},w_j,w_{j+1}$ are $(A,A,B,B)$, $(A,B,B,A)$, $(B,A,A,B)$, $(B,B,A,A)$, with probability $\frac{1}{4}$ for each.
Then, the total expected discrepancy under the rearranged arrival sequence $(x_{\sigma(1)},\ldots,x_{\sigma(T)})$ can be expanded as
\begin{align} 
\label{eq:disrearranged}
\bE_{\bm{w}}[\Dis((x_{\sigma(1)},\ldots,x_{\sigma(T)}),\bm{w})] = \frac{1}{|\bm{w}_{-i,-j}|} & \sum_{\bm{w}_{-i,-j}} \Big( \frac{1}{4}\Dis((x_{\sigma(1)},\ldots,x_{\sigma(T)}),(\bm{w}_{-i,-j},(A,A,B,B))) \nonumber \\
& + \frac{1}{4}\Dis((x_{\sigma(1)},\ldots,x_{\sigma(T)}),(\bm{w}_{-i,-j},(A,B,B,A))) \nonumber \\
& + \frac{1}{4}\Dis((x_{\sigma(1)},\ldots,x_{\sigma(T)}),(\bm{w}_{-i,-j},(B,A,A,B))) \nonumber \\
& + \frac{1}{4}\Dis((x_{\sigma(1)},\ldots,x_{\sigma(T)}),(\bm{w}_{-i,-j},(B,B,A,A))) \Big).
\end{align} 

By the construction of our rearrangement, we know all other subjects receive the same treatment assignments $\bm{w}_{-i,-j}$.
Also note that, 
and therefore, $\Dis((x_1,\ldots,x_T),(\bm{w}_{-i,-j},(A,B,B,A)))=\Dis((x_{\sigma(1)},\ldots,x_{\sigma(T)}),(\bm{w}_{-i,-j},(A,B,B,A)))$ and $\Dis((x_1,\ldots,x_T),(\bm{w}_{-i,-j},(B,A,A,B)))=\Dis((x_{\sigma(1)},\ldots,x_{\sigma(T)}),(\bm{w}_{-i,-j},(B,A,A,B)))$. 

To obtain our main goal, which is  $E_V[\Dis((x_{\sigma(1)},\ldots,x_{\sigma(T)}),V)] \geq E_V[\Dis((x_1,\ldots,x_T),V)]$, we show that for any realization $\bm{w}_{-i,-j}$, $\Dis((x_{\sigma(1)},\ldots,x_{\sigma(T)}),(\bm{w}_{-i,-j},(A,A,B,B))) \geq \Dis((x_1,\ldots,x_T),(\bm{w}_{-i,-j},(A,B,A,B)))$ and $\Dis((x_{\sigma(1)},\ldots,x_{\sigma(T)}),(\bm{w}_{-i,-j},(B,B,A,A))) \geq \Dis((x_1,\ldots,x_T),(\bm{w}_{-i,-j},(B,A,B,A)))$. As these two statements are symmetric, we will only show the first one. 

To show that for any realization $\bm{w}_{-i,-j}$, $\Dis((x_{\sigma(1)},\ldots,x_{\sigma(T)}),(\bm{w}_{-i,-j},(A,A,B,B))) \geq \Dis((x_1,\ldots,x_T),(\bm{w}_{-i,-j},(A,B,A,B)))$, under the arrival instance $(x_1,\ldots,x_T)$ and realization $(\bm{w}_{-i,-j},(A,B,A,B))$, we let $(a_1,a_2,\ldots,a_{\frac{T}{2}})$ as subjects assigned to group A and $(b_1,b_2,\ldots,b_{\frac{T}{2}})$ as subjects assigned to group B, where $a_1 \leq a_2 \leq \ldots \leq a_{\frac{T}{2}}$ and $b_1 \leq b_2 \leq \ldots \leq b_{\frac{T}{2}}$. Under the arrival instance $(x_{\sigma(1)},\ldots,x_{\sigma(T)})$ and realization $(\bm{w}_{-i,-j},(A,A,B,B))$, we let $(a_1',a_2',\ldots,a_{\frac{T}{2}}')$ as subjects assigned to group A and $(b_1',b_2',\ldots,b_{\frac{T}{2}}')$ as subjects assigned to group B, where $a_1' \leq a_2' \leq \ldots \leq a_{\frac{T}{2}}'$ and $b_1' \leq b_2' \leq \ldots \leq b_{\frac{T}{2}}'$. Then, by definition of discrepancy, we have 
\[
\Dis((x_1,\ldots,x_T),(\bm{w}_{-i,-j},(A,B,A,B))) = \sum_{i=1}^{\frac{T}{2}}\vert a_i-b_i \vert, 
\]
\[
\Dis((x_{\sigma(1)},\ldots,x_{\sigma(T)}),(\bm{w}_{-i,-j},(A,A,B,B)))=\sum_{i=1}^{\frac{T}{2}}\vert a_i'-b_i' \vert
\]

We denote $a_i$ and $b_i$ as matched pair. Next, we assume that $x_i=a_u$, $x_{i+1}=b_v$, $x_j=a_w$, and $x_{j+1}=b_x$, the smallest subjects in $(a_1,a_2,\ldots,a_{\frac{T}{2}})$ which is greater than or equal to $x_{\frac{T}{2}}$ is $a_y$, and the smallest subjects in $(b_1,b_2,\ldots,b_{\frac{T}{2}})$ which is greater than or equal to $x_{\frac{T}{2}}$ is $b_z$. By definition, we have $a_u,b_v \leq \min\{a_y,b_z\} \leq a_w,b_x$. 

Now, under $(x_{\sigma(1)},\ldots,x_{\sigma(T)})$ and realization $(\bm{w}_{-i,-j},(A,A,B,B))$, $x_{i+1}=b_v$ is assigned into group $A$, we let $b_v$ is the $s^{\text{th}}$ smallest number in group A. Also, $x_j=a_w$ is assigned into group $B$, we let $a_w$ is the $t^{\text{th}}$ smallest number in group B. Then, we split the discussion into several cases based on the matching possibility under $(x_1,\ldots,x_T)$ and realization $(\bm{w}_{-i,-j},(A,B,A,B))$: (we may put this as an additional lemma)
\begin{enumerate}
    \item $x_i$ and $x_{i+1}$ are matched pair, and $x_j$ and $x_{j+1}$ are matched pair. In this case $u=v$ and $w=x$, if we put $b_v$ in group A and $a_w$ in group B, we have
 $\Dis((x_{\sigma(1)},\ldots,x_{\sigma(T)}),(\bm{w}_{-i,-j},(A,A,B,B)))=\vert a_u- a_w\vert + \vert b_v-b_x \vert + \sum_{i \neq u,w}\vert a_i'-b_i' \vert \geq \sum_{i=1}^{\frac{T}{2}}\vert a_i-b_i \vert$, where the inequality is because $a_u,b_v \leq x_{\frac{T}{2}} \leq a_w,b_x$.
    \item $x_i$ is matched with $b_u<b_v$ and $x_j$ is matched with $b_w<b_x$. In this case, if $b_v$ is in group A and $a_w$ is in group B, $a_u$ is still matched with $b_u$, and $a_x$ is still matched with $b_x$, but 
    $a_v$ and $b_w$ become unmatched. Therefore, we have $\Dis((x_{\sigma(1)},\ldots,x_{\sigma(T)}),(\bm{w}_{-i,-j},(A,A,B,B)))=\vert b_v- b_w\vert + \vert a_v-a_w \vert + \sum_{i \neq v,w}\vert a_i'-b_i' \vert \geq \sum_{i=1}^{\frac{T}{2}}\vert a_i-b_i \vert$.
    \item $x_i$ is matched with $b_u>b_v$ and $x_j$ is matched with $b_w>b_x$. In this case, we have $a_v<a_u<b_v<b_u<a_x<a_w<b_x<b_w$ or $a_v<a_u<b_v<a_x<b_u<a_w<b_x<b_w$.  
    If $b_v$ is in group A and $a_w$ is in group B, we have 
    $\Dis((x_{\sigma(1)},\ldots,x_{\sigma(T)}),(\bm{w}_{-i,-j},(A,A,B,B)))=\vert a_v- b_u\vert + \vert a_w-a_w \vert + \vert b_v- b_x\vert+ \vert a_x- b_w\vert +  \sum_{i \neq u,v,w,x}\vert a_i'-b_i' \vert \geq \sum_{i=1}^{\frac{T}{2}}\vert a_i-b_i \vert$.
    \item For all other cases, we can use symmetric statements with the previous three cases to validate the statement.
\end{enumerate}

Therefore, we have for any $\bm{w}_{-i,-j}$, we have $\Dis((x_{\sigma(1)},\ldots,x_{\sigma(T)}),(\bm{w}_{-i,-j},(A,A,B,B))) \geq \Dis((x_1,\ldots,x_T),(\bm{w}_{-i,-j},(A,B,A,B)))$. This validates the statement in Lemma~\ref{lem:adjustment}.
\Halmos
\endproof

\subsection{Proof of Proposition~\ref{prop:OnePigeonhole}}

\proof{Proof of Proposition~\ref{prop:OnePigeonhole}.}
If the adversary is non-adaptive, it has to choose (i) the value of all covariates $(x_1,\ldots,x_T)$; (ii) the sequence of arrival into this pigeonhole.

Lemma~\ref{lem:adjustment} provides the foundation for our claim that, following complete adjustment, any arrival instance retains an expected discrepancy identical to that of ${x_1, x_3, ..., x_{T-1}} = {x_1', ..., x_{T/2}'}$ and ${x_2, x_4, ..., x_{T}} = {x_{T/2+1}', ..., x_T'}$. It is essential to note that we do not enforce any specific order here. This phenomenon occurs due to the intrinsic design of the pigeonhole: each pair of subjects $(x_{2m-1},x_{2m})$ operates independently of any other pair $(x_{2n-1},x_{2n})$, thereby allowing us to interchange the indices within each pair.

Moreover, for a pair of subjects $(\tilde{x}_{2m-1},\tilde{x}_{2m})$ such that $\tilde{x}_{2m-1} \in \{x_{T/2+1}', ..., x_T'\}$ and $\tilde{x}_{2m} \in \{x_1', ..., x_{T/2}'\}$, we can generate a coupling with the pair $x_{2m-1} \in \{x_1', ..., x_{T/2}'\}$ and $x_{2m} \in \{x_{T/2+1}', ..., x_T'\}$ in the following manner: if $\tilde{x}_{2m-1} = 1$, then $x_{2m-1} = 0$; alternatively, if $\tilde{x}_{2m-1} = 0$, then $x_{2m-1} = 1$. This arrangement is easily validated as a coupling, considering that the pigeonhole design assigns each odd index subject to either the control or treated group with a probability of $\frac{1}{2}$.

Let $\mathcal V$ be the realization set of $\{x_1,x_2,\ldots,x_T\}$. By the definition of pigeonhole design, for each $V \in \mathcal V$, the occurrence probability is $(\frac{1}{2})^{\frac{T}{2}}$ because each of $\{x_1, x_3, ..., x_{T-1}\}$ has probability $\frac{1}{2}$ to be either in A or B, and after that, the realization of $\{x_2, x_4, ..., x_{T}\}$ is determined. Therefore, we have 
\begin{align*}
E_V[\Dis((x_1,\ldots,x_T),V)]=\frac{1}{(\frac{1}{2})^{\frac{T}{2}}} \sum_{V \in \mathcal{V}}\Dis((x_1,\ldots,x_T),V).
\end{align*}
Now we wish to expand
\begin{align*}
\sum_{V \in \mathcal{V}}\Dis(\bm{x},V) = \sum_{t=1}^T \xi_t x_t,
\end{align*}
where $\xi_t \in \mathbb{Z}$ is an abstract notation that counts how many times (out of a total of $(\frac{1}{2})^{\frac{T}{2}}$ many selections of $V$) subject $t$ is matched to another subject with a larger index (negative number stands for the number of time matched to another subject with a smaller index).
Note that we have ordered $x_1 \leq x_2 \leq \ldots \leq x_T$.
For any $V$, denote $\sigma_s(V) = \left\{t \left| (T-t+1) \in V \right. \right\}$.
Then for any $t \in [T]$, whenever subject $t$ is matched to a subject with a larger index under $V$, we know that subject $(T-t+1)$ is matched to a subject with a smaller index under $\sigma_s(V)$.
As a result, for any $t \in [T], \xi_t=-\xi_{T+1-t}$.

Therefore, we have
\begin{align*}
E_V[\Dis((x_1,\ldots,x_T),V)] &=\frac{1}{(\frac{1}{2})^{\frac{T}{2}}}\sum_{t=1}^T \xi_t x_t \\
& = \frac{1}{(\frac{1}{2})^{\frac{T}{2}}} \sum_{t=1}^{\frac{T}{2}}\vert \xi_t \vert (x_{T-t+1} - x_t) \\
& \leq \frac{1}{(\frac{1}{2})^{\frac{T}{2}}} \sum_{t=1}^{\frac{T}{2}}\vert \xi_t \vert(1-0) \\
& = \frac{1}{(\frac{1}{2})^{\frac{T}{2}}} \sum_{t=1}^{\frac{T}{2}}\vert \xi_t \vert.
\end{align*}
The equality sign holds if and only if $x_t=1$ for all $t > \frac{T}{2}$ and $x_t = 0$ for all $t \leq \frac{T}{2}$.
Under this case, $(x_1,x_2,\ldots,x_T)=(0,0,\ldots,0,1,1,\ldots,1)$, where we have $\frac{T}{2}$ zeros and $\frac{T}{2}$ ones. 

Finally, as the pigeonhole design will select $H \sim \mathrm{Bin}(\frac{T}{2},\frac{1}{2})$ zeros in the instance, we can apply Lemma \ref{lem:lowerfluid} to get that the expected discrepancy is on the order of $O(T^{\frac{1}{2}})$. 
\Halmos
\endproof

\section{Proof of Lemma \ref{lem:Pigeonhole:p=1:LB}} \label{sec:Proof:Pigeonhole:p=1:LB}

\proof{Proof of Lemma \ref{lem:Pigeonhole:p=1:LB}}
Let $0 = l_0 < l_1 < l_2 < ... < l_K = 1$ be the inputs.
For each $k \in [K]$, define $L_k = l_K - l_{K-1}$, and define $n_k = \frac{L_k^2}{\sum_{i=1}^K L_i^2} T$.
For any pigeonhole design that is parameterized by $\cP(\cS)$ = $\big\{[0, l_1)$, $[l_1,l_2)$, ..., $[l_{K-1},l_K] \big\}$, consider the following arrival instance,
\begin{multline*}
(x_1, x_2, ..., x_T) = \\
\big(\underbrace{0, l_1-\epsilon, 0, l_1-\epsilon, ..., 0, l_1-\epsilon}_{\frac{n_1}{2} \ \text{many} \ (0, l_1-\epsilon)}, \underbrace{l_1,l_2-\epsilon,l_1,l_2-\epsilon, ...,l_1,l_2-\epsilon}_{\frac{n_2}{2} \ \text{many} \ (l_1, l_2-\epsilon)}, ..., \underbrace{l_{K-1}, l_K, l_{K-1}, l_K, ..., , l_{K-1}, l_K}_{\frac{n_K}{2} \ \text{many} \ (l_{K-1}, l_K)}\big),
\end{multline*} 
where for each $k \in [K]$, $n_k$ is the total number of subjects in each pigeonhole.

For each pigeonhole $k \in [K]$, the pigeonhole design assigns each $l_{k-1}$ to either control group or treated group with probability $\frac{1}{2}$.
Now denote $H_k$ to be the number of $l_{k-1}$ subjects that are assigned to the control group.
We know that $H_k$ follows a Binomial distribution, i.e., $H_k \sim \mathrm{Bin}(\frac{n_k}{2},\frac{1}{2})$. 
Due to Lemma \ref{lem:lowerfluid}, there exists some constant $c_k$ such that the expected discrepancy is at least $c_k \cdot L_k \cdot (n_k)^{\frac{1}{2}}$.
We denote $\underline{c} = \min_k c_k$. 

Due to Cauchy-Schwarz inequality, the discrepancy from all the pigeonholes is at least 
\begin{align*}
\underline{c} \cdot \sum_{k=1}^K \cdot L_k \cdot (n_k)^\frac{1}{2} = \underline{c} \cdot \sqrt{\Big(\sum_{k=1}^K L_k^2\Big) \Big(\sum_{k=1}^K n_k\Big)} = \underline{c} \sqrt{T \sum_{k=1}^K L_k^2}.
\end{align*}

Next, since $L_k^2$ is convex in $L_k$, by Jensen's inequality, 
\begin{align*}
\sqrt{T \sum_{k=1}^K L_k^2} \geq \sqrt{T} \sqrt{K (\frac{1}{K})^2} = \sqrt{\frac{T}{K}},
\end{align*}
where the inequality takes equality if and only if $L_k = \frac{1}{K}$ for any $k \in [K]$. 
If there are at most $K=O(T^{\frac{1}{2}})$ pigeonholes, the above expression yields a discrepancy on the order of $\underline{c} \cdot \sqrt{\frac{T}{K}} = \Omega(T^{\frac{1}{4}})$.

On the other hand, if there are at least $\vert \cS \vert=\Omega(T^{\frac{1}{2}})$ pigeonholes, we consider the following arrival instance such that the number of subjects in each pigeonhole is an odd number.
Due to the pigeonhole design, finally there will be one remaining subject unmatched in each pigeonhole. 
In the proof of Lemma \ref{lem:Pigeonhole:p=1}, component 2, by conditional coupling with the completely randomized design, we know that if there are $\vert \cS \vert$ subjects unmatched, the total expected discrepancy generated from matching such unmatched subjects is on the order of $\Theta(\vert \cS \vert^{\frac{1}{2}} )=\Omega(T^{\frac{1}{4}})$.
Combining both cases we finish the proof.
\Halmos
\endproof

\section{Proofs of Theorem~\ref{thm:MatchedPair:p=0}, Theorem~\ref{thm:CompletelyRandomized:p=0}, and Theorem~\ref{thm:Pigeonhole:p=0}} \label{sec:Proof:Thm123}

\proof{Proof of Theorem \ref{thm:MatchedPair:p=0}}
Suppose that we don't have any continuous dimension, and we have $q$ discrete dimensions. Recall that for any $i \in [q]$, $S_i$ is the finite supports along the $i$-th dimension, and the number of supports in this dimension is $\vert S_i \vert = m_i$. The total number of supports is $\prod_{i\in[q]} m_i$. 

Consider any arriving input $(x_1,x_2,\ldots,x_T)$, where $x_j \in \Cross_{i \in [q]} S_i$ for any $j \in [T]$. Then, for any support, if there exists at least two subjects arriving on this support, we match as many pairs as we can. By doing this, each pair being matched leads to $0$ discrepancy, and there are at most $\prod_{i\in[q]} m_i$ subjects remaining unmatched. As the diameter of $q$ dimension cube is $\sqrt{q}$, for any matching strategy between these remaining subjects, the discrepancy is bounded by $\frac{1}{2}\sqrt{q}\prod_{i\in[q]} m_i=\Theta(1)$.
\Halmos
\endproof

\proof{Proof of Theorem \ref{thm:CompletelyRandomized:p=0}}
We first show that the expected discrepancy is on the order of $\Omega(T^{\frac{1}{2}})$. We randomly pick two support, $s_1$ and $s_2$. Then, we take an arrival instance $(x_1,x_2,\ldots,x_T)$ as $(s_1,s_1,\ldots,s_1,s_2,s_2,\ldots,s_2)$, which contains $\frac{T}{2}$ $s_1$ and $\frac{T}{2}$ $s_2$. 
By Lemma \ref{lem:lowerfluid}, the expected discrepancy is on the order of $\Theta(\Vert s_2-s_1 \Vert_2 T^{\frac{1}{2}})$ under this arrival instance. As $\Vert s_2-s_1 \Vert_2 \leq \sqrt{q}$, we have the expected discrepancy is $\Theta(T^{\frac{1}{2}})$, which implies that the expected discrepancy under all instances is on the order of $\Omega(T^{\frac{1}{2}})$.

We second show that the expected discrepancy is on the order of $O(T^{\frac{1}{2}})$. Suppose that there are $N_i$ subjects on the $i^{\text{th}}$ support. Let $H_i$ be the random variable which represents the number of subjects on the $i^{\text{th}}$ support being assigned into the control group by the complete randomized policy. We have $H_i \sim \mathrm{Bin}(N_i,\frac{1}{2})$. Then, the number of subjects on the $i^{\text{th}}$ support being assigned into the treated group by the complete randomized policy is $N_i-H_i$. 
By Lemma \ref{lem:lowerfluid}, we have $\bE[|H_i-(N_i-H_i)|]=\Theta(\sqrt{N_i})$, which means that the unmatched subjects on $i^{\text{th}}$ support is on the order of $\Theta(\sqrt{N_i})$. As the Euclidean distance between any two supports is upper bounded by $\sqrt{q}$, the total discrepancy is then upper bounded by $\sqrt{q}\sum_{i=1}^{\prod_{j\in[q]} m_j}\bE[|H_i-(N_i-H_i)|]=\Theta(\sqrt{T})$ because $\sum_{i=1}^{\prod_{j\in[q]} m_j} N_i = T$.
\Halmos
\endproof

\proof{Proof of Theorem \ref{thm:Pigeonhole:p=0}}
For any arriving input $(x_1,x_2,\ldots,x_T)$, where $x_j \in \cup_{i \in [q]} S_i$ for any $j \in [T]$, as we take any support as a pigeonhole, for any support, if there exists at least two subjects arriving on this support, we match as many pairs as we can. By doing this, each pair being matched leads to $0$ discrepancy, and there are at most $\prod_{i\in[q]} m_i$ subjects remaining unmatched. As the diameter of $q$ dimension cube is $\sqrt{q}$, for any matching strategy between these remaining subjects, the discrepancy is bounded by $\frac{1}{2}\sqrt{q}\prod_{i\in[q]} m_i=\Theta(1)$.
\Halmos
\endproof

\section{Proofs of Theorem~\ref{thm:MatchedPair:p=1}, Theorem~\ref{thm:CompletelyRandomized:p=1}, and Theorem~\ref{thm:Pigeonhole:p=1}}
\label{sec:Proof:Thm456}

\proof{Proof of Theorem \ref{thm:MatchedPair:p=1} }
When $q > 0$, for any discrete support, if we apply the matched-pair design on its corresponding continuous dimension, by Lemma \ref{lem:MatchedPair:p=1}, the discrepancy is less than or equal to $1$. As there are $\prod_{i=1}^{q}m_i$ different supports on $q$ discrete dimensions, we have the total discrepancy is bounded by $1 \cdot \prod_{i=1}^{q}m_i = \Theta(1)$.
\Halmos
\endproof

\proof{Proof of Theorem \ref{thm:CompletelyRandomized:p=1} }
When $q > 0$, for any discrete support, if we apply the pigeonhole design on its corresponding continuous dimension, by Lemma\ref{lem:CompletelyRandomized:p=1} , the expected discrepancy is on the order of $\Theta(T^{\frac{1}{2}})$. Therefore, if the adversary selects an instance with all of the subjects on the same discrete support, the expected discrepancy is on the order of $\Theta(T^{\frac{1}{2}})$, which implies that the expected discrepancy under all possible instances is $\Omega(T^{\frac{1}{2}})$.

Next, we show that the expected discrepancy is $O(T^{\frac{1}{2}})$ for $p = 1, q > 0$. Suppose that there are $N_i$ subjects on the $i^{\text{th}}$ support, with $\sum_{i=1}^{\prod_{j\in[q]} m_j} N_i = T$. If we perform the complete randomized design on all $T$ subjects, we have for any $N_i$ subjects, if we denote $H_i$ as the random variable which represents the number of subjects among $N_i$ subjects being assigned into the control group by the complete randomized policy, we have $H_i \sim \mathrm{Bin}(N_i,\frac{1}{2})$. Also, 
the number of subjects being assigned into the treated group by the complete randomized policy is $N_i-H_i$. By Lemma \ref{lem:lowerfluid}, we have $\bE[|H_i-(N_i-H_i)|]=\Theta(\sqrt{N_i})$, which means that the unmatched subjects on $i^{\text{th}}$ support is on the order of $\Theta(\sqrt{N_i})$. 

By the union bound, we have the total expected discrepancy is bounded by the summation of expected discrepancy on the continuous dimension corresponding to each discrete support. Therefore, we have the total expected discrepancy is bounded by $1 \cdot \sum_{i=1}^{\prod_{j\in[q]} m_j}\bE[|H_i-(N_i-H_i)|]=\Theta(\sqrt{T})$.
\Halmos
\endproof

\proof{Proof of Theorem \ref{thm:Pigeonhole:p=1} }
When $q > 0$, for any discrete support, if we apply the pigeonhole design on its corresponding continuous dimension, by Lemmas \ref{lem:Pigeonhole:p=1} and \ref{lem:Pigeonhole:p=1:LB}, the discrepancy is on the order of $\Theta(T^{\frac{1}{4}}(\log{T})^\frac{1}{2})$. As there are $\prod_{j=1}^{q}m_j$ different supports on $q$ discrete dimensions, and we are giving pigeonhole on each of the discrete support, we have the total discrepancy is bounded by $\Theta(T^{\frac{1}{4}}(\log{T})^\frac{1}{2}) \cdot \prod_{i=1}^{q}m_i = \Theta(T^{\frac{1}{4}}(\log{T})^\frac{1}{2})$.
\Halmos
\endproof

\section{Proof of Theorem~\ref{thm:MatchedPair:p>=2}}
\label{sec:Proof:MatchedPair:p>=2}

\proof{Proof of Theorem~\ref{thm:MatchedPair:p>=2}}
To show the total discrepancy of matched-pair design is $\Theta\big(T^{\frac{p-1}{p}}\big)$, we only need to prove the $\Omega\big(T^{\frac{p-1}{p}}\big)$ part.
The $O\big(T^{\frac{p-1}{p}}\big)$ part is proved as Theorem~4.2 in \citet{bai2021inference}, which we describe again in Lemma~\ref{lem:MatchedPair:p>=2:UB}.

We prove by constructing an arrival sequence such that the discrepancy on this sequence is $\Omega\big(T^{\frac{p-1}{p}}\big)$.
Consider the following sequence. 
We first pick one support from the finitely many of discrete supports among the $q$ discrete dimensions.
Then, we fix the support in the $q$ discrete dimensions, i.e., all the subjects take the same value among the $q$ discrete dimensions.
In the $p$ continuous dimensions, we evenly split the covariate space $[0,1]^p$ into $T$ smaller hypercubes, such that each hypercube has edge length $T^{-\frac{1}{p}}$.
Finally, we let the continuous covariates of each subject to be at the center of each smaller hypercube, respectively.

By construction, the distance between the covariates of any two subjects is at least $T^{-\frac{1}{p}}$.
There are a total of $T/2$ pairs of subjects.
Therefore, the discrepancy is at least $T^{-\frac{1}{p}} \cdot T = T^{\frac{p-1}{p}}$.
So the matched-pair design has a discrepancy on the order of $\Omega\big(T^{\frac{p-1}{p}}\big)$.
Combining with Lemma~\ref{lem:MatchedPair:p>=2:UB} we finish the proof.
\Halmos
\endproof

\section{Proof of Lemma~\ref{lem:CompletelyRandomized:p>=2}}
\label{sec:Proof:lem:CompletelyRandomized:p>=2}

\proof{Proof of Lemma~\ref{lem:CompletelyRandomized:p>=2}}
Recall that we construct the family of arrival sequences as follows.
First, we split the p-dimensional unit hypercube to $T$ smaller hypercubes, each with edge-length $T^{-\frac{1}{p}}$.
Second, let there be exactly one subject in each smaller hypercube.
Mathematically, let the family of arrival sequences be
\begin{multline*}
\mathcal{X}^{(T)} = \bigg\{ (\bm{x}_1, \bm{x}_2, ..., \bm{x}_T) \big| \forall \mathbb{X} \in \left\{[0, T^{-\frac{1}{p}}), [T^{-\frac{1}{p}},2T^{-\frac{1}{p}}), \ldots, [1-T^{-\frac{1}{p}},1] \right\}^p, \\
\exists t, s.t.  \ \bm{x}_t \in \mathbb{X}, \text{ and } \forall t' \ne t, \bm{x}_t' \notin \mathbb{X} \bigg\}.
\end{multline*}

First, we show that when $p=2$, for any $T$, the following holds:
\begin{align}
\label{eqn:inductionp=2}
\forall \bm{x}^{(T)} \in \mathcal{X}^{(T)}, \ALG^C(\bm{x}^{(T)}) \leq \sqrt{2} \sqrt{T}\log T\sqrt{\log T}.
\end{align}
We prove this for any $T = 2^{2i}$ by induction on $i \in \mathbb{N}$.
First, when $T = 2^2$, we have
$\ALG^C(\bm{x}) \leq 2\sqrt{2}  \leq \sqrt{2} \sqrt{4}\log 4+\sqrt{4\log 4}$.
The first inequality is because there are $2$ pairs and each pair contributes at most $\sqrt{2}$ discrepancy.

Suppose \eqref{eqn:inductionp=2} holds for $i = i_0 \in \mathbb{N}$, i.e., \eqref{eqn:inductionp=2} holds for $2^{2i_0 } = T_0$.
We show that for $i = i_0 + 1$, i.e., for $2^{2(i_0+1) } = 2^2 T_0$, the inequality \eqref{eqn:inductionp=2} also holds.
Consider the new family of arrival sequences $\mathcal{X}^{(2^2 T_0)}$, and one arrival sequence $\bm{x}^{(2^2 T_0)} \in \mathcal{X}^{(2^2 T_0)}$.
There are a total of $2^2 T_0$ subjects.
And there are $2^2 T_0$ small squares each with edge length $\frac{1}{2} T_0^{-\frac{1}{2}}$.
There is exactly one subject in each small square.

For the unit square with edge length $1$, we cut it by half in each dimension.
Then, there are $2^{2}$ squares each with edge length $\frac{1}{2}$, which we refer to as the ``good-squares.''
In each such good-square, the arrival sequence $\bm{x}^{(2^2 T_0)}$ has $T_0$ subjects that are evenly distributed.
If we apply completely randomized design in each good-square, we have that the expected discrepancy in such good-square is $\frac{1}{2}\sqrt{2} \sqrt{T_0}\log T_0$.
This is because all distances are shrunken by a half.
However, as we apply completely randomized design in the unit square, it does not necessarily guarantee that the number of subjects between control and treated group are the same in each good-square.
By Lemma \ref{lem:hoeffding}, we know that with probability at least $(1-\frac{2}{T_0^2})^{2^2}$, the absolute difference between the number of control and treated subjects is bounded by $\sqrt{T_0 \log T_0}$ in every good-square.

Conditional on the low probability event, which happens with probability at most $\frac{2}{T_0^2} \cdot 2^2$ due to union bound, the discrepancy can be upper bounded by $\frac{2^2 T_0}{2}$.
So the contribution to the total expected discrepancy can be upper bounded by $\frac{2}{T_0^2} \cdot 2^2 \cdot \frac{2^2 T_0}{2} = \frac{2^{4}}{T_0} \leq 2^2 = 4$, where the inequality is due to $T_0 \geq 2^2$.

Next, conditional on the high probability event, suppose in good-square $i, i \in \{1,2,3,4\}$, there is an imbalanced number of control and treated subjects of no more than $\sqrt{T_0 \log T_0}$ in every good-square.
Then, we randomly pick no more than $\sqrt{T_0 \log T_0}$ subjects in the group with more subjects and denote as $R_i$.
For all the subjects that are not in $R_i$ but in good-square $i$, we make a minimum weight perfect matching, whose discrepancy is upper bounded by the discrepancy of a completely randomized design.
In each good-square, the expected discrepancy on the number of subjects that are matched can be upper bounded by $\frac{1}{2}\sqrt{2} \sqrt{T_0}\log T_0\sqrt{\log T_0}$.
There are $4$ such squares, and so the expected discrepancy is 
\[
4 \cdot \frac{1}{2}\sqrt{2} \sqrt{T_0}\log T_0\sqrt{\log T_0}=2\sqrt{2} \sqrt{T_0}\log T_0\sqrt{\log T_0}.
\]

Note that we still have no more than $\sqrt{T_0 \log T_0}$ subjects in each $R_i$, we match them randomly.
The total discrepancy of subjects in $R_i$ can be bounded by $\frac{2^{2}\sqrt{T_0 \log T_0}}{2} \cdot \sqrt{2}$.
Therefore, the total discrepancy of completely randomized design under $\bm{x}^{(2^2 T_0)}$ is upper bounded by
\begin{align*}
2\sqrt{2} \sqrt{T_0}\log T_0\sqrt{\log T_0} + 2\sqrt{2} \sqrt{T_0\log T_0} + 4 & = 2\sqrt{2} \sqrt{T_0 \log T_0}(\log T_0 +1) + 4\\
& \leq 2\sqrt{2} \sqrt{T_0 \log T_0}(\log T_0 +\log 4) \\
& = 2\sqrt{2} \sqrt{T_0 \log T_0}\log 4T_0 \\
& \leq \sqrt{2} \sqrt{4T_0 \log 4T_0}\log 4T_0.
\end{align*}

Second, we show that when $p \geq 3$, for any $T$, the following holds:
\begin{align}
\label{eqn:induction}
\forall \bm{x}^{(T)} \in \mathcal{X}^{(T)}, \ALG^C(\bm{x}^{(T)}) \leq \sqrt{p} T^{\frac{p-1}{p}}+\sqrt{T\log T}\log T.
\end{align}
We prove this for any $T = 2^{ip}$ by induction on $i \in \mathbb{N}$.
First, when $T = 2^p$, we have
$\ALG^C(\bm{x}) \leq \sqrt{p} 2^{p-1} = \sqrt{p} T^{\frac{p-1}{p}} \leq \sqrt{p} T^{\frac{p-1}{p}} + \sqrt{T\log T}\log T$.
The first inequality is because there are $2^{p-1}$ many pairs and each pair contributes at most $\sqrt{p}$ discrepancy.

Suppose \eqref{eqn:induction} holds for $i = i_0 \in \mathbb{N}$, i.e., \eqref{eqn:induction} holds for $2^{i_0 p} = T_0$.
We show that for $i = i_0 + 1$, i.e., for $2^{(i_0+1) p} = 2^p T_0$, the inequality \eqref{eqn:induction} also holds.
Consider the new family of arrival sequences $\mathcal{X}^{(2^p T_0)}$, and one arrival sequence $\bm{x}^{(2^p T_0)} \in \mathcal{X}^{(2^p T_0)}$.
There are a total of $2^p T_0$ subjects.
And there are $2^p T_0$ small hypercubes each with edge length $\frac{1}{2} T_0^{-\frac{1}{p}}$.
There is exactly one subject in each small hypercube.

For the unit hypercube with edge length $1$, we cut it by half in each dimension.
Then, there are $2^{p}$ hypercubes each with edge length $\frac{1}{2}$, which we refer to as the ``good-cubes.''
In each such good-cube, the arrival sequence $\bm{x}^{(2^p T_0)}$ has $T_0$ subjects that are evenly distributed.
If we apply completely randomized design in each good-cube, we have that the expected discrepancy in such good-cube is $\frac{1}{2}\sqrt{p}T_0^{\frac{p-1}{p}}+\frac{1}{2}\sqrt{T_0\log T_0}\log T_0$.
This is because all distances are shrunken by a half.
However, as we apply completely randomized design in the unit hypercube, it does not necessarily guarantee that the number of subjects between control and treated group are the same in each good-cube.
By Lemma \ref{lem:hoeffding}, we know that with probability at least $(1-\frac{2}{T_0^2})^{2^p}$, the absolute difference between the number of control and treated subjects is bounded by $\sqrt{T_0 \log T_0}$ in every good-cube.

Conditional on the low probability event, which happens with probability at most $\frac{2}{T_0^2} \cdot 2^p$ due to union bound, the discrepancy can be upper bounded by $\frac{2^p T_0}{2}$.
So the contribution to the total expected discrepancy can be upper bounded by $\frac{2}{T_0^2} \cdot 2^p \cdot \frac{2^p T_0}{2} = \frac{2^{2p}}{T_0} \leq 2^p$, where the inequality is due to $T_0 \geq 2^p$.

Next, conditional on the high probability event, suppose in good-cube $i, i \in \{1,2,...,2^p\}$, there is an imbalanced number of control and treated subjects.
Then, we randomly pick no more than $\sqrt{T_0 \log T_0}$ subjects in the group with more subjects and denote as $R_i$.
For all the subjects that are not in $R_i$ but in good-cube $i$, we make a minimum weight perfect matching, whose discrepancy is upper bounded by the discrepancy of a completely randomized design.
In each good-cube, the expected discrepancy on the number of subjects that are matched can be upper bounded by $\frac{1}{2}\sqrt{p}T_0^{\frac{p-1}{p}}+\frac{1}{2}\sqrt{T_0\log T_0}\log T_0$.
There are $2^p$ such cubes, and so the expected discrepancy is 
\[
2^{p-1}\sqrt{p}T_0^{\frac{p-1}{p}}+2^{p-1}\sqrt{T_0\log T_0}\log{T_0} = \sqrt{p}(2^pT_0)^{\frac{p-1}{p}}+2^{p-1}\sqrt{T_0\log T_0}\log T_0.
\]

Note that we still have no more than $\sqrt{T_0 \log T_0}$ subjects in each $R_i$, we match them randomly.
The total discrepancy of subjects in $R_i$ can be bounded by $\frac{2^{p}\sqrt{T_0 \log T_0}}{2} \cdot \sqrt{p}$.
Therefore, the total discrepancy of completely randomized design under $\bm{x}^{(2^p T_0)}$ is upper bounded by
\begin{align*}
& \sqrt{p}(2^pT_0)^{\frac{p-1}{p}}+2^{p-1}\sqrt{T_0\log T_0}\log T_0+2^{p-1}\sqrt{p}\sqrt{T_0 \log T_0} + 2^p \\
= & \sqrt{p}(2^pT_0)^{\frac{p-1}{p}}+2^{p-1}\sqrt{T_0\log T_0}\log T_0+2^{p-1}\left(\sqrt{p}\sqrt{T_0 \log T_0} + 2\right) \\
\leq & \sqrt{p}(2^pT_0)^{\frac{p-1}{p}}+2^{p-1}\sqrt{T_0\log (2^pT_0)}\log T_0+2^{p-1}\log 2^p\sqrt{T_0 \log (2^pT_0)} \\
= & \sqrt{p}(2^pT_0)^{\frac{p-1}{p}}+\sqrt{2^pT_0\log (2^pT_0)}\log (2^pT_0),
\end{align*}
where the inequality holds when $p \geq 3$.
\Halmos
\endproof

\section{Proof of Theorem~\ref{thm:Pigeonhole:p>=2}}
\label{sec:Proof:Pigeonhole:p>=2}

\proof{Proof of Theorem~\ref{thm:Pigeonhole:p>=2}.}

We first consider the case when $q=0$.
Fix $c$ to be any positive real number that is greater than $1$, i.e., $c > 1$.
Consider the case when the length of each pigeonhole is $c^{\frac{1}{p}}T^{-\phi}$ and there are $\frac{1}{c}T^{p\phi}$ many pigeonholes, i.e., $\cP(\cS) = \left\{[0, c^{\frac{1}{p}}T^{-\phi}), [T^{-\phi},2c^{\frac{1}{p}}T^{-\phi}), \ldots, [1-c^{\frac{1}{p}}T^{-\phi},1] \right\}^p$.
Following an execution of Algorithm~\ref{alg:Pigeonhole:p>=2}, there must be a moment that either the control or the treated group reaches $\frac{T}{2}$ in size.
We define a stopping time based on the above.

First, for any subject $t \in [T]$, let $n^0_t$ and $n^1_t$ be the total number of control and treated subjects after subject $t$ is assigned, respectively.
Denote $n^0_t = n^1_t = 0$ to reflect that no subject was assigned yet at the beginning of the entire horizon.
With the above definitions, define
\begin{align*}
\tau = \min \Bigg\{ t \in [T] \bigg\vert n^0_t = \frac{T}{2} \ \text{or} \ n^1_t = \frac{T}{2} \Bigg\}.
\end{align*}

So far until subject $\tau$, we have not entered the balancing periods, i.e., the ``if'' condition in Line~4 of Algorithm~\ref{alg:Pigeonhole:p>=2} has not been triggered.
As a result, each pigeonhole must have at most one unmatched subject.
We analyze the expected discrepancy that comes from the following two components.

\noindent \textbf{Component 1}:
Consider the first $\tau$ subjects.
Denote the set of time indices for the matched subjects to be $\mathcal{T}_M$; denote the set of time indices for the unmatched subjects to be $\mathcal{T}_U$.
Component 1 provides an upper bound on the discrepancy generated from $\mathcal{T}_M$ the matched subjects.

Since the diameter of each pigeonhole is $c^{\frac{1}{p}}\sqrt{p}T^{-\phi}$, the discrepancy generated from the matched subjects $\mathcal{T}_M$ can be upper bounded by
\begin{align*}
\frac{\tau}{2} \cdot c^{\frac{1}{p}}\sqrt{p}T^{-\phi} \leq \frac{1}{2} c^{\frac{1}{p}}\sqrt{p}T^{1-\phi}.
\end{align*}
To conclude Component 1, the discrepancy generated from $\mathcal{T}_M$ is on the order of $O(T^{1-\phi})$.

\noindent \textbf{Component 2}:
Next, we turn our attention to $\mathcal{T}_U$, the unmatched subjects, together with the last $(T-\tau)$ subjects.

Recall that each pigeonhole must have at most one unmatched subject.
Now focus on the pigeonholes such that there is exactly one unmatched subject in that pigeonhole.
Let $H^0_\tau$ and $H^1_\tau$ be the number of pigeonholes such that there is exactly one unmatched subject in the control group and in the treated group, respectively.
Denote $H = H^0_\tau + H^1_\tau$.
It is obvious to see that $H$ is upper bounded by the number of pigeonholes, i.e., $H \leq \frac{1}{c} T^{p\phi}$.

Next, note that 
\begin{align*}
\vert H^0_\tau - H^1_\tau \vert = T - \tau,
\end{align*}
as the difference between the number of control subjects and the number of treated subjects is exactly equal to the number of pigeonholes with one unmatched control subject and the number of pigeonholes with one unmatched treated subject.

To understand the distribution of $(T - \tau)$, we characterize the distributions of $H^0_\tau$ and $H^1_\tau$, respectively.
For each pigeonhole that has exactly one unmatched subject, the assignment of this unmatched subject is control with probability $\frac{1}{2}$, and treated with probability $\frac{1}{2}$.
Since the adversary is non-adaptive (also referred to as oblivious adversary), the assignment of control or treated of each unmatched subject is independent of others, and independent of the total number of pigeonholes which has exactly one unmatched subject.
As a result, conditional on $H = H^0_\tau + H^1_\tau$, we know that
\begin{align*}
H^0_\tau \sim \textrm{Bin}\big(H, \frac{1}{2}\big), && H^1_\tau = H - H^0_\tau.
\end{align*}
Then, $H^0_\tau - H^1_\tau = 2 H^0_\tau - H$, and we know that $\bE[H^0_\tau - H^1_\tau] = 0$.

Define the following event conditional on any $1 \leq H \leq \frac{1}{c}T^{p\phi}$,
\begin{align*}
\mathcal{E}_0 = \Bigg\{ \big\vert H^0_\tau - H^1_\tau \big\vert \leq 2 \sqrt{H \log{H}} \Bigg\}.
\end{align*}
Then, due to Lemma~\ref{lem:hoeffding},
\begin{align*}
\Pr(\bar{\mathcal{E}}_0) = \Pr\Big( \big\vert H^0_\tau - H^1_\tau \big\vert > 2 \sqrt{H \log{H}} \Big) \leq \frac{2}{H^2}.
\end{align*}
We next distinguish the low probability event case and the high probability event case.

\noindent \textbf{Case 1} (low probability): For any $H$ and conditional on the low probability event $\bar{\mathcal{E}}_0$, we have that the discrepancy generated from $\mathcal{T}_U$ the unmatched subjects together with the last $(T-\tau)$ subjects can be upper bounded by
\begin{align*}
(T-\tau) \cdot 1 = \big\vert H^0_\tau - H^1_\tau \big\vert \leq H^0_\tau + H^1_\tau = H.
\end{align*}
So this low probability event contributes at most $\frac{2}{H^2} \cdot H \leq 2$ discrepancy in expectation.

\noindent \textbf{Case 2} (high probability): In this case, we can no longer crudely upper bound the discrepancy by the total number of unmatched subjects because we can no longer take advantage of the small enough probability.
However, we can take advantage of the fact that there is only a small difference between the number of unmatched subjects in the control and treated groups.

For any $H$ and conditional on the high probability event $\mathcal{E}_0$, we define a coupling of two stochastic processes.
First, we define a stochastic process $Z$ as follows.
For every $i \in \{1,2,...,H\}$, $Z_i=1$ with probability $\frac{1}{2}$, and $Z_i=0$ with probability $\frac{1}{2}$.
The assignment of all the unmatched subjects follows $Z$.
Second, we define a stochastic process $Y$ as follows.
Conditional on $H^0_\tau$ and $H^1_\tau$, we randomly ``select'' $\big(2\min\{H^0_\tau, H^1_\tau\}\big)$ indices from $\{1,2,...,H\}$.
On the selected indices, $Y$ follows a completely randomized design, i.e., we randomly pick $\min\{H^0_\tau, H^1_\tau\}$ indices and assign $Y_i=1$, and the other $\min\{H^0_\tau, H^1_\tau\}$ indices are assigned $Y_i=0$.
On the remaining $\vert H^0_\tau - H^1_\tau \vert$ many indices that are not ``selected,'' we set $Y_i=1$ with probability $\frac{1}{2}$, and $Y_i=0$ with probability $\frac{1}{2}$.
Conditional on any pair of $H^0_\tau$ and $H^1_\tau$, $Y$ is a coupling process of $Z$.
This is because a completely randomized design is a Bernoulli design conditional on the numbers of control and treated subjects being fixed.

Now that we have the coupling, we can view the assignments of the $H$ many unmatched subjects as if they were generated from the process $Y$ as we have defined above.
There are two components of subjects that we analyze separately.
First, in process $Y$, there are $\big( 2 \min\{H^0_\tau, H^1_\tau\} \big)$ many subjects that are assigned based on the completely randomized design.
Due to Lemma~\ref{lem:CompletelyRandomized:p>=2}, we know that the total discrepancy between these subjects can be upper bounded by either the following quantity in the $p=2$ case,
\begin{align*}
\sqrt{2} \Big(\min\{H^0_\tau, H^1_\tau\}\Big)^{\frac{p-1}{p}} \Big(\log{\big(\min\{H^0_\tau, H^1_\tau\}\big)}\Big)^\frac{3}{2} \leq & \sqrt{2} \Big(\frac{1}{2}H\Big)^{\frac{p-1}{p}} \Big(\log(\frac{1}{2}H)\Big)^\frac{3}{2} \\
\leq & \sqrt{2} \Big(\frac{1}{2c}T^{p\phi}\Big)^{\frac{p-1}{p}} \Big(\log(\frac{1}{2c}T^{p \phi})\Big)^\frac{3}{2},
\end{align*}
which is on the order of $O\left(T^{(p-1)\phi} (\log{T})^\frac{3}{2}\right)$; or upper bounded by the following quantity in the $p \geq 3$ case,
\begin{align*}
& \sqrt{p}\Big(\min\{H^0_\tau, H^1_\tau\}\Big)^{\frac{p-1}{p}} + \Big(\min\{H^0_\tau, H^1_\tau\}\Big)^{\frac{1}{2}} \Big(\log{\big(\min\{H^0_\tau, H^1_\tau\}\big)}\Big)^\frac{3}{2} \\
\leq & \sqrt{p} \Big(\frac{1}{2}H\Big)^{\frac{p-1}{p}} + \Big(\frac{1}{2}H\Big)^{\frac{1}{2}} \Big(\log{(\frac{1}{2}H)}\Big)^\frac{3}{2}\\
\leq & \sqrt{p} \Big(\frac{1}{2c}T^{p\phi}\Big)^{\frac{p-1}{p}} + \Big(\frac{1}{2c}T^{p\phi}\Big)^{\frac{1}{2}} \Big(\log(\frac{1}{2c}T^{p \phi})\Big)^\frac{3}{2},
\end{align*}
which is on the order of $O\left(T^{(p-1)\phi}\right)$.

Second, in process $Y$, there are $\vert H^0_\tau - H^1_\tau \vert$ many subjects that are assigned into either the control group or the treatment group with probability $\frac{1}{2}$ each.
On these subjects, together with the last $(T-\tau)$ subjects, we match them in any arbitrary fashion.
The discrepancy between each pair is no more than $1$.
Therefore, the total discrepancy can be upper bounded by 
\begin{align*}
\frac{1}{2}(\vert H^0_\tau - H^1_\tau \vert + (T-\tau)) \leq 2 \sqrt{H \log{H}} \leq \frac{2}{c} T^\frac{p\phi}{2} \big( \log{(c^{-1}T^{p\phi})} \big)^\frac{1}{2}
\end{align*}
where the first inequality is because we are conditional on the high probability event. 
The above expression is on the order of $O(T^\frac{p\phi}{2} (\log{T})^\frac{1}{2})$.

So this high probability event contributes an expected discrepancy that is on the order of $O(T^{(p-1)\phi} (\log{T})^\frac{3}{2})$ when $p=2$ and on the order of $O(T^{(p-1)\phi})$ when $p \geq 3$.
Combining the low probability event and the high probability event, the expected discrepancy is on the order of $O(T^{(p-1)\phi} (\log{T})^\frac{3}{2})$ when $p=2$ and on the order of $O(T^{(p-1)\phi})$ when $p \geq 3$ for Component 2.

Finally, combining both Component 1 and Component 2, the total expected discrepancy is on the order of $O\left( T^{1-\phi} + T^{(p-1)\phi} (\log{T})^{\frac{3}{2}} \right)$ when $p = 2$, and on the order of $O\left( T^{1-\phi} + T^{(p-1)\phi} \right)$ when $p \geq 3$.

To conclude the proof, we turn to the general case where $q \geq 0$.
We have proved the case when $q = 0$.
When $q \geq 1$, for any discrete support, the discrepancy is on the order of $O\left( T^{1-\phi} + T^{(p-1)\phi} (\log{T})^{\frac{3}{2}} \right)$ when $p = 2$, and on the order of $O\left( T^{1-\phi} + T^{(p-1)\phi} \right)$ when $p \geq 3$. 
As there are $\prod_{i=1}^{q}m_i$ different supports on the $q$ discrete dimensions, and we are devising the pigeonholes on each of the discrete support, we have the total discrepancy is bounded by $O\left( T^{1-\phi} + T^{(p-1)\phi} (\log{T})^{\frac{3}{2}} \right)$ when $p = 2$, and on the order of $O\left( T^{1-\phi} + T^{(p-1)\phi} \right)$ when $p \geq 3$, because $\prod_{i=1}^{q}m_i$ does not scale with $T$.
\Halmos
\endproof

\proof{Proof of Corollary~\ref{coro:Pigeonhole:p>=2}.}
From Theorem~\ref{thm:Pigeonhole:p>=2}, when $p=2$, the total expected discrepancy is on the order of $O\left( T^{1-\phi} + T^{(p-1)\phi} (\log{T})^{\frac{3}{2}} \right)$; when $p\geq 3$, the total expected discrepancy is on the order of $O\left( T^{1-\phi} + T^{(p-1)\phi} \right)$.
If we choose $\phi$ such that 
\begin{align*}
T^{1-\phi}=T^{(p-1)\phi},
\end{align*}
then we have $\phi=\frac{1}{p}$.
So when $p=2$, the pigeonhole design has an expected discrepancy on the order of $O\left(T^{\frac{p-1}{p}} (\log{T})^\frac{3}{2}\right)$; when $p\geq 3$, the pigeonhole design has an expected discrepancy on the order of $O\big(T^{\frac{p-1}{p}}\big)$. 
\Halmos
\endproof

\section{Proof of Theorem \ref{thm:Pigeonhole:clustered}} \label{sec:Proof:thm:Pigeonhole:clustered}

\proof{Proof of Theorem \ref{thm:Pigeonhole:clustered}}
1. We show that if $\gamma$ is known, selecting $\zeta = \frac{1}{p} + \frac{p-1}{p} \gamma$ can yield an expected discrepancy on the order of $O\big(T^{\frac{p-1}{p}(1-\gamma)}\big)$.

To start, we follow the stopping time analysis in Theorem~\ref{thm:Pigeonhole:p>=2}.
Let $\tau$ be the moment such that, after assigning subject $\tau$, either the control or the treated group reaches $\frac{T}{2}$ in size.
Among the first $\tau$ arriving subjects, we sequentially match subjects at the same pigeonhole in pairs, leaving at most one unmatched subject from each pigeonhole.
Denote the set of matched subjects to be $\mathcal{T}_M$; denote the set of unmatched subjects to be $\mathcal{T}_U$.

First, we focus on the discrepancy generated from $\mathcal{T}_M$.
Since the diameter of each pigeonhole is $c^{\zeta}T^{-\zeta}$, the discrepancy generated from the matched subjects $\mathcal{T}_M$ can be upper bounded by $c^{\zeta}\sqrt{p}T^{-\zeta}\tau \leq c^{\zeta}\sqrt{p} \ T^{1-\zeta} = O\big(T^{\frac{p-1}{p}(1-\gamma)}\big)$.

Next we turn our attention to the discrepancy generated from $\mathcal{T}_U$ and from the last $(T-\tau)$ subjects.
Note that, since each cluster has a diameter $T^{-\gamma}$.
Each cluster can overlap with at most $2^p \cdot \frac{T^{-\gamma}}{c^{\zeta}T^{-\zeta}} = 2^p c^{-\zeta} T^{\zeta - \gamma}$ many pigeonholes.
As a short hand notation, denote $\nu = \zeta - \gamma$.
Due to the proof of Theorem~\ref{thm:Pigeonhole:p>=2}, the discrepancy generated from $\mathcal{T}_U$ and from the last $(T-\tau)$ subjects can be upper bounded by
\begin{align*}
\sqrt{p} (\frac{1}{2c}T^{p\nu})^{\frac{p-1}{p}} + (\frac{1}{2c}T^{p\nu})^{\frac{1}{2}} \left(\log(\frac{1}{2c}T^{p \nu})\right)^\frac{3}{2}+\frac{1}{\sqrt{c}}T^{\frac{p\nu}{2}}\sqrt{\log \left(\frac{1}{c}T^{p\nu} \right)} + \frac{1}{2} T^{\frac{p \nu}{2}}.
\end{align*}
Using Landau notations, the above quantity can be written as
\begin{align*}
O\big(T^{\frac{p-1}{p}(1-\gamma)} \big(\log{T}\big)^\frac{3}{2}\big)
\end{align*}
when $p = 2$, and
\begin{align*}
O\big(T^{\frac{p-1}{p}(1-\gamma)}\big)
\end{align*}
when $p \geq 3$.
Finally, combining both cases, when $p=2$, the total expected discrepancy in the $\gamma$ known case is on the order of $O\big(T^{\frac{p-1}{p}(1-\gamma)} \big(\log{T}\big)^\frac{3}{2}\big)$; when $p \geq 3$, the total expected discrepancy in the $\gamma$ known case is on the order of $O\big(T^{\frac{p-1}{p}(1-\gamma)}\big)$.

2. We show that if $\gamma$ is unknown, selecting $\zeta = \frac{1}{p} + \frac{p-1}{p} \underline{\gamma}$ can yield an expected discrepancy on the order of $O\big(T^{\frac{p-1}{p}(1-\underline{\gamma})}\big)$.

If each cluster has a diameter $T^{-\underline{\gamma}}$, we replace $\gamma$ by $\underline{\gamma}$ in the first claim, and we can get the expected discrepancy on the order of $O\big(T^{\frac{p-1}{p}(1-\underline{\gamma})}\big)$. 
If each cluster has a diameter $T^{-\gamma}$ with $\gamma > \underline{\gamma}$, we have $\bigcup_{n=1}^N \mathcal{C}(\bm{v}_n, T^{-\gamma}) \subset \bigcup_{n=1}^N \mathcal{C}(\bm{v}_n, T^{-\underline{\gamma}})$. 
As the arrival sequence come from a smaller domain which is only a subset of the domain when $\gamma=\underline{\gamma}$, we have the discrepancy is at most on the order of $O\big(T^{\frac{p-1}{p}(1-\underline{\gamma})} \big(\log{T}\big)^\frac{3}{2}\big)$ when $p = 2$; at most on the order of $O\big(T^{\frac{p-1}{p}(1-\underline{\gamma})}\big)$ when $p \geq 3$.
\Halmos
\endproof

\section{Robustness Check: Average Treatment Effect Estimation Using Yahoo! Data}
\label{sec:simu:Robustness}

Since we have specified only one data generating process in Section~\ref{sec:Simulations:ATE}, we conduct more simulations in this section for robustness check.

Similar to the specified data generating process in Section~\ref{sec:Simulations:ATE}, we first build a linear regression model based on the dataset to predict the click probabilities.
But unlike the specification in Section~\ref{sec:Simulations:ATE}, we increase the magnitude of the $10$ largest coefficients in the linear regression model.
Here, we use the $10$ largest coefficients instead the $5$ largest coefficients.

We then use this linear model to generate the click probabilities under ``control,'' which yields a $35.94\%$ probability on average.
We then add $U[0, 35.94\%]$ independent and identically distributed uniform noises to the above probabilities, and generate the click probabilities under ``treatment,'' which yields a $53.80\%$ probability on average.
If a probability is either negative or greater than $1$, we trim the probability number to fall between $[0,1]$.

Finally, we generate independent Bernoulli variables using such probabilities for both control and treatment.
Since we generate both outcomes under control and treatment, we can calculate the average treatment effect as
\begin{align*}
\tau = \frac{1}{T} \sum_{t=1}^T Y_t(1) - \frac{1}{T} \sum_{t=1}^T Y_t(0) = 53.97\% - 36.00\% = 17.97\%.
\end{align*}
Note that, since we have trimmed the probability to be between $[0,1]$, the $53.97\%$ and $36.00\%$ are different from the $35.94\%$ and $53.80\%$, respectively.

\begin{figure}[!tb]
\center
\includegraphics[width=0.7\textwidth]{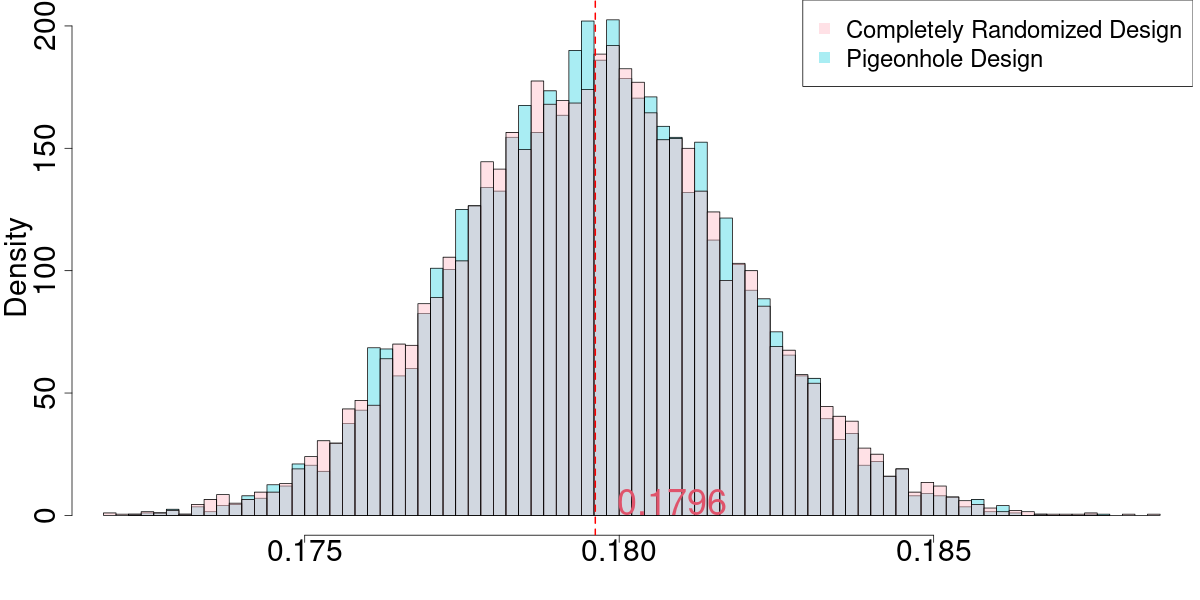}
\caption{Robustness check: empirical distributions of the estimators under two different designs.}
\label{fig:HistogramNew}
\end{figure}

We use the same experimental design and use the same difference-in-means estimator.
Now we compare the performances of $\widehat{\tau}^\mathrm{PhD}$ and $\widehat{\tau}^\mathrm{CRD}$.
We present their empirical distributions in Figure~\ref{fig:HistogramNew}.
In Figure~\ref{fig:HistogramNew}, the pink histogram stands for the distribution of $\widehat{\tau}^\mathrm{CRD}$ while the blue one stands for the distribution of $\widehat{\tau}^\mathrm{PhD}$.
The red vertical line indicates the average treatment effect $\tau$.
Figure~\ref{fig:HistogramNew} shows that the pigeonhole design yields an estimator that is more concentrated around $\tau$ than the completely randomized design.
Quantitatively, $\bE[\widehat{\tau}^\mathrm{CRD}] = \bE[\widehat{\tau}^\mathrm{PhD}] = 17.96\% \approx \tau$. 
So both estimators are unbiased.
Moreover, $\Var[\widehat{\tau}^\mathrm{CRD}] = 4.93 \cdot 10^{-6}$ and $\Var[\widehat{\tau}^\mathrm{PhD}] = 4.62 \cdot 10^{-6}$.
The pigeonhole design reduces $6.25\%$ variance compared to the completely randomized design.

\end{document}